\documentclass[12pt]{article}
\pdfoutput=1

\usepackage{array} 
\usepackage{amssymb}
\usepackage{graphics,graphpap}
\usepackage{graphicx}
\usepackage{color}
\usepackage{graphicx}
\usepackage{dcolumn}
\usepackage{epsfig}
\usepackage{epstopdf}
\usepackage{bbm}
\usepackage{amsmath}
\usepackage{amsfonts}
\usepackage{textcomp}
\usepackage{breqn}
\usepackage{setspace}
\usepackage{slashed}
\usepackage{url}
\usepackage{booktabs}
\usepackage[babel]{inputenc}
\usepackage[parfill]{parskip}
\usepackage{subcaption}
\usepackage{cite}

\newcommand{\equref}[1]{Eq.~\eqref{#1}}

\newcommand{\Figref}[1]{Fig.~\ref{#1}}

\newcommand{\Secref}[1]{Sec.~\ref{#1}}

\newcommand{\Appref}[1]{Appendix~\ref{#1}}

\usepackage{soul}

\newcommand{\sub}[2]{#1_{\mathrm{#2}}} 						
\newcommand{\ssscript}[3]{#1_{\mathrm{#2}}^{\mathrm{#3}}} 	
\newcommand{\ssscriptupper}[3]{#1_{#2}^{\mathrm{#3}}}

\newcommand{\diffd}{\mathrm{d}}													
\newcommand{\dd}[1]{\frac{\mathrm{d}}{\mathrm{d} #1}}							
\newcommand{\DD}[2]{\frac{\mathrm{d} #1}{\mathrm{d} #2}}						

\newcommand{\partiald}[1]{\frac{\partial}{\partial #1}}
\newcommand{\partialdd}[2]{\frac{\partial #1}{\partial #2}}

\newcommand{\invps}[1]{\frac{\diffd ^3 #1}{\left(2\pi\right)^3 2 E_{#1}}}

\newcommand{\Lag}{\mathcal{L}}											
\newcommand{\LagArg}[1]{\mathcal{L}_{\mathrm{#1}}}											
\newcommand{\delslashed}{\slashed{\partial}}
\newcommand{\Msquared}{\left| \mathcal{M} \right|^2}
\newcommand{\orderofunit}[2]{\mathcal{O} \left(#1 \, \unitonly{#2} \right)}	

\newcommand{\av}[1]{\ensuremath{\left\langle #1 \right\rangle}}

\newcommand{\hc}{\mathrm{h.c.}} 						

\newcommand{\deltadist}[1]{\delta \left(#1\right)}
\newcommand{\deltadistn}[2]{\delta^{\left(#1\right)} \left(#2\right)}

\newcommand{\twopin}[1]{\left( 2 \pi \right)^#1}
\newcommand{\erf}[1]{\ensuremath{\mathrm{erf}\left(#1\right)}}

\newcommand{\dmf}[1]{dark matter\footnote{}}

\newcommand{\unit}[2]{#1 \, \mathrm{#2}}
\newcommand{\unitonly}[1]{\mathrm{#1}}

\newcommand{\code}[1]{\textalltt{#1}}
\makeatletter
\newcommand*{\textalltt}{}
\DeclareRobustCommand*{\textalltt}{%
	\begingroup
	\let\do\@makeother
	\dospecials
	\catcode`\\=\z@
	\catcode`\{=\@ne
	\catcode`\}=\tw@
	\verbatim@font\@noligs
	\@vobeyspaces
	\frenchspacing
	\@textalltt
}
\newcommand*{\@textalltt}[1]{%
	#1%
	\endgroup
}
\makeatother

\newcommand{\beq}{\begin{eqnarray}}
\newcommand{\eeq}{\end{eqnarray}}

\newcommand{\bmp}{\noindent\begin{minipage}{16cm}}
\newcommand{\emp}{\end{minipage}\vskip 7mm} 

\def\drawbox#1#2{\hrule height#2pt
        \hbox{\vrule width#2pt height#1pt \kern#1pt
              \vrule width#2pt}
              \hrule height#2pt}

\def\Asym#1#2{\vcenter{\vbox{\drawbox{#1}{#2}
              \kern-#2pt 
              \drawbox{#1}{#2}}}}

\def\simge{\mathrel{%
   \rlap{\raise 0.511ex \hbox{$>$}}{\lower 0.511ex \hbox{$\sim$}}}}

\def\simle{\mathrel{
   \rlap{\raise 0.511ex \hbox{$<$}}{\lower 0.511ex \hbox{$\sim$}}}}

\def\s#1{\setbox0=\hbox{$#1$}%
\rlap{\ifdim\wd0>.7em\kern.22\wd0\else\kern.1\wd0\fi /}#1}

\newcommand{\CHP}{\ensuremath{\sub{\mathcal{C}}{HP}}}
\newcommand{\CGamma}{\ensuremath{\sub{\mathcal{C}}{\Gamma}}}

\usepackage{fancyhdr}
\usepackage[english]{babel}
\usepackage{blindtext}

\begin{document}

\begin{titlepage}
\title{\vspace*{-3.0cm}
\hfill {\small MPP-2015-13}\\[20mm]
\vspace*{-1.5cm}
\bf\Large
keV Sterile Neutrino Dark Matter from Singlet Scalar Decays: Basic Concepts and Subtle Features
\\[5mm]\ }

\author{
Alexander Merle\thanks{email: \tt amerle@mpp.mpg.de}~~~and~~Maximilian Totzauer\thanks{email: \tt totzauer@mpp.mpg.de}
\\ \\
{\normalsize \it Max-Planck-Institut f\"ur Physik (Werner-Heisenberg-Institut),}\\
{\normalsize \it F\"ohringer Ring 6, 80805 M\"unchen, Germany}\\
}
\date{\today}
\maketitle
\thispagestyle{empty}

\begin{abstract}
\noindent
We perform a detailed and illustrative study of the production of keV sterile neutrino Dark Matter (DM) by decays of singlet scalars in the early Universe. In the current study we focus on providing a clear and general overview of this production mechanism. For the first time we study all regimes possible on the level of momentum distribution functions, which we obtain by solving a system of Boltzmann equations. These quantities contain the full information about the production process, which allows us to not only track the evolution of the DM generation but to also take into account all bounds related to the spectrum, such as constraints from structure formation or from avoiding too much dark radiation. In particular we show that this simple production mechanism can, depending on the regime, lead to strongly non-thermal DM spectra which may even feature more than one peak in the momentum distribution. These cases could have particularly interesting consequences for cosmological structure formation, as their analysis requires more refined tools than the simplistic estimate using the free-streaming horizon. Here we present the mechanism including all concepts and subtleties involved, for now using the assumption that the effective number of relativistic degrees of freedom is constant during DM production, which is applicable in a significant fraction of the parameter space. This allows us to derive analytical results to back up our detailed numerical computations, thus leading to the most comprehensive picture of keV sterile neutrino DM production by singlet scalar decays that exists up to now.
\end{abstract}

\end{titlepage}

\section{\label{sec:Intro}Introduction}

Despite great advances in our understanding, our Universe holds mysteries that are yet to be resolved. One of the biggest questions is about the identity of the so-called Dark Matter (DM) which -- in terms of cosmic energy density -- outweighs ordinary matter by a factor of about five~\cite{Ade:2013zuv,Planck:2015xua}. While historically (motivated by, e.g., supersymmetry) our best guess for DM was a Weakly Interacting Massive Particle (WIMP) with a mass of a few hundred GeV and roughly weak interaction strength, by now we have unfortunately not seen a clear signal of such a particle. Even worse, attempts for direct detection disfavour big parts of the parameter space that was deemed ``natural''~\cite{Aprile:2012nq,Akerib:2013tjd,Agnese:2014aze,Angloher:2014myn}. While WIMPs cannot be excluded, we are nevertheless at a point where we should seriously think of alternatives~\cite{Baer:2014eja}.

After all, there are several possibilities left for what DM could be. Alternative ideas range from very light scalars such as axions~\cite{Duffy:2009ig} over non-standard fermions in supersymmetry (such as gravitinos~\cite{Steffen:2006hw} or axinos~\cite{Choi:2013lwa}) up to very heavy exotic options like WIMPzillas~\cite{Kolb:1998ki}. In this work, we concentrate on a candidate motivated by neutrino physics, a sterile neutrino $\nu_s$ of (typically) a mass of a few keV. While our natural guess would be for the sterile neutrino mass to be much larger, it is in reality not bound and there exist several consistent theoretical frameworks in which sterile neutrinos can be very light~\cite{Abazajian:2012ys,Merle:2013gea}. Sterile neutrino DM has originally been proposed by Dodelson and Widrow (DW)~\cite{Dodelson:1993je} who suggested that, although sterile neutrinos would not have interactions strong enough to keep them in thermal equilibrium in the early Universe, they could nevertheless be produced gradually by the thermal plasma due to their admixtures to active neutrinos. Although the DW mechanism was at that time consistent with all bounds~\cite{Colombi:1995ze}, by now we know that it is excluded by structure formation: it produces too hot DM~\cite{Boyarsky:2008xj}. Taking the sterile neutrino mass to larger values is not possible due to the X-ray bound on the DM decay $\nu_s \to \nu + \gamma$, where $\nu$ is any active neutrino (see Ref.~\cite{Canetti:2012kh} for a comprehensive discussion and Ref.~\cite{Merle:2013ibc} for probably the most recent collection of limits).\footnote{In 2014 two groups have independently reported a \emph{tentative} signal of an X-ray line at $3.5$~keV~\cite{Bulbul:2014sua,Boyarsky:2014jta} which would, if interpreted as originating from sterile neutrino decay, indicate a particle with a mass of $7.1$~keV and an active-sterile mixing angle $\theta$ of roughly $\sin^2 (2\theta) \sim 7\cdot 10^{-11}$. However, up to this date there is still a discussion on-going in the literature about whether or not this line is in fact a real signal, see Refs.~\cite{Riemer-Sorensen:2014yda,Anderson:2014tza,Boyarsky:2014ska,Jeltema:2014qfa,Boyarsky:2014paa,Bulbul:2014ala,Malyshev:2014xqa}. Given that the signal, if it exists, is at the moment still not at a statistically highly significant level, we do not take any side here but instead suggest to wait until more data is collected, to hopefully either strongly confirm or clearly refute a line signal.}

Several other production mechanisms for keV sterile neutrino DM have been proposed. If a primordial lepton asymmetry is present in the early Universe, the active-sterile transitions could be resonantly enhanced, as proposed by Shi and Fuller (SF)~\cite{Shi:1998km}. This mechanism produces a relatively cold DM component in addition to the thermal DW part, which leads to a colder spectrum in accordance with all bounds~\cite{Canetti:2012kh,Abazajian:2014gza,Popa:2015eta} -- however, note that also this mechanism is in some tension with data if the $3.5$~keV line is taken seriously~\cite{Merle:2014xpa}. In theories with an extended gauge group, the sterile neutrinos could equilibrate and be produced via generic freeze-out, if the resulting overabundance is corrected by a sufficient amount of entropy production~\cite{Bezrukov:2009th,Nemevsek:2012cd} -- however this scenario, even though not fully excluded, is on quite general grounds threatened by big bang nucleosynthesis~\cite{King:2012wg}.

An alternative is non-thermal production of keV sterile neutrinos via particle decays. This possibility is discussed extensively in the literature, with the decaying particle in most cases being a scalar: variants range from inflaton decay~\cite{Shaposhnikov:2006xi,Bezrukov:2009yw} over a general scalar singlet that could itself either freeze-out~\cite{Kusenko:2006rh,Petraki:2007gq} or freeze-in~\cite{Merle:2013wta,Adulpravitchai:2014xna,Kang:2014cia} to the case where the scalar is electrically charged~\cite{Frigerio:2014ifa}. More general possibilities exist as well, for example the production from pion decays~\cite{Lello:2014yha}, from Dirac fermions~\cite{Abada:2014zra}, or from light vector bosons~\cite{Boyanovsky:2008nc,Shuve:2014doa}. A benefit of this mechanism is that the velocity spectrum of the keV neutrinos produced by decays is quite generally known to have a tendency to be colder than that from other mechanisms~\cite{Merle:2014xpa,Petraki:2007gq,Merle:2013wta,Petraki:2008ef,Bezrukov:2014qda}.

While several cases of decay production are discussed, the treatments available in the literature involve some crude estimates and approximations. Even though a non-thermal DM spectrum could have interesting and unexpected consequences for cosmological structure formation, many references work on the level of rate equations and only \emph{estimate} the spectrum, if at all. Instead, given that the \emph{distribution function} is the most decisive quantity and that it can be computed from first principles with reasonable effort, its determination should be put on more solid grounds. This is partially done for the production of keV neutrinos by a general singlet scalar entering thermal equilibrium~\cite{Petraki:2007gq}, however, only in the approximation where the effective number of relativistic degrees of freedom $g_*$ is constant. While this approximation is not usable in all of the parameter space, it is nevertheless a good approximation for large enough production temperatures and, after all, without this assumption it would not be possible to obtain analytical estimates. Yet, ultimately a fully numerical treatment will be necessary.

In this paper, we will start this endeavour by giving a comprehensive discussion of the production of keV sterile neutrino DM via the decays of general singlet scalars, which themselves are produced via a Higgs portal. As shown in Refs.~\cite{Petraki:2007gq,Merle:2013wta,Adulpravitchai:2014xna}, depending on the coupling there exist different regimes where the scalar itself could e.g.\ be a WIMP or a feebly interacting massive particle and decay in or out of thermal equilibrium. We will generalise the previous treatments and derive the full set of approximate formulas for all regimes possible, but again under the assumption of $g_*=$ const. We furthermore perform a fully numerical study of the allowed regions in the parameter space and obtain distribution functions for all relevant parameter points, which in principle allows us to determine all relevant DM properties for a given point in the parameter space. We determine the regions where the correct DM abundance is obtained for a given sterile neutrino mass, thereby taking into account all relevant bounds.

Structure formation properties of DM candidates can be estimated using the so-called free-streaming horizon $\lambda_{\rm FS}$. With this quantity, we show that different estimates are possible which may lead to quite different results. We clearly illustrate that $\lambda_{\rm FS}$, in spite of being a standard estimator, can lead to inconsistent results depending on the approximations used to calculate it.

The current paper serves as an illustration of the general principles and its purpose is \emph{to give a clear overview of the relatively complicated and subtle details involved}. As such, some aspects lie beyond the scope of the current work. For example, \emph{we neglect active-sterile mixing} and thus completely disregard DW production, which in reality cannot be switched off for non-zero active-sterile mixing (and which on the contrary is desired for the related X-ray signal), in favour of analytical results. Our approximation is still good for very small active-sterile mixing, but it is nevertheless too simplified and will be dropped (as well as the assumption $g_*=$const.) in a future purely numerical investigation of all possible cases~\cite{Merle:Proj:RefinedScalarProd&DW}. Also the full set of implications for cosmological structure formation cannot be obtained using $\lambda_{\rm FS}$, so that a numerical computation of the structure formation properties is needed. While we could in principle add this to the current work, it would lead away from our main point and furthermore prolong the paper even more, so that we have decided to decouple this study, too~\cite{Merle:Proj:StructureFormation}. Our goal is to ultimately deliver a fully comprehensive study of scalar decay production of keV sterile neutrino DM, in order to set the stage to discriminate it from alternative mechanisms by future data. Current bounds indicate that this may be possible in the not-too-far future~\cite{Merle:2014xpa}, so that this endeavour should be pursued in particular if the $3.5$~keV line signal survives.

This paper is organised as follows. We start with an illustrative discussion in Sec.~\ref{sec:illustrative}, which is supposed to give an overview of our considerations and results at a relatively non-technical level. We then introduce the main equations to be solved in Sec.~\ref{sec:KineticEquations}. In Sec.~\ref{sec:AnalyticalResults}, we present analytical approximations for all cases where they can be obtained. After a discussion of the aspects relevant for cosmic structure formation, Sec.~\ref{sec:StructureFormation}, we finally present the numerical results of our study in Sec.~\ref{sec:Results}, along with a discussion of all bounds and of the validity of our considerations. We conclude in Sec.~\ref{sec:Conclusions}. Throughout the paper we try to keep the discussion on a minimally technical level, however, all technical details relevant for an inclined reader are exposed in Appendices~A and~B.

\section{\label{sec:illustrative}Illustrative discussion of the setting and overview of the results}

This section mainly serves the purpose of describing in simple terms what we have done in our study. Before entering the technical details, we give an overview not only of the setting we are working in but also of some illustrative results. This will hopefully make the paper more accessible and prevent the reader from getting lost in the technical details.

Our basic setting consists of a real singlet scalar field $S$ which must somehow couple to the right-handed neutrino fields $N$. The most generic coupling doing this job is a Yukawa term $\frac{y}{2} S\overline{N^c}N$ with coupling strength $y$ which, if the scalar develops a non-zero vacuum expectation value (VEV) $\langle S \rangle$, leads to a Majorana mass $m_N=y\left\langle S \right \rangle$, where we have assumed only one generation of right-handed neutrinos for simplicity. Thus, the minimal set of ingredients we need in addition to the standard model (SM) consists of exactly these fields. Our minimal Lagrangian is
\begin{equation}
 \Lag = \LagArg{SM} + \left[i\overline{N} \delslashed N + \frac{1}{2}\left(\partial_\mu S\right)\left(\partial^\mu S\right) - \frac{y}{2}S\overline{N^c}N +\hc \right] - \sub{V}{scalar} + \Lag_\nu \,,
 \label{eq:ModelLagrangian}
\end{equation}
which is very similar to what had been used previously~\cite{Petraki:2007gq,Merle:2013wta,Adulpravitchai:2014xna}. Here, $\Lag_\nu$ denotes the part of the Lagragian giving mass to active neutrinos. The details of this mass generation are in fact not very important for our DM production mechanism, as is the number of fermion generations. However, in the current work, we make the additional assumption of \emph{vanishing active-sterile mixing}. In most settings, there will be couplings between active and sterile neutrinos that in reality cannot be switched off. But, since in this paper we want to present analytical results wherever possible, we take this mixing to be exactly zero and in turn have \emph{no DM production from the DW mechanism}.\footnote{Given that the astrophysical X-ray bounds push the active-sterile mixing down to tiny values~\cite{Canetti:2012kh,Merle:2013ibc}, this approximation is not necessarily bad. However, an additional contribution from the DW production can modify some of the results obtained here, which is why we will drop this assumption in future works and turn to a purely numerical treatment~\cite{Merle:Proj:RefinedScalarProd&DW,Merle:Proj:StructureFormation}.} The scalar potential $\sub{V}{scalar}$ takes the most general form compatible with an assumed global $\mathbb{Z}_4$-symmetry:\footnote{For possible issues related to the breaking of this symmetry by a non-vanishing VEV $\av{S}$, see \cite{Merle:2013wta} and references therein. Note also that giving up the $\mathbb{Z}_4$ symmetry allows for extra terms in the Lagrangian, resulting in more processes that can contribute to equilibrating the scalar. In this paper, we stick to the symmetry in order to obtain a minimal parameter space for our exploratory analysis. Considerations of taking into account terms linear and cubic in $S$ in the potential can be found in \cite{Petraki:2007gq}.}
\begin{equation}
 \sub{V}{scalar} = - \mu_H^2 H^\dagger H - \frac{1}{2} \mu_S^2 S^2 + \lambda_H \left(H^\dagger H\right)^2 + \frac{\lambda_S}{4} S^4 +2\lambda \left(H^\dagger H\right)S^2 \,.
 \label{eq:ScalarPotentialLagrangian}
\end{equation}
This potential could easily lead to a VEV $\left \langle S \right \rangle \approx \unitonly{GeV}-\unitonly{TeV}$, thus motivating a relatively small Yukawa coupling $y\sim 10^{-9}$--$10^{-5}$ in order for the mass of the sterile neutrino to be in the $\unitonly{keV}$-range.

Not only can the above Yukawa coupling lead to a sterile neutrino mass, but it will also be responsible for sterile neutrino DM production. Several processes could be thought of: the leading contributions are the reactions $SS\leftrightarrow hh$ and $S\rightarrow NN$, the first of which couples the scalar field $S$ to the thermal plasma and the second of which produces sterile neutrinos from the decay of $S$. In principle, also processes like $SS\rightarrow NN$ would be possible, but the corresponding rate is proportional to $y^4$ which is negligible for a sufficiently small Yukawa coupling $y$. We also neglect the inverse reaction $NN \rightarrow S$ which is suppressed due to the heavily suppressed phase space originating in the kinematics of any $2$-to-$1$ process.

Let us add that, in general, there will be mixing between the scalar $S$ and the SM Higgs field after electroweak symmetry breaking, which we have completely ignored. However, given that this mixing is proportional to the generally small Higgs portal coupling $\lambda$ and that it is also suppressed if the singlet scalar is considerably heavier than the Higgs, which is the case that we are investigating here (cf.\ Sec.~\ref{sec:AnalyticalResults}), taking into account the mixing would not at all change our results. Note that this would change if one used singlet scalar masses very close to or below the Higgs mass~\cite{Adulpravitchai:2014xna}, which is why this simplifying assumption must be dropped if we are to extend our considerations to lighter singlets. On the other hand, in that limit we would need to drop further assumptions used in this work (in particular the assumption that $g_*$ is constant), so that it makes sense to postpone these considerations to future work~\cite{Merle:Proj:RefinedScalarProd&DW}.

With these assumptions, it is clear that every scalar present in the early Universe will either decay into two sterile neutrinos or undergo pairwise annihilation into pairs of Higgs bosons. Depending on the exact values of the couplings, there are different regimes possible. For example, if the Higgs portal coupling  $\lambda$ is small enough and the initial abundance\footnote{We usually use the term \emph{abundance} to denote a particle number density. Depending on the context, the term \emph{relic abundance} will be used for the particle number density or the corresponding energy density after the process discussed is complete.} of the scalars is zero, then the scalar will never enter thermal equilibrium but it will only be produced occasionally from the plasma. In a more modern language, this mechanism would be called \emph{freeze-in}~\cite{McDonald:2001vt,Hall:2009bx}, and the corresponding particle would be called a \emph{feebly interacting massive particle} (FIMP). In this regime, the annihilation into Higgs bosons can be neglected since its reaction rate will be suppressed by the square of the (tiny) scalar density. Hence, the frozen-in abundance of scalars will ultimately be translated into a relic abundance of sterile neutrinos with a particle number just twice the one of the scalars at freeze-in (in a co-moving volume) -- irrespective of the size of the Yukawa coupling $y$ as long as it is large enough that all scalars have decayed by now. Another example would be the case where the scalar couples to the Higgs strongly enough to be in thermal equilibrium. Then the argument of doubling the number of particles would still be valid if the decay width of the scalar is sufficiently small for the decay to become effective only after freeze-out. However, if the decay proceeds already while the scalar is in equilibrium, further contributions will be present making the whole picture more complicated. We will discuss all cases in detail in \Secref{sec:AnalyticalResults}, where we present analytical estimates for the momentum distribution functions $f$ and yields $Y$ wherever possible.

\begin{figure}[p]
 \vspace{-2.5cm}
 \hspace{-0.5cm}
 \includegraphics[width=18cm,keepaspectratio]{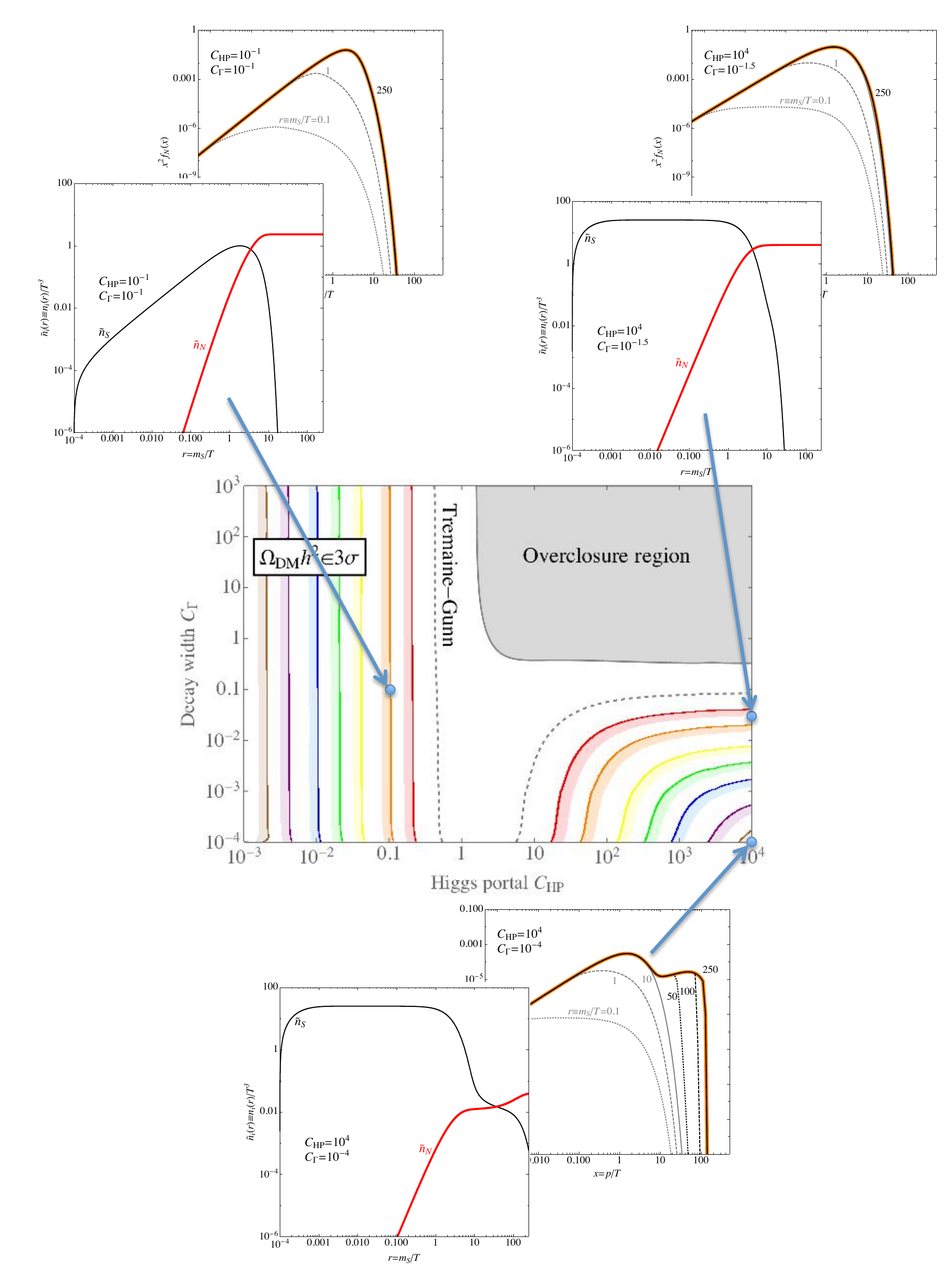}
 \caption{\label{fig:ClosureRegionsCGammaCHP}Lines of correct abundance with example distributions and evolutions.}
\end{figure}

We later on turn to a numerical computation of the DM production. To arrive at the final DM relic abundance, we need to solve a system of Boltzmann equations to compute the momentum distribution function $f(p,t)$ of the DM particle. Ultimately this distribution function contains all relevant information: one can e.g.\ use it to compute the final DM relic abundance, but it also encodes the information about the evolution of the momentum spectrum with time, i.e., how many particles exist per momentum interval at any given temperature of the Universe. This allows to not only track the course of DM generation in detail, but it furthermore gives information about the velocities of the DM particles at the epochs relevant for the formation of structures (i.e., galaxies and galaxy clusters) in space.

To give a snapshot of the results to be presented, we show in Fig.~\ref{fig:ClosureRegionsCGammaCHP} a plot of the lines of correct DM abundance for several different sterile neutrino masses, in a plane of the parameters $\mathcal{C}_{\rm HP}$ and $\mathcal{C}_\Gamma$ which, as we will explain later, are nothing else than an effective Higgs portal coupling and an effective decay width in convenient units. We have augmented the plot by some example evolutions of the DM production and the underlying distribution functions for some specific parameter points marked in the central figure, in order to illustrate what is behind our computations. The purpose of Fig.~\ref{fig:ClosureRegionsCGammaCHP} is to give a graphical illustration of what will be presented in this paper. All the plots displayed will be discussed in great detail in Sec.~\ref{sec:Results}, where also the terminology, colour-codes, and labelling will be carefully outlined so that, while advancing with the paper, the reader will ultimately be enabled to get a full understanding of Fig.~\ref{fig:ClosureRegionsCGammaCHP}.

\section{\label{sec:KineticEquations}The kinetic equations in the early Universe}

The fundamental embodiment of every particle species in the early Universe is their distribution function $f (\vec{p}, t )$ in momentum space. This quantity does contain all the information we need to deduce the cosmological impact of the species under consideration, so that the determination of $f$ is paramount. Due to isotropy, we will always assume that the distribution functions $f$ only depend on the moduli of the associated $3$-momenta. To compute a distribution function, we need to solve the corresponding Boltzmann equation:
\begin{equation}
\hat{L} \left[f\right] = C\left[f\right].
\label{eq:BoltzmannAbstract}
\end{equation}
The left-hand side of this equation contains the so-called Liouville operator $\hat{L}$:
\begin{equation}
 \hat{L} = \partiald{t} - H p \partiald{p} \,,
 \label{eq:LiouvilleFRW}
\end{equation}
where $p$ is the modulus of the particle's 3-momentum and $H$ is the Hubble function. The collision term $C\left[f\right]$ on the right-hand side can be interpreted as a source term, encoding the interaction of the species of interest with itself and with the other particle species present in the plasma. This term contains all the information about the processes which contribute to the production of the species under consideration, and which accordingly shape the resulting momentum distribution function. Collision terms can be relatively lengthy, which is why we report the explicit form of $C\left[f\right]$ only in \Appref{app:DetailsKineticEquations}. In this section, however, we prefer to give a more intuitive explanation of how to obtain the distribution functions of the sterile neutrinos.

Since DM production happens in the very early Universe, we only need to consider the era of radiation dominance. During this epoch, the Hubble function can be written as $H\left(T\right) = T^2 / M_0 $, where $T$ is the temperature of the plasma and $M_0$ is a function that implicitly depends on time via the evolution of the degrees of freedom $g_*$:
\begin{equation}
 M_0 = \left(\frac{45 \sub{M}{Pl}^2}{4 \pi^3 g_*} \right)^{1/2} = \unit{7.35 g_*^{-1/2} \times 10^{18}}{GeV} \,.
 \label{eq:DefM0}
\end{equation}
Introducing dimensionless variables, 
\begin{equation}
 x \equiv p/T\ \ \ \ {\rm and}\ \ \ \ r \equiv m_S/T \,,
 \label{eq:variables}
\end{equation}
the Liouville operator from \equref{eq:LiouvilleFRW} can be recast into the following form:
\begin{equation}
 \hat{L}= \DD{r}{T} \DD{T}{t} \partiald{r} +  H x \left(\frac{\frac{r}{\sqrt{g_*}} \dd{r}\sqrt{g_*} }{1-\frac{r}{\sqrt{g_*}} \dd{r}\sqrt{g_*} } \right) \partiald{x} \,.
 \label{eq:LiouvilleOperator_r_x}
\end{equation}

Throughout this work, we will stick to the assumption that the numbers of effective degrees of freedom ($g_*,g_{*S}$) are \emph{constant until the production of sterile neutrinos is completed}, as done e.g.\ in~\cite{Petraki:2007gq,Merle:2013wta}. This assumption is \emph{absolutely necessary to obtain analytical results} and, in fact, it is not a too bad approximation in a large fraction of the relevant parameter space, as we will illustrate in \Secref{subsec:AssessingEvolutionDOF}. Nevertheless we will drop it in later works where we are going to present more realistic and purely numerical studies~\cite{Merle:Proj:RefinedScalarProd&DW,Merle:Proj:StructureFormation}.

If the number of effective degrees of freedom does not change during the period of interest, the advantage of the variables $x$ and $r$ becomes obvious. Accordingly, the dynamics of the scalar and the sterile neutrino are given by the following set of equations:
\begin{align}
 \partialdd{f_S}{r} &= \DD{T}{r} \DD{t}{T} \left(C_{hh\rightarrow SS}^{S}+C_{SS\rightarrow hh}^{S}+C_{S\rightarrow NN}^{S} \right)\,,  \label{eq:DynamicsScalar} \\
 \partialdd{f_N}{r} &= \DD{T}{r} \DD{t}{T} \;  C_{S\rightarrow NN}^{N} \,.
 \label{eq:DynamicsSN}
\end{align}
In Eqs.~\eqref{eq:DynamicsScalar} and~\eqref{eq:DynamicsSN}, the upper indices on the collision terms mark the species the kinetic equation of which is governed by this term, while the subscripts describe the actual process. Thus, $C_{SS\rightarrow hh}^{S}$ describes the depletion of scalars due to annihilations into pairs of Higgses while $C_{hh\rightarrow SS}^{S}$ describes the reverse process. In turn, $C_{S\rightarrow NN}^{S}$ describes the depletion of scalars due to decays and $C_{S\rightarrow NN}^{N}$ encodes the creation of sterile neutrinos from the decays of scalars. Note that the collision terms contain information about the kinematics, too, so that $C_{S\rightarrow NN}^{N}$ and $C_{S\rightarrow NN}^{S}$ differ by more than just a sign. For a detailed derivation and explicit expressions, see \Appref{app:DetailsKineticEquations}.

Since we approximate $g_*$ as constant during the time of production and since the matter-radiation equality only takes place at temperatures of $\orderofunit{1}{eV}$, i.e., long after DM production is completed, we consistently use the time-temperature relation $\DD{T}{t} = -HT$ in \equref{eq:DynamicsSN}. Using this as well as the explicit form of the collision terms as derived in \Appref{subapp:DetailsKineticEquationsScalar}, we find the kinetic equation for the scalar:
\begin{dmath}
 \partialdd{f_S\left(x,r\right)}{r} = \frac{1}{\sqrt{x^2+r^2}} \left[\frac{1}{4\pi} \CHP \exp \left(-\sqrt{x^2+r^2}\right) \mathcal{F}\left(x,r\right) - \CGamma r^2 f_S\left(x,r\right) -\frac{1}{4\pi} \CHP f_S\left(x,r\right) \,  2 \pi \int\limits_0^\infty \diffd \hat{x} \, \hat{x}^2 \int\limits_{-1}^{\sub{\alpha}{max}} \diffd \cos \theta f_S\left(\hat{x},r\right) \mathcal{G} \left(\hat{x},\cos \theta,r\right)\right] \\
 \equiv \mathcal{Q}\left(x,r\right) - \mathcal{P}\left(x,r\right) f_S -\mathcal{R}\left(x,r\right) \, \mathcal{I}_r\left[f_S\right] \, f_S \,,
\label{eq:KineticEquationScalarComplete}
\end{dmath}
where we have defined
\begin{align}
\mathcal{Q}\left(x,r\right) &\equiv \frac{\CHP\exp \left(-\sqrt{x^2+r^2}\right) \mathcal{F}\left(x,r\right) }{4\pi \sqrt{x^2+r^2}}, \label{eq:DefQ} \\
\mathcal{P}\left(x,r\right)&\equiv \frac{\CGamma r^2}{\sqrt{x^2+r^2}} \label{eq:DefP}, \\
\mathcal{R}\left(x,r\right) &\equiv \frac{\CHP}{4\pi \sqrt{x^2+r^2}} \label{eq:DefR}, \\
\mathcal{I}_r\left[f_S\right] &\equiv 2\pi \int\limits_0^\infty \diffd\hat{x}\, \hat{x}^2 \int\limits_{-1}^{\sub{\alpha}{max}} \diffd \cos \theta f_S\left(\hat{x},r\right) \mathcal{G}\left(\hat{x},\cos \theta,r\right).
\end{align}
For the explicit forms of the kinematic functions $\mathcal{F}$ and $\mathcal{G}$ and the definition of $\sub{\alpha}{max}$, see \Appref{app:DetailsKineticEquations}.

In Eqs.~\eqref{eq:KineticEquationScalarComplete} to~\eqref{eq:DefR}, we have used two important dimensionless auxiliary quantities:
\begin{equation}
 \left\{
 \begin{matrix}
 \textrm{\emph{the effective decay width}:} \hfill \hfill \hfill & & & \CGamma \equiv \frac{M_0}{m_S}\frac{\Gamma}{m_S}\,, \hfill \hfill \hfill\\
 \textrm{\emph{the effective (squared) Higgs portal}:} & & & \CHP \equiv \frac{M_0}{m_S} \frac{\lambda^2}{16\pi^3} \,.
 \end{matrix}\right.
 \label{eq:DefCGammaAndCHP}
\end{equation}
In the remainder of this work it will turn out convenient to use these quantities to span the parameter space of our setting. Note that, during the DM production process, we assume $M_0$ to be constant by virtue of \equref{eq:DefM0}. Hence the interpretations of $\CHP$ as an effective Higgs portal coupling and of $\CGamma$ as an effective decay width are appropriate, in the sense that the dependence of $M_0$ on $g_*$ does not change their behaviour and hence they indeed play practically the same roles as the fundamental Lagrangian parameters behind them. For simplicity, we may often refer to \CHP\ as \emph{Higgs portal} and to \CGamma\ as \emph{decay width}.

Turning to the sterile neutrino distribution function, which is our main quantity of interest, we can use the assumption of neglecting the back-reaction $NN\rightarrow S$ to find a very simple form for the corresponding kinetic equation. Using Eqs.~\eqref{eq:DynamicsScalar} and~\eqref{eq:DynamicsSN}, see \Appref{subapp:DetailsKineticEquationsSN} for details, one can derive a very intuitive \emph{master equation} for the distribution function of the sterile neutrino in terms of that of the scalar:
\begin{align}
 f_N \left(x,r\right)=\int_{0}^{r}{\diffd r' \, 2 \CGamma \frac{{r'}^2}{x^2} \int_{\sub{\hat{x}}{min}}^{\infty}{\diffd \hat{x}\, \frac{\hat{x}}{\sqrt{\hat{x}^2 + {r'}^2}} f_S\left(\hat{x},r'\right)}}
 \label{eq:KineticEquationSNDoubleIntegral} \,,
\end{align}
with $\sub{\hat{x}}{min} = \left\|x-{r'}^2/\left(4x\right)\right\|$. It is this equation that will be absolutely central for us: once we have managed to determine the distribution function $f_S$ of the scalar from first principles, we only need to plug it into Eq.~\eqref{eq:KineticEquationSNDoubleIntegral} to determine the distribution function $f_N$ of the sterile neutrino. As to be expected, the distribution function $f_S$ of the scalar will look differently depending on which regime we are looking at (e.g., the scalar being an early decaying WIMP compared to a late one). This will be translated by Eq.~\eqref{eq:KineticEquationSNDoubleIntegral} into different resulting sterile neutrino distribution functions. Note that, transforming the momentum variable into an energy, Eq.~\eqref{eq:KineticEquationSNDoubleIntegral} perfectly coincides with \cite[Eq.~(8)]{Shaposhnikov:2006xi} and \cite[Eq.~(10)]{Petraki:2007gq}. Note also that the master equation for the sterile neutrino distribution is decoupled from the one for the scalar only because of our assumption that the reaction $NN\rightarrow S$ is negligible.

\section{\label{sec:AnalyticalResults}Analytical results for the distribution functions and relic abundance}

In this section, we present the limiting cases that can be treated analytically. Some of the results can be found in the literature, see in particular Ref.~\cite{Petraki:2007gq}, while others are completely new. Basically there exist two main cases, the scalar itself can either enter thermal equilibrium in the early Universe and thus act similarly to a WIMP or it can be only very feebly interacting, thus undergoing freeze-in like a FIMP:
\begin{enumerate}

\item The \emph{FIMP-regime}: This limiting case is characterised by a small Higgs portal $\CHP$ and an (almost) arbitrary decay constant $\CGamma$. The scalars that are produced via freeze-in subsequently decay into sterile neutrinos. However, for the abundance itself the time of the decay is not very relevant as long as it happens before matter-radiation equality, which is why the exact value of $\CGamma$ does not matter much in this regime.\footnote{Note that, for this regime, we will always assume a zero initial abundance for the scalars (and trivially for the sterile neutrinos). If there was a non-negligible initial abundance, this would change our results since the primordial scalars would then add to the ones produced via freeze-in. However, given that there is no reason for such an initial abundance to be present and that we do not see much value in speculating how it could possibly have been produced, we stick to the conservative viewpoint and only produce scalars from the freeze-in mechanism itself.\\ On the other hand, one could argue that sterile neutrinos and/or singlet scalar fields could quite generically couple to the inflaton field (see Refs.~\cite{Shaposhnikov:2006xi,Bezrukov:2009yw} for examples concerning the former case). In such scenarios, assuming that inflation is the correct theory in the first place, our scenario might be modified considerably. However, such couplings are only compulsory if the SM gauge group and all other low-energy symmetries are the only ones up to very high scales and even then, from a model building point of view, there exist various possibilities to strongly suppress certain couplings, e.g.\ by locating the various fields on different branes. While our considerations could in principle be extended to include this point, this would add further complications of which it is however unclear whether they exist, or not. We thus stick to the most minimal setting and disregard any primordial abundance of sterile neutrinos and/or the singlet scalar, as well as the related coupling to the inflaton field.} It will, however, play a role for the spectrum itself, cf.\ Sec.~\ref{sec:Results}.

\item The \emph{WIMP-regime}: For large enough Higgs portals $\CHP$ the scalar will thermalise, i.e., it enters thermal equilibrium and any information about its initial abundance is lost since the abundance will always be the thermal one. The particle remains in equilibrium until the interaction rate for the process $hh\leftrightarrow SS$ drops below the expansion rate of the Universe and the scalar decouples from the thermal bath. During the course of these events, depending on the exact value of the decay width $\CGamma$, the scalar can decay into sterile neutrinos at various stages:

\begin{enumerate}
    
\item \emph{In-equilibrium decay}: If the decay width $\CGamma$ is large enough and if the Higgs portal $\CHP$ is large, too, the sterile neutrinos are produced already very early from the decays of equilibrated scalars. The sterile neutrinos arising from the decays of the few residual scalars that are present after freeze-out can then practically be neglected since the large Higgs portal guarantees a small relic abundance of frozen-out scalars.

\item \emph{Out-of-equilibrium decay}: The scalar couples to SM particles strongly enough to enter thermal equilibrium. However, if the decay width is sufficiently small, the production of sterile neutrinos during the time of equilibrium is negligible and it is sufficient to only take into account the late decays of the scalars after their freeze-out. In this regime, the scalar itself acts as a DM-like species, but it is unstable and decays before it can significantly contribute to the energy budget of the Universe.

\item \emph{Intermediate regime}: For intermediate values of the decay width \CGamma, neither of the above limiting cases is applicable and it is a priori not clear how well the combination of both of them can correctly describe the situation. On the other hand, this case can be of particular interest since it results into a distribution function with \emph{two} intrinsic momentum scales. This may open up interesting possibilities to tackle the well-known small scale problems of cosmological structure formation~\cite{Klypin:1999uc,BoylanKolchin:2011dk,Navarro:1996gj,Ferrero:2011au}. We will therefore treat an explicit example for this case numerically in \Secref{subsec:FreeStreamingFailingExample}.

\end{enumerate}
\end{enumerate}

We will now discuss these different regimes one by one. In all cases, we will take the limit $\xi \equiv m_h/m_S \rightarrow0$, cf.\ Appendix~\ref{subapp:DetailsKineticEquationsScalar}. Even for $\xi \approx 0.3$, the net effect on the distribution function is of the order of a few per mille. We want to keep the mass $m_S$ of the scalar far from the electroweak (EW) phase transition to ensure that scalars $S$ are only produced by Higgs bosons. For masses $m_S < \sub{v}{EW}$, a new range of interaction becomes important like the production of EW gauge bosons from scalars $S$, $SS \leftrightarrow VV$. This scenario is discussed in~\cite{Adulpravitchai:2014xna} on the level of abundances, and we will take these modifications into account on a further more technical study where the focus will be put on even more realistic numerical results~\cite{Merle:Proj:RefinedScalarProd&DW}.

\subsection{\label{subsec:DeepFIMPLimit}The FIMP-regime}

For sufficiently small Higgs portals $\CHP$ (i.e., for $\lambda\ll 10^{-6}$~\cite{Petraki:2007gq,Merle:2013wta}), the scalar never enters equilibrium and hence one can -- for vanishing initial abundance -- neglect in \equref{eq:KineticEquationScalarComplete} the term $\mathcal{R}\left(r,x\right) \, \mathcal{I}_r\left[f_S\right] \, f_S$  which scales as $f_S^2$. The remaining kinetic equation is an ordinary differential equation which can be solved analytically. For a vanishing initial abundance of the scalar, the resulting distribution function is given by:
  \begin{equation}
   f_S\left(r,x\right) = \CHP \int_{0}^{r} \diffd \rho\ \rho K_1\left(\rho\right) \frac{\exp\left(-\sqrt{\rho^2 +x^2}\right)}{\sqrt{\rho^2 +x^2}} \left[\frac{e^{\rho \sqrt{\rho^2+x^2}}}{e^{r \sqrt{r^2+x^2}}} \left(\frac{\rho + \sqrt{\rho^2+x^2}}{r + \sqrt{r^2+x^2}}\right)^{-x^2}\right]^{\mathcal{C}_\Gamma/2} \,,
  \label{eq:ScalarDistributionFunctionDeepFIMP_xi0}
  \end{equation}
where $K_1$ is the first modified Bessel function of second kind, cf.~\equref{eq:KinematicIntegral_hhToSS_xi0}. Plugging \equref{eq:ScalarDistributionFunctionDeepFIMP_xi0} into \equref{eq:KineticEquationSNDoubleIntegral} in order to get an analytical result for the distribution function of the sterile neutrino is not very instructive, since there is no simple form of that expression. However, the abundance of the sterile neutrino can be computed analytically for late times, $r\rightarrow \infty$. Setting $\CGamma$ to zero corresponds to a stable scalar. With this choice, \equref{eq:ScalarDistributionFunctionDeepFIMP_xi0} can be integrated rather easily and one obtains the (hypothetical) relic abundance of the stable scalar. Since all frozen-in scalars will however ultimately decay into two sterile neutrinos for a non-vanishing decay width \CGamma, the abundance of sterile neutrinos will then just be twice the abundance of the would-be-stable scalar~\cite{Merle:2013wta}. The result for the yield $Y = n/s$, with $n$ and $s$ the particle number and entropy densities, respectively, is given by
\begin{equation}
 Y_N\left(r\rightarrow \infty\right) = \frac{135}{64\pi^2} \frac{\CHP}{g_*\left(\sub{T}{prod}\right)} \,.
 \label{eq:YieldSterileNeutrinoDeepWIMP_xi0}
\end{equation}
Here, $\sub{T}{prod}$ denotes the temperature at the time of production.\footnote{Of course the time of production is subject to some ambiguities in its definition. Both freeze-in/freeze-out of the scalar and its subsequent decay are continuous processes, the time scales of which are determined by \CHP\ and \CGamma. It is hence convenient to define the production time as the point when the abundance of sterile neutrinos has passed some threshold fraction of the final abundance, which we take to be $95$\%.}

\subsection{\label{subsec:WIMPLimit}The WIMP-regime}

\paragraph{In-equilibrium-decay}
If both the Higgs portal \CHP\ and the decay width \CGamma\ are large, sterile neutrinos are efficiently produced from the decays of scalars already while being in equilibrium. This case is covered at length in~\cite{Petraki:2007gq}. However, our analytical expression differs by a constant factor, which is why we will sketch the most important steps needed to derive the result. If the scalar is in equilibrium, we know its distribution function exactly. Accordingly, the authors of~\cite{Petraki:2007gq} use a Bose-Einstein (BE) distribution to capture the quantum nature of the scalar. Since the whole set of equations governing the dynamics was however derived using the Maxwellian approximation in the Boltzmann equation, one might wonder whether it would be more consistent to again use a Maxwell-Boltzmann (MB) distribution for the scalar. We will exemplify both cases in order to illustrate that the difference between them is irrelevant.

Plugging the BE and MB distributions into~\eqref{eq:CollisionTerm_SToNN_N} yields:
\begin{equation}
 f_N\left(x,r\right) = \begin{cases}
                         8 \CGamma \int\limits_{1}^{z_r}{\diffd z\, x \sqrt{z-1} \log\left(\frac{1}{1-e^{-x z}}\right)}\quad \text{(BE)} \\
                         8 \CGamma \int\limits_{1}^{z_r}{\diffd z\, x \sqrt{z-1} e^{-xz}} = \frac{e^{-x}\sqrt{\pi}\, \erf{\sqrt{x\left(z_r-1\right)}}}{2\sqrt{x}} -e^{-x z_r}\sqrt{z_r-1}\quad \text{(MB)}
                        \end{cases}\,,
 \label{eq:SterileNeutrinoDistributionDeepWIMP}
\end{equation}
where we have introduced the variable $z_r\equiv r^2/\left(4x^2\right)+1$ for convenience.\footnote{Note that, as we had already mentioned, we have neglected the mixing between the two physical scalars, which is a very good approximation in our case. However, in order to simplify the comparison of our results to the ones obtained in Ref.~\cite{Petraki:2007gq}, it is of course necessary to apply the same approximation to the results from that reference.} Integrating $f_N\left(x,r\right)$ over $\diffd^3 x$ and again taking the limit $r\rightarrow \infty$ allows to calculate the yield for late times:
\begin{equation}
 Y_N\left(r\rightarrow \infty\right) = \begin{cases}
                                   \frac{135}{4\pi^3}\zeta\left(5\right) \frac{\CGamma}{g_*\left(\sub{T}{prod}\right)} \quad \text{(BE)}\\
                                   \frac{135}{4\pi^3} \frac{\CGamma}{g_*\left(\sub{T}{prod}\right)} \quad \text{(MB)}
                                  \end{cases}\,.
 \label{eq:SterileNeutrinoYieldDeepWIMP}
\end{equation}
Both results only differ by a factor of $\zeta\left(5\right)\approx 1.0369$, which justifies the use of either distribution. Our result in the BE case is however larger by a factor of $5/2$ compared to the one reported in~\cite{Petraki:2007gq}. While one may easily forget powers of two in these computations, we could not trace any step where a factor of $5$ could possibly be introduced, making us confident that our results are correct.

\paragraph{Out-of-equilibrium decay}
This limiting case describes a scenario where the scalar is in equilibrium and ultimately freezes out. The decay width \CGamma~is so small that practically no sterile neutrinos are produced before the scalar decouples from the plasma. Only after the scalar has frozen-out, the production of the sterile neutrinos sets in. As in~\cite{Petraki:2007gq}, we make the approximation that the scalar has a thermal distribution until it freezes out instantaneously at $r=\sub{r}{FO}$. In this case the kinetic equation, \equref{eq:KineticEquationScalarComplete}, can be solved: 
\begin{equation}
 f_S\left(x,r>\sub{r}{FO}	\right) = f_{\rm eq}\left(x,\sub{r}{FO}\right) \left(\frac{r + \sqrt{r^2 +x^2}}{\sub{r}{FO} + \sqrt{\sub{r}{FO}^2+x^2}} \right)^{\CGamma x^2/2} e^{-\CGamma \left(r \sqrt{x^2 +r^2} - \sub{r}{FO} \sqrt{x^2 +\sub{r}{FO}^2}\right)/2} \,,
 \label{eq:ScalarDistributionFunctionEarlyFreezeOut2}
\end{equation}
where $f_{\rm eq}\left(x,\sub{r}{FO}\right)$ is the equilibrium distribution of $S$ at freeze-out. It could again be taken to be BE or, more consistently, MB. In the case of BE, our expression coincides with \cite[Eq.~(43)]{Petraki:2007gq}. The final abundance of sterile neutrinos can in this limiting case again be calculated from doubling the abundance of scalars at freeze-out. Given as a function of $\sub{r}{FO}$, the expression for the yield is
\begin{equation}
 Y_N\left(r\rightarrow \infty\right) = \frac{45}{4\pi^4 g_*\left(\sub{T}{prod}\right)} \int\limits_{\sub{r}{FO}}^{\infty}{\diffd \epsilon\, \epsilon\frac{\sqrt{\epsilon^2-\sub{r}{FO}^2}}{e^\epsilon-\delta}} \,\,\ \left( =  \frac{45 \sub{r}{FO}^2 K_2(\sub{r}{FO})}{4\pi^4 g_*\left(\sub{T}{prod}\right)} \textrm{  for MB}\right)\,,
  \label{eq:SterileNeutrinoYieldOutOfEquilibrium}
\end{equation}
with $\delta=1$ ($\delta=0$) for the BE (MB) case. Also here the numerical difference between both versions is fairly small for realistic values of $\sub{r}{FO}$.

\paragraph{Intermediate regime}
If $\CGamma$ is in an intermediate regime, neither the production \emph{before} nor the one \emph{after} freeze-out completely dominates the sterile neutrino distribution function, such that no simple analytical treatment is possible. In this instant we have to rely on a purely numerical treatment. We will present a sample case for this regime in \Secref{sec:Results}. Nevertheless this intermediate regime may be of special interest as it features \emph{two} intrinsic scales.

\subsection{\label{subsec:abundance}Calculating the relic abundance}

So far we have shown formulae to compute the yield at $r\rightarrow \infty$. To convert this into the commonly quoted closure parameters, we first have to multiply the yield by today's entropy density of $s_0 =2891.2~{\rm cm}^{-3}$~\cite{Agashe:2014kda} and then compare the DM energy density today to the critical density of the Universe. We can write
\begin{align}
\sub{\Omega}{DM} h^2 = \frac{m_N Y_N (r\to \infty) s_0}{\sub{\rho}{crit}/h^2} \,,
\label{eq:ClosureParameter}
\end{align}
where $\sub{\rho}{crit}/h^2 = \unit{1.054 \times 10^{-2}}{MeV\ cm^{-3}}$~\cite{Agashe:2014kda}. With this conversion formula at hand, it is straightforward to transform the analytical estimates for the yield $Y_N$ in Secs.~\ref{subsec:DeepFIMPLimit} and~\ref{subsec:WIMPLimit} into expressions for the DM relic abundance, which can then be compared to the observed $1\sigma$ range obtained by the Planck collaboration, $\sub{\Omega}{DM} h^2 = 0.1188 \pm 0.0010$~\cite{Planck:2015xua}.

\section{\label{sec:StructureFormation}Aspects of structure formation}

Up to now, we have only been concerned with the DM relic abundance, i.e., the \emph{amount} of DM in the Universe. However, for a viable DM candidate it is not only important to be sufficiently abundant in the Universe, but it must also allow for structures such as galaxies to form. Clearly, this imposes constraints on the momentum distribution function $f(p,t)$ of the DM candidate under consideration: it must not be hot, i.e., it is not allowed to be too relativistic at the time of structure formation when galaxies form (more precisely, the allowed amount of hot DM is at most a tiny $1$\% of all DM in the Universe~\cite{Abazajian:2004zh,dePutter:2012sh}).

The form of the distribution function $f(p,t)$, i.e.\ the spectrum of the DM candidate, can be constrained from the observed matter distribution in the cosmos: the evolution of spatial inhomogeneities depends on the spectrum of the DM particles and therefore it ultimately constrains the DM production mechanism. Since it is computationally impossible to run a simulation of structure formation for any possible distribution function, a commonly used indicator that can be calculated easily for a given $f(p,t)$ is the so-called \emph{free-streaming horizon} $\lambda_{\rm FS}$. This quantity describes the average distance a DM particle would have travelled after production without collisions and not subject to gravitational clustering. In fact this quantity can serve as a good estimator for a length below which the formation of structures is heavily suppressed. For that reason, the free-streaming horizon $\sub{\lambda}{FS}$ is commonly used in the literature to discard DM models with a spectrum that is too hot to explain the filamentary structure of the large scale matter distribution in the Universe by preventing galaxy-sized objects of being formed.

\subsection{\label{subsec:DefinitionFreeStreaming}Treatment of the free-streaming horizon}

In this section, we will show that the free-streaming horizon itself is subject to substantial uncertainties in its definition, which will make clear why we later present some of our results in two different versions. Moreover we will demonstrate that even our simple one-component DM model can produce highly non-thermal momentum distribution functions, which may even feature \emph{two} intrinsic momentum/velocity scales. In such a case, the average momentum (and hence the free-streaming horizon) cannot be expected to lead to sensible conclusions.

The free-streaming horizon is defined by~\cite{Boyarsky:2008xj}:
\begin{equation}
 \sub{\lambda}{FS} \equiv \int_{\sub{T}{prod}}^{T_0}{\frac{\av{v\left(T\right)}}{a\left(T\right)} \DD{t}{T} \diffd T} \,,
 \label{eq:DefRFS}
\end{equation}
where $\sub{T}{prod}$ is the temperature at which production can be seen as complete (in the sense discussed before) and $T_0$ is today's temperature. Here, $\av{v\left(T\right)}$ is the average velocity of the sterile neutrinos that can be calculated from the distribution function:
\begin{align}
	\av{v\left(T\right)} = \frac{\int\limits_0^\infty \diffd x\ \frac{x^3}{\sqrt{x^2 + r^2 \left(m_N/m_S\right)^2}} f_N\left(x,r\right)}{\int_0^\infty \diffd x\ x^2 f_N\left(x,r\right)} \,.
	\label{eq:DefVMean}
\end{align}
From \equref{eq:DefVMean} it becomes clear that $\av{v}$ converges to unity, i.e., to the speed of light, as long as $f_N$ is concentrated around values of $x = p/T \gg \sqrt{r^2 m_N^2/m_S^2} = m_N/T$, just as expected. Since $f_N\left(x\right)$ does not change after the production process is complete, the factor of $r^2 m_N^2/m_S^2$ in the square root increases. Note also that, once $r^2 m_N^2/m_S^2$ is greater than the value(s) of $x$ around which $f$ is concentrated, \equref{eq:DefVMean} converges to the non-relativistic expression, $\av{v} \rightarrow T/m_N \av{x} \equiv \av{p}/m_N$. 
  
For our numerics, we follow the approximations usually found in the analytical approach~\cite{Boyarsky:2008xj}, namely we will assume the Universe to be completely radiation dominated until some temperature $T_{\rm eq}$ (``matter-radiation equality''), where the Universe switches to being completely matter dominated. The last epoch of vacuum dominance is irrelevant: even though this period dominates the past of the Universe on an absolute time scale (the matter-vacuum equality being located at a redshift of $z=\Omega_\Lambda/\sub{\Omega}{m}\approx 2.2$, corresponding to a time of roughly $3\times 10^9~{\rm a}$~\cite{Carmeli:2005if}), the velocities of the sterile neutrinos in this epoch are so tiny that the resulting contribution to the free-streaming horizon is negligible. Commonly, the evolution of the degrees of freedom (d.o.f.)~is implemented by an additional dilution factor of $\xi^{1/3} = [g_{*S}(\sub{T}{prod})/g_{*S}(T_0)]^{1/3}\approx \left(106.75/3.36\right)^{1/3}\approx 3.17$, by which $\sub{\lambda}{FS}$ is rescaled to account for entropy dilution (cf.~\cite{Petraki:2007gq,Merle:2013wta}). Note that, although we approximate $g_{*S}\approx {\rm const.}$ during DM production, we nevertheless have to take into account the entropy dilution until today, since there is no justification for the above assumption to be valid through the entire history of the Universe. Departing from the above approximation also during DM production would modify the dilution factor $\xi^{1/3}$, but not too drastically because of the presence of the third root. We will later on perform an estimate of the validity of our approximation, cf.~\Secref{subsec:AssessingEvolutionDOF}. Note that, however, in this formalism the dependence on $g_{*S}\left(\sub{T}{prod}\right)$ is quite mild, such that it is safe to use the SM number of d.o.f.\ $g_{*S}=106.75$ even though the new particles contribute as well to some degree, depending on how strongly suppressed their true distribution functions are compared to a thermal one.
  
There is, however, an issue with the analytical approach. The above treatment does not capture the physical fact that, in reality, the evolution of the d.o.f.\ also enters in the time-temperature relation inside the integral in Eq.~\eqref{eq:DefRFS}, but this can be taken into account rather easily in a numerical evaluation of the integral. We therefore compute $\lambda_{\rm FS}$ in a second (numerical) version, in order to take the full evolution of the d.o.f.\ into account. Thereby, our numerics uses a set of fitted analytical formulae for the evolution of the d.o.f.~\cite{PhysRevD.82.123508}. For more technical details on this numerical integration, see \Appref{app:DetailsrFS}.

In order to compare to the results already present in the literature, we will follow both approaches in parallel. To get an idea about whether the DM can be classified as cold, warm, or hot for a certain set of parameters, we take $\sub{\lambda}{FS} \stackrel{!}{=} \unit{0.1}{Mpc}$ to mark the boundary between hot and warm DM. This choice is relatively common in the literature (see, e.g., Refs.~\cite{Colin:2000dn,Lin:2000qq,Viel:2005qj,Das:2010ts}), and it corresponds to the size of a typical dwarf satellite galaxy, which yields a sensible physical motivation. The boundary between warm and cold DM in turn is smooth but it is clear that $\sub{\lambda}{FS}$ should be considerably smaller in this case, so that we choose $\sub{\lambda}{FS} \stackrel{!}{=} \unit{0.01}{Mpc}$, i.e., one order of magnitude smaller than for the hot/warm boundary. Of course there is some arbitrariness involved in these choices, but given that the free-streaming horizon in itself can only yield an indication, the actual error introduced is not as serious as it may seem at first sight. In general the free-streaming horizon can only serve as an order-of-magnitude estimate, and it clearly should not be used to prematurely discard unclear results. Ultimately, only more advanced computations of structure formation can assess whether scenarios with borderline free-streaming horizons should be discarded, or not~\cite{Merle:Proj:StructureFormation}. For recent developments beyond the commonly used free-streaming approach, see e.g.~\cite{Schneider:2014rda}.

\subsection{\label{subsec:FreeStreamingFailingExample}Free-streaming horizon failing -- an explicit example featuring twin peaks}

As mentioned before, even our simple one-component DM model can feature a distribution function with \emph{two} intrinsic scales, namely in the case where the relic abundance of sterile neutrinos is produced from the decay of equilibrated scalars and the decay of frozen-out scalars in comparable amounts. We present here the exemplary case of $\CHP=10^4$ and $\CGamma = 10^{-3.5}$, which yields the correct relic abundance for a (relatively large) sterile neutrino mass of about $\unit{73}{keV}$. Fixing the scalar mass to be $\unit{1}{TeV}$ and the number of degrees of freedom at high tempertures to $g_*=106.5$, this would correspond to values of the Lagrangian couplings of $\left(y, \lambda \right) = \left(4.7 \times 10^{-9}, \, 8.3 \times 10^{-5}\right)$.

\begin{figure}[t]
   \centering
   \includegraphics[width= 0.6 \textwidth]{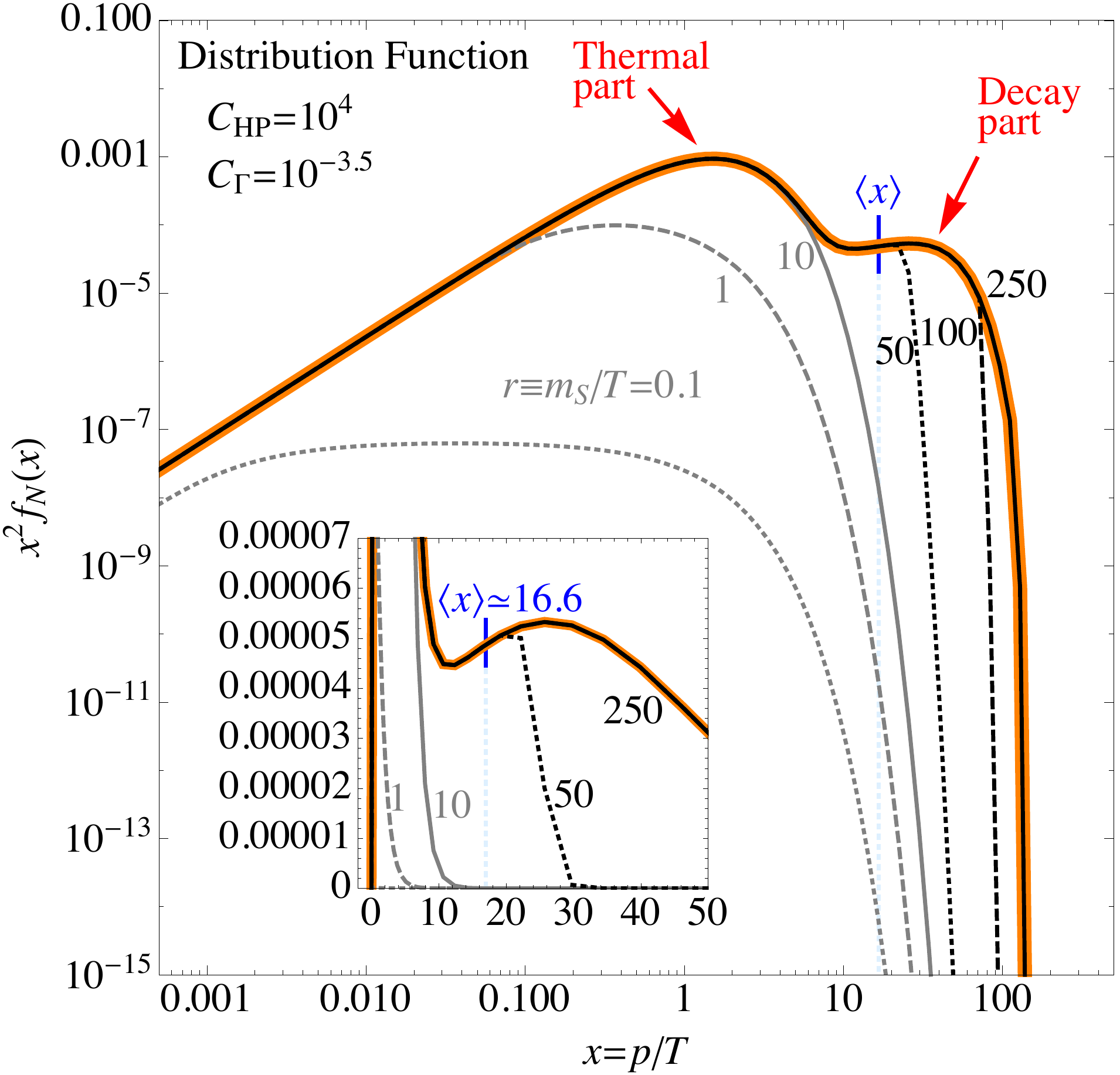}
   \caption{Example of the evolution of a distribution function of sterile neutrinos. One can clearly distinguish two momentum scales (global maximum at $x_1\approx1.5$ and second, local maximum at $x_2\approx26$). The mean comoving momentum $\left\langle x\right\rangle$ is located at $\left\langle x\right\rangle\approx 16.6$. The standard deviation $\sqrt{\left\langle x^2\right\rangle-\left\langle x\right\rangle^2}$ is approximately $26$, which illustrates that the mean value $\langle x \rangle$ contains practically no information.}
   \label{fig:DoublePeakExample}
\end{figure}

\Figref{fig:DoublePeakExample} shows the distribution function of the sterile neutrino for different values of the time parameter $r$. One can clearly see how early production from the plasma populates the lower comoving momenta (dubbed \emph{Thermal part} -- although the resulting distribution may not be perfectly thermal), while the late contributions mainly originate from the decay of the frozen-out scalars (\emph{Decay part}). The inset in the plot shows the enlarged region between the two peaks.

It is obvious that in this case the  average momentum does not at all capture the characteristics of the distribution function. According to our estimates using $\lambda_{\rm FS}$ and the average value $\langle x \rangle \equiv \av{p}/T \simeq 16.6$, this point in the parameter space corresponds to a scenario where the sterile neutrinos are on the borderline between being hot and being warm (cf.~\Secref{sec:Results}). However, in fact the low momentum (``thermal'') part with $\langle x \rangle_{\rm low} \approx 2.5$ contributes practically as strongly as the high momentum (``decay'') part with $\langle x \rangle_{\rm high} \approx 35.7$, where in both cases we have approximated the respective peaks with the individual results from Eqs.~\eqref{eq:SterileNeutrinoDistributionDeepWIMP} and~\eqref{eq:ScalarDistributionFunctionEarlyFreezeOut2}, respectively. Note that this splitting introduces some numerical uncertainties, since the expression in \equref{eq:ScalarDistributionFunctionEarlyFreezeOut2} is quite sensitive to small deviations in $\sub{r}{FO}$, which in turn suffers from some arbitrariness in the exact definition (due to freeze-out being a process with a small but finite temporal extent rather than an immediate effect).

Estimating the two corresponding free-streaming horizons, cf.\ Eq.~\eqref{eq:DefRFS},
\begin{align*}
 \left(\ssscript{\lambda}{FS,thermal}{numerical} ,\ssscript{\lambda}{FS,thermal}{estimate} \right) &\sim \unit{\left(0.05,0.01\right)}{Mpc}\,,\\
\left(\ssscript{\lambda}{FS,decay}{numerical} ,\ssscript{\lambda}{FS,decay}{estimate} \right) &\sim \unit{\left(0.7,0.1\right)}{Mpc} \,,
\end{align*}
the left peak tends to yield cold/warm DM while the right one would indicate hot DM in both cases. The full distribution function corresponds to $\left(\ssscript{\lambda}{FS,total}{numerical} ,\ssscript{\lambda}{FS,total}{estimate} \right) \approx \unit{\left(0.27,0.05\right)}{Mpc}$, which is perfectly in between the two individual estimates and consistently indicates a case at the borderline of warm/hot DM. Even though the splitting into two distinct parts introduces some extra numerical uncertainty with respect to the values for the complete distribution function, these values clearly illustrate the issue with using the free-streaming horizon as an estimator.

This is precisely one of the cases where more detailed studies about structure formation have to be performed in order to obtain a final answer. With the simple estimate of the free-streaming horizon alone, such a scenario should not be prematurely discarded. In any case, it is worthwhile noting that a one-component DM model can have two similarly important momentum scales in its distribution function, which may open up new possibilities to tackle the small-scale problems from structure formation simulations.

\section{\label{sec:Results}Results and bounds}

In this section we will present our detailed results and we will also address possible bounds on the scenario as well as the validity of our considerations.

\subsection{\label{subsec:ResutlsAbundance}The Dark Matter abundance}

\begin{figure}[t]
 \centering
 \includegraphics[width=0.8 \textwidth]{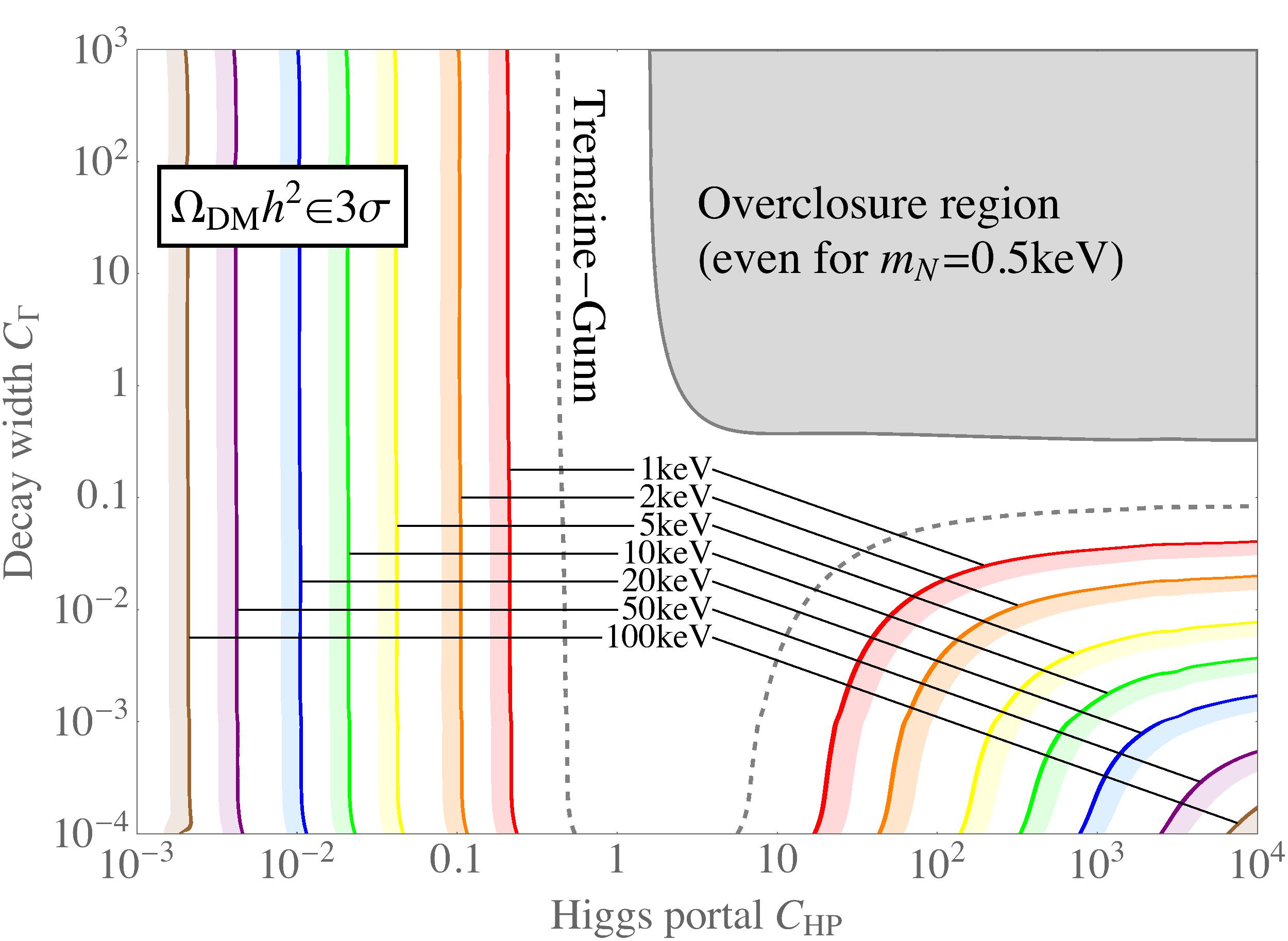}
 \caption{\label{fig:Abundances}Lines of correct abundance (dark colours) and of sizeable but insufficient abundance (faint colours) in the $\CHP$-$\CGamma$ plane for different values of the sterile neutrino mass. In addition, the Tremaine-Gunn and overclosure bounds are displayed, see text for details.}
\end{figure}

Let us first discuss the DM abundance, which is displayed in \Figref{fig:Abundances} (similarly to the one shown in \Secref{sec:illustrative}). In this plot, the dark coloured bands mark the regions in the $\CHP$-$\CGamma$ plane where a sterile neutrino of a given mass yields an abundance in accordance with the $3 \sigma$ range from the Planck 2015 data~\cite{Planck:2015xua}. For example, the dark red lines correspond to a sterile neutrino with a mass of $m_N = 1$~keV. Here, the lines on the left correspond to the FIMP regime while the ones on the right correspond to the WIMP regime. We also mark, by the neighbouring fainter lines, the regions where a sizeable but not sufficiently large abundance is generated.

There are two important bounds displayed in \Figref{fig:Abundances}. Let us start with the so-called Tremaine-Gunn (TG) bound~\cite{Tremaine:1979we}, which is based on the idea that any collection of identical fermions must have a certain minimum phase space density. Applying this bound to the observed dwarf satellite galaxies leads to a lower bound of roughly $m_N > 0.5$~keV on the sterile neutrino mass~\cite{Boyarsky:2008ju}. Thus, smaller masses are ultimately forbidden by the Pauli exclusion principle. In our plots this bound is marked by the gray dashed line around the white area which in particular cuts into the parameter space for values of \CHP\ close to one. It basically indicates where the relic density for $m_N = 0.5$~keV equals the upper $3\sigma$ bound~\cite{Planck:2015xua}. The second bound comes from the fact that the DM should not ``overclose'' the Universe, i.e., its energy density fraction $\Omega_{\rm DM}$ should be smaller than one. Since in the figures we marked the lines of correct abundance for different masses $m_N$, the resulting forbidden regions do in fact also depend on the mass of the DM particle. However, given that there is a model-independent lower value for the mass from the TG bound, at least for this smallest mass of $m_N = 0.5$~keV the overclosure region marks an absolute bound. Not too surprisingly, this forbidden region is smaller than that excluded by the TG bound, and in our plots it is marked by the gray patches in the upper right corners.

\begin{figure}[t]
 \centering
 \begin{tabular}{lr}
 \includegraphics[width=8cm]{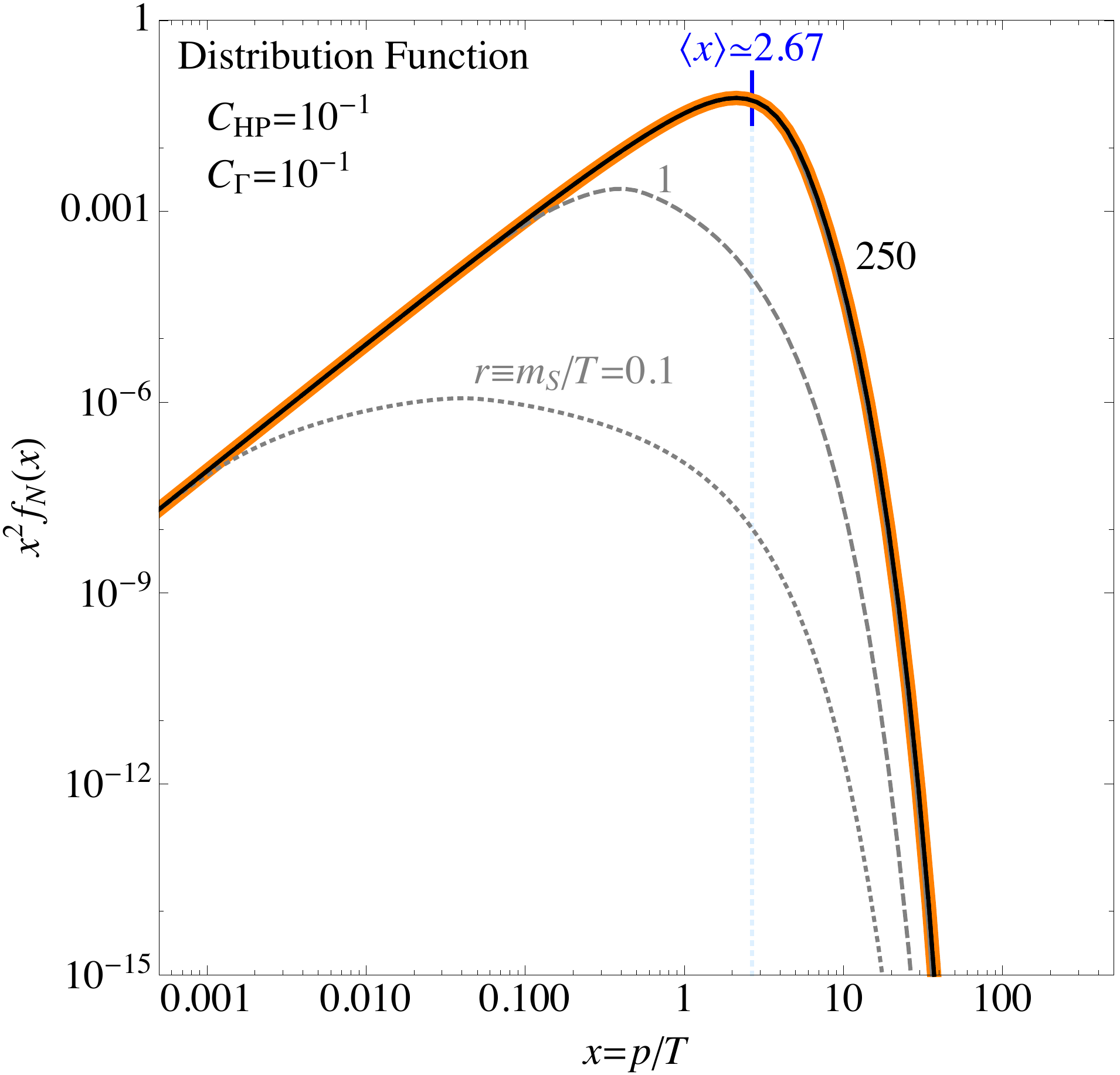} & \includegraphics[width=7.8cm]{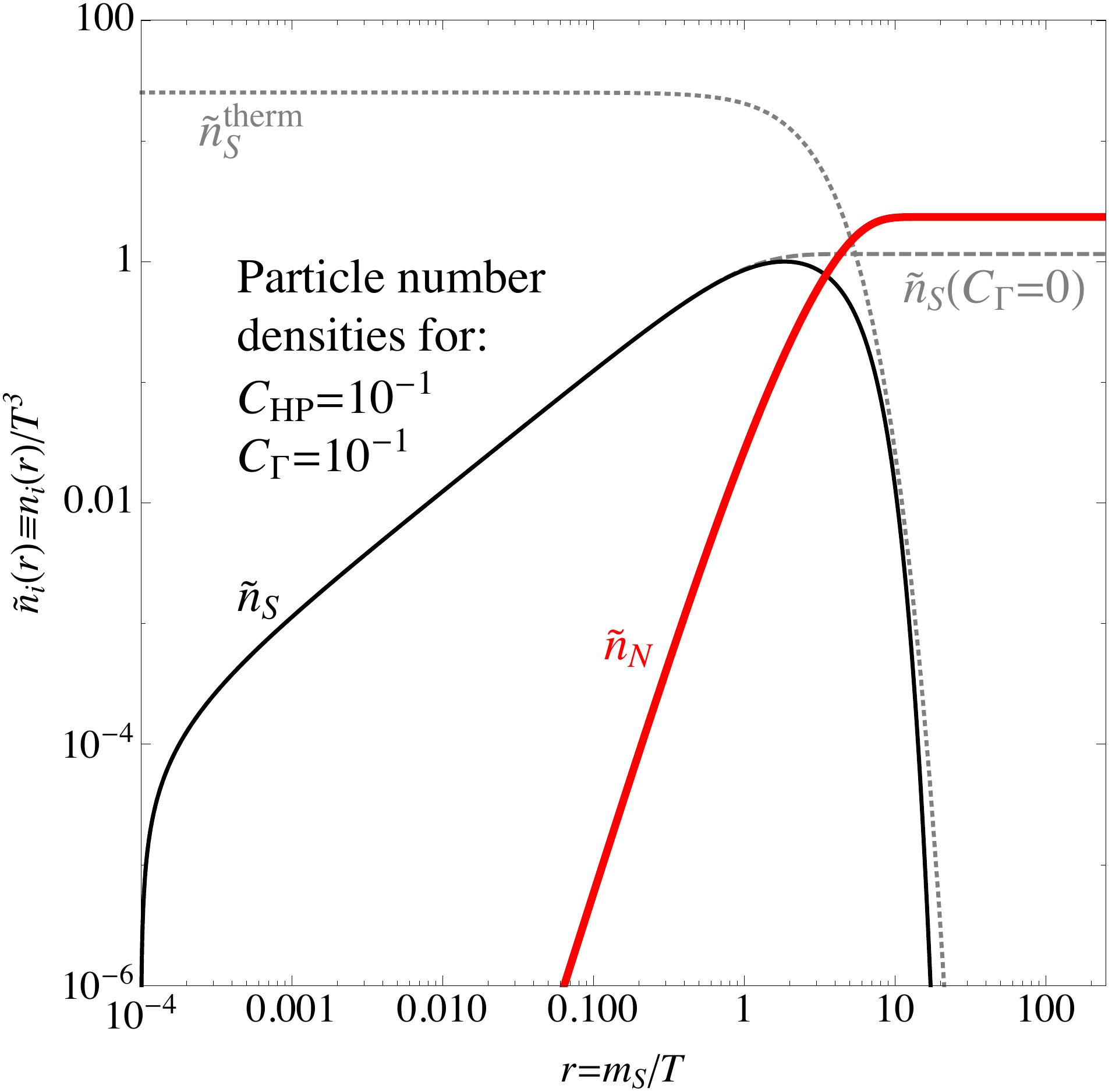}
 \end{tabular}
 \caption{\label{fig:snapshot_FIMP}The evolutions of the distribution function (left) and of the abundances (right) for a point corresponding to the scalar being a FIMP.}
\end{figure}

As already indicated in Secs.~\ref{sec:illustrative} and~\ref{sec:AnalyticalResults}, depending on the exact values of the parameters different regimes are possible. Let us discuss a few numerical examples. Starting with the FIMP case, in Fig.~\ref{fig:snapshot_FIMP} we illustrate an example point $(\CHP,\CGamma) = (10^{-1},10^{-1})$ which corresponds to this regime. Taking again the benchmark value $m_S = \unit{1}{TeV}$ and using $g_*=106.5$ at high temperatures, the effective couplings translate into $\left(y, \lambda \right) = \left(8.4 \times 10^{-8}, \, 2.6 \times10^{-7}\right)$ on the Lagrangian level. On the left panel, we depict the evolution of the sterile neutrino distribution function $f_N(r)$ with the time parameter $r$. As one can see, most of the abundance is produced around the time $r \sim 1$. Soon afterwards the production ceases such that even for very late times, the distribution hardly changes (as soon as $r\sim 10$, the distribution is practically identical to the final one). The distribution exhibits a clear peak whose maximal value is very close to the mean momentum over temperature, $\langle x \rangle \simeq \langle p/T \rangle \simeq 2.67$. This means that this sterile neutrino distribution is \emph{colder} than a thermally produced one, for which this number would be equal to $3.15$~\cite{Shaposhnikov:2006xi}. However, this point nevertheless turns out to be in the hot DM region, cf.\ Sec.~\ref{subsec:ResultsFreeStreaming}. This is mainly due to the fact that this point in parameter space requires a mass of the sterile neutrino of $m_N \approx \unit{2}{keV}$ in order to fulfil the relic abundance constraint. This low mass in turn leads to a long time of highly relativistic free-streaming.

The evolution of the abundances $\tilde n_i$, i.e., the integrals over the distribution function divided by $T^3$, is depicted on the right panel of Fig.~\ref{fig:snapshot_FIMP}. We display both the abundances $\tilde n_S$ of the scalar and $\tilde n_N$ of the sterile neutrino. Starting with the scalar (solid black line) we can see that, as to be expected from a generic FIMP, the abundance of the scalar is gradually built up with increasing time parameter $r$. However, it never reaches a thermal abundance, as can be seen by comparing the black curve to the hypothetical one for a scalar in thermal equilibrium (dotted gray line). If the scalar was stable, its abundance would not anymore change after the freeze-in is completed, cf.\ dashed gray line, and it would in practice act as some type of DM. However, given that the scalar is unstable, once it does decay its abundance decreases and instead a sizeable abundance $\tilde n_N$ of sterile neutrinos is built up (red solid line). Indeed, because of each scalar decaying into two sterile neutrinos, the final abundance of sterile neutrinos is exactly twice the one of the would-be-stable scalar for late times (numerically we obtain $\tilde n_N(r=250)/\tilde n_S(\CGamma =0, r=250) \simeq 2.03$, in excellent agreement with the expectation). Furthermore, we can use Eq.~\eqref{eq:YieldSterileNeutrinoDeepWIMP_xi0} to cross-check our numerics, and both results agree nearly perfectly with each other, within a deviation of only 2.8\% in this case. Note that this value is \emph{not} a measure of the quality of our numerical methods since we do not know a priori in which part of the parameter space the analytical results approximate the exact result to a desired accuracy. We expect the deviation to become smaller as $\CGamma$ further decreases. In fact, on the edge of our parameter space where $\CGamma = 10^{-4}$, the analytical result is reproduced with deviations well below $1\%$. 

\begin{figure}[th]
 \centering
 \begin{tabular}{lr}
 \includegraphics[width=8cm]{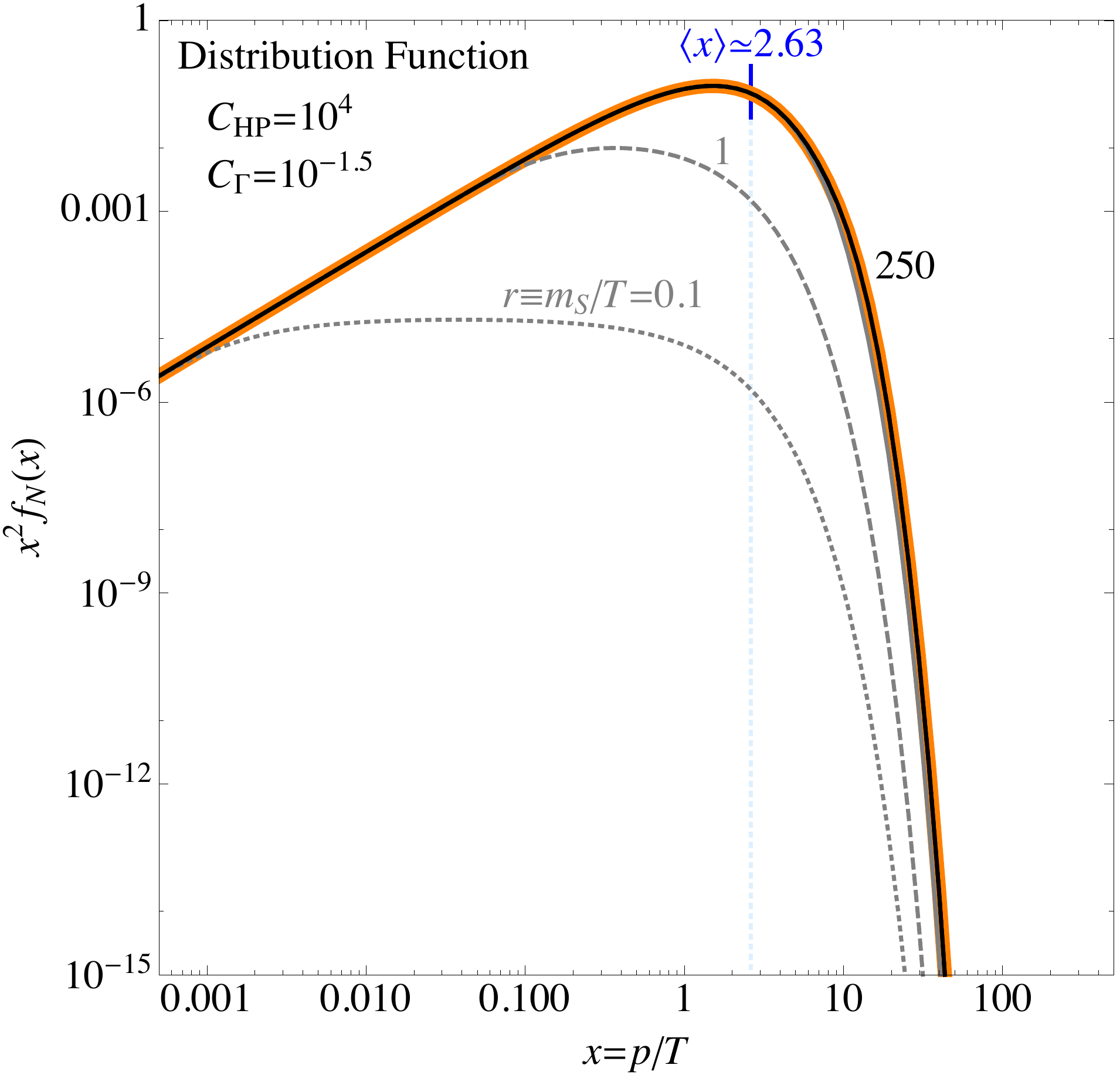} & \includegraphics[width=7.8cm]{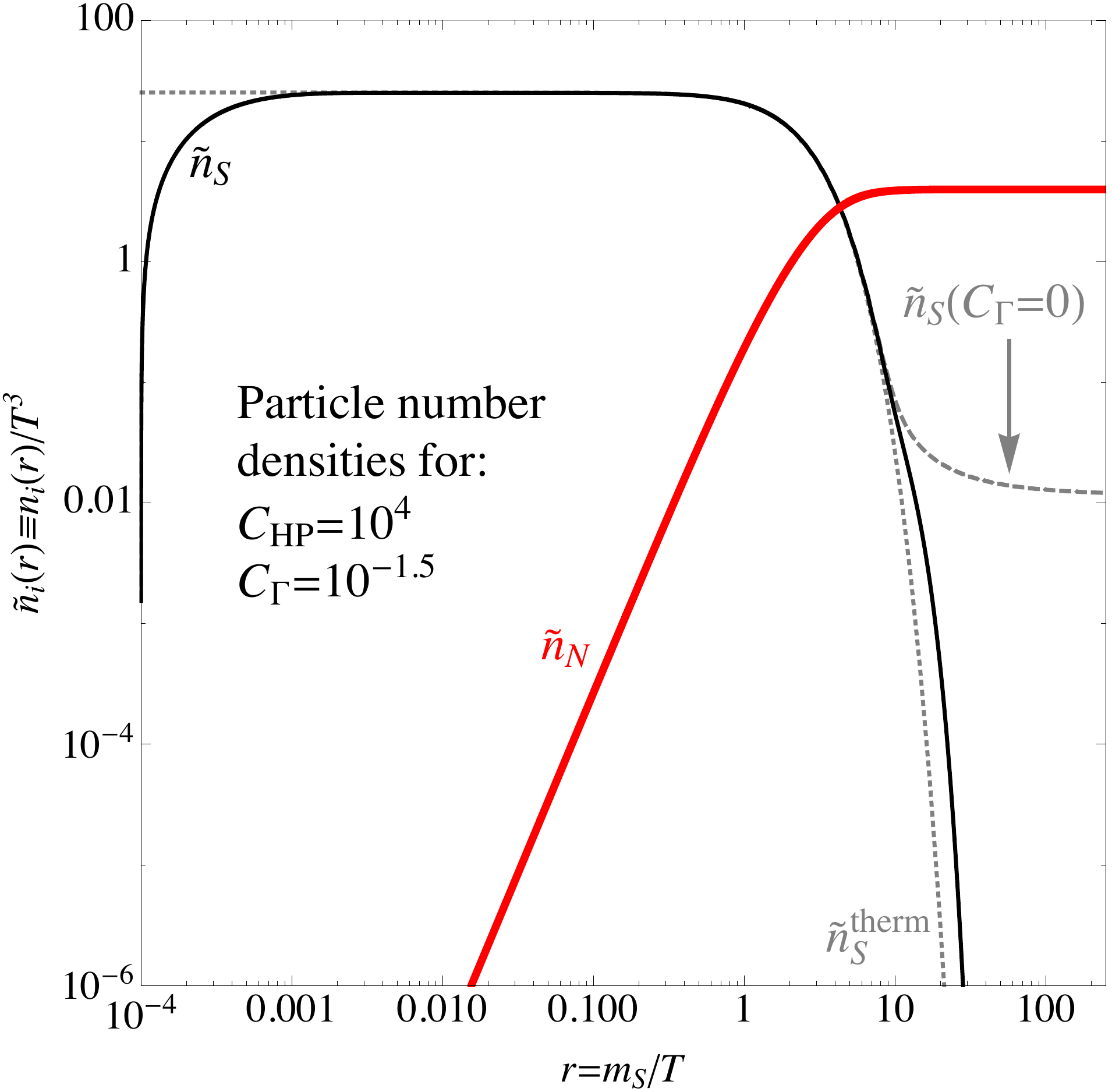}
 \end{tabular}
 \caption{\label{fig:snapshot_WIMP_IE}The same as Fig.~\ref{fig:snapshot_FIMP}, but for the WIMP case with decay in equilibrium.}
\end{figure}

Let us now turn to the WIMP cases. For a large enough decay width, an equilibrated scalar can decay while being in thermal equilibrium. This regime is in fact only realised in a relatively small corner of the parameter space, but one point which is a good example for this behaviour is $(\CHP,\CGamma) = (10^4,10^{-1.5})$ -- corresponding to $\left(y, \lambda\right) = \left(4.72 \times 10^{-8},\, 8.3 \times 10^{-5}\right)$ for our standard reference values of $m_S=\unit{1}{TeV}$ and $g_*=106.5$ -- as depicted in Fig.~\ref{fig:snapshot_WIMP_IE}. Glancing at the distribution function (left panel) first, the evolution appears to be relatively similar to the FIMP example just discussed -- although the distribution looks slightly flatter at early times. This distribution also seems to be slightly colder than a thermal one, as to be expected~\cite{Merle:2014xpa,Petraki:2007gq,Merle:2013wta,Petraki:2008ef,Bezrukov:2014qda}. However, also this point will turn out to correspond to hot DM.

The evolution of the abundances reveals the difference to the FIMP case, cf.\ right panel of Fig.~\ref{fig:snapshot_WIMP_IE}. Here, it is clearly visible that the scalar (solid black line), although starting with a vanishing initial abundance, equilibrates very quickly and then follows the thermal curve (dotted gray line).\footnote{We have explicitly checked that this happens independently of the initial abundance, as it should.} During this time, the scalar is highly relativistic and thus its decay is in fact not very efficient. However, given that its abundance is thermal and thus very large, the few occasional decays are sufficient to gradually built up a sizeable abundance of sterile neutrinos.\footnote{In fact, one could equally well interpret this case simply as the sterile neutrino itself being a FIMP.} The scalar remains in equilibrium for a relatively long time, until $r\sim 5$, and if it was stable it would just resemble the generic behaviour for frozen-out WIMPs (cf.\ dotted gray curve). However, given that the scalar decays relatively quickly, the frozen-out abundance does not survive and is converted into sterile neutrinos (solid red line). 

We can again compare the abundances for late times, which in this case gives $\tilde n_N(r=250)/\tilde n_S(\CGamma =0, r=250) \simeq 327.6$. This is vastly different from the previous result, but this behaviour is easy to understand: as long as the scalar is in equilibrium, it may decay more or less arbitrarily fast without its abundance being affected, because it is constantly re-generated by the thermal plasma. This happens very efficiently, so that a very large number of sterile neutrinos is produced while the scalar is still in thermal equilibrium. Of course, for the frozen-out abundance alone, the factor of two would again be present -- but this part makes up only about $0.6\%$ of the final sterile neutrino abundance. Hence, this case indeed corresponds very well to the limit of only having scalar decays in equilibrium. Using the approximation obtained in Eq.~\eqref{eq:SterileNeutrinoYieldDeepWIMP}, we indeed obtain a final abundance which agrees with the numerical result within 4.5\%. 

\begin{figure}[th]
 \centering
 \begin{tabular}{lr}
 \includegraphics[width=8cm]{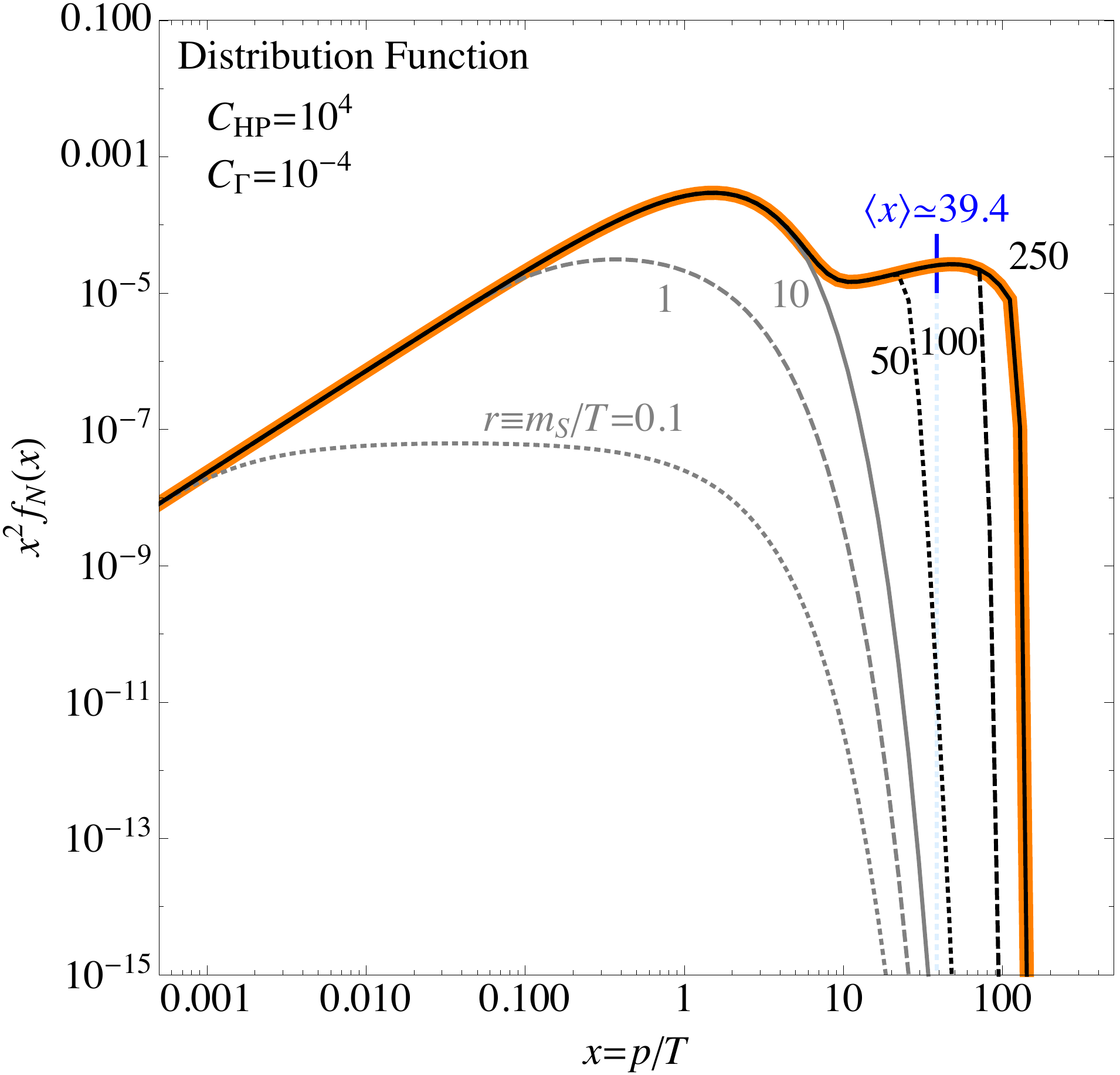} & \includegraphics[width=7.8cm]{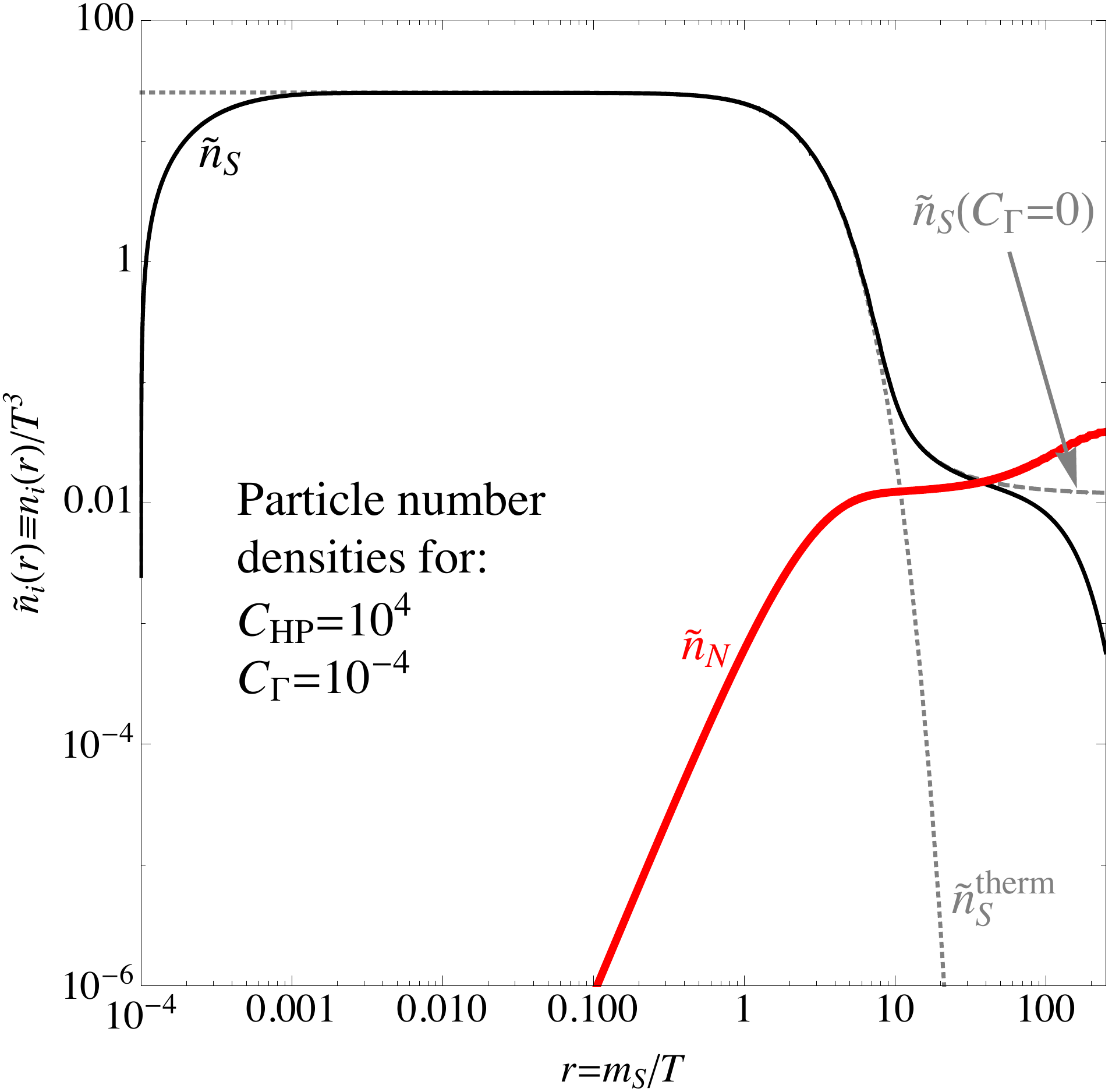}
 \end{tabular}
 \caption{\label{fig:snapshot_WIMP_inter}The same as Fig.~\ref{fig:snapshot_FIMP}, but for the intermediate WIMP case.}
\end{figure}

When we go down to smaller values of $\CGamma$, we will reach the intermediate WIMP regime where some scalars decay in equilibrium while another non-negligible fraction decays only after freeze-out, thereby contributing to the final sterile neutrino abundance in similar amounts. The example displayed in Fig.~\ref{fig:snapshot_WIMP_inter} is $(\CHP,\CGamma) = (10^4,10^{-4})$, which corresponds to the bottom right corner of the parameter space considered, and to $\left(y, \lambda\right) = \left(2.7 \times 10^{-9},\, 8.3 \times 10^{-5}\right)$ for the reference values for $m_S$ and $g_*$. Having a look at the distribution function first, see left panel, one can see that a twin peak structure is visible -- just as seen in the example for the free-streaming horizon failing, cf.\ Fig.~\ref{fig:DoublePeakExample}. Looking at the evolution with the time parameter $r$, it is visible that for early times the left (lower momentum) peak gradually builds up, while the right (higher momentum) peaks is only generated much later. This behaviour already suggests that the left peak arises from decays in equilibrium while the right peak is generated by late decays with the scalar $S$ already having frozen out and thus being out of equilibrium. This notion is supported by the late peak being the one corresponding to higher momenta: the freeze-out of the scalar happens at a temperature close to its mass, while the decay is a gradual process which takes some time to happen after the scalar has become non-relativistic. Thus, some energy is ``stored'' in the scalar mass and, once the scalars decay, the characteristic energy scale of the resulting sterile neutrinos will be of the order of the scalar mass -- which may be considerably larger than the temperature of the Universe at that time.

Glancing at the right panel we can see a matching behaviour in the abundances. While the scalar is in equilibrium, it gradually builds up a sizeable abundance of sterile neutrinos. However, this process ceases to be efficient once the scalar turns non-relativistic, thereby dropping in abundance and ultimately freezing out. Although the scalar abundance is much smaller than it was during the equilibrium time, now that the particles are non-relativistic the decays become efficient and completely translate the abundance of frozen-out scalars into sterile neutrinos, where the particle number is again doubled (but only for the frozen-out part).

\begin{figure}[th]
 \centering
 \begin{tabular}{lr}
 \includegraphics[width=8cm]{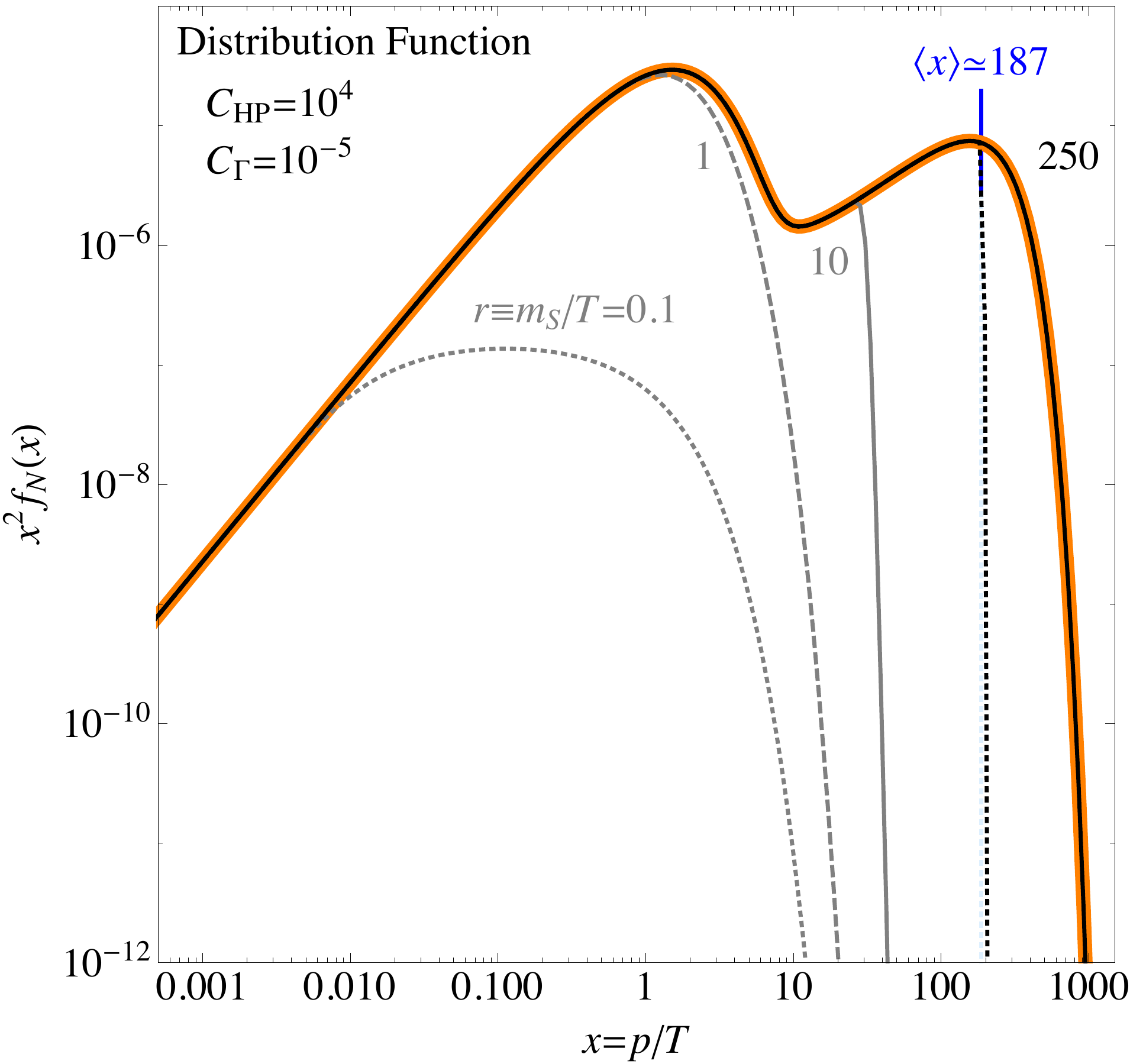} & \includegraphics[width=7.8cm]{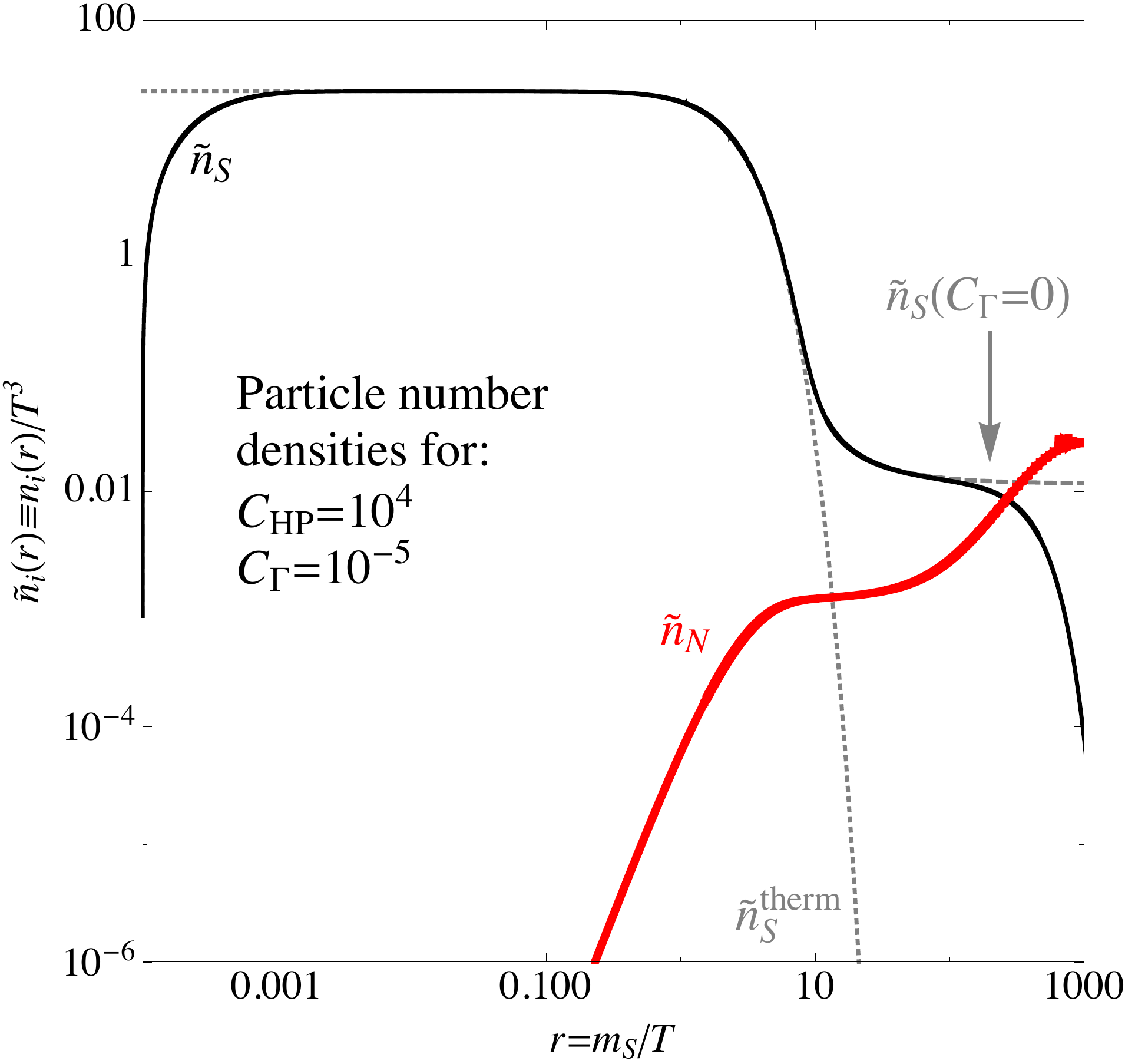}
 \end{tabular}
 \caption{\label{fig:snapshot_WIMP_OoE}The same as Fig.~\ref{fig:snapshot_FIMP}, but for the Out-of-equilibrium WIMP case.}
\end{figure}

Finally, we have in Sec.~\ref{subsec:WIMPLimit} also discussed the case where the scalar practically fully decays out of equilibrium. On the other hand, given that the case just discussed was already located at the very edge of the parameter space, this final case does not seem to make sense at all. However, this does not mean that the decay solely out of equilibrium does not exist, but only that it is not very accurate to treat it under the assumption $g_*\simeq {\rm const.}$, as we will demonstrate in Sec.~\ref{subsec:AssessingEvolutionDOF}. We still want to briefly discuss a toy example of this case, for parameter values of $\CHP =10^4$ and $\CGamma = 10^{-5}$, which is depicted in \Figref{fig:snapshot_WIMP_OoE} and which corresponds to Lagrangian level couplings of $\left(y, \lambda\right)=\left(2.4 \times 10^{-10},\, 8.3 \times10^{-5}\right)$ in our benchmark scenario.

Even though the double logarithmic scale in both panels might be misleading, the main contribution now comes from the late decay of frozen-out scalars. This can be seen easily when considering the sterile neutrino abundance $\tilde{n}_N$ in the right panel. While the scalar is in equilibrium, a small abundance of sterile neutrinos builds up until a plateau is reached for $r \sim 10$. Subsequently, the decay sets in and the relic abundance of scalars is converted into sterile neutrinos at high momenta. The final abundance exceeds the value of the intermediate plateau by more than a factor of $20$. Also, the average momentum $\av{x}$ is strongly dominated by the high momenta, just as expected. Since the production during equilibrium is proportional to $\CHP$, we would have to lower this parameter by several orders of magnitude to make the first peak vanish in a log-log plot. As already argued, such a case would be far beyond the validity of our assumption of a constant $g_*$ during production and will hence not be considered in this work.

Up to now, we have mainly been concerned with the DM abundance, but we have not yet shown whether the DM produced is in accordance with structure formation. A discussion of this point will be given in the next subsection.

\subsection{\label{subsec:ResultsFreeStreaming}Results for the free-streaming horizon}

\begin{figure}
\centering
 \begin{subfigure}{0.7 \textwidth}
  \centering
  \includegraphics[width=\textwidth]{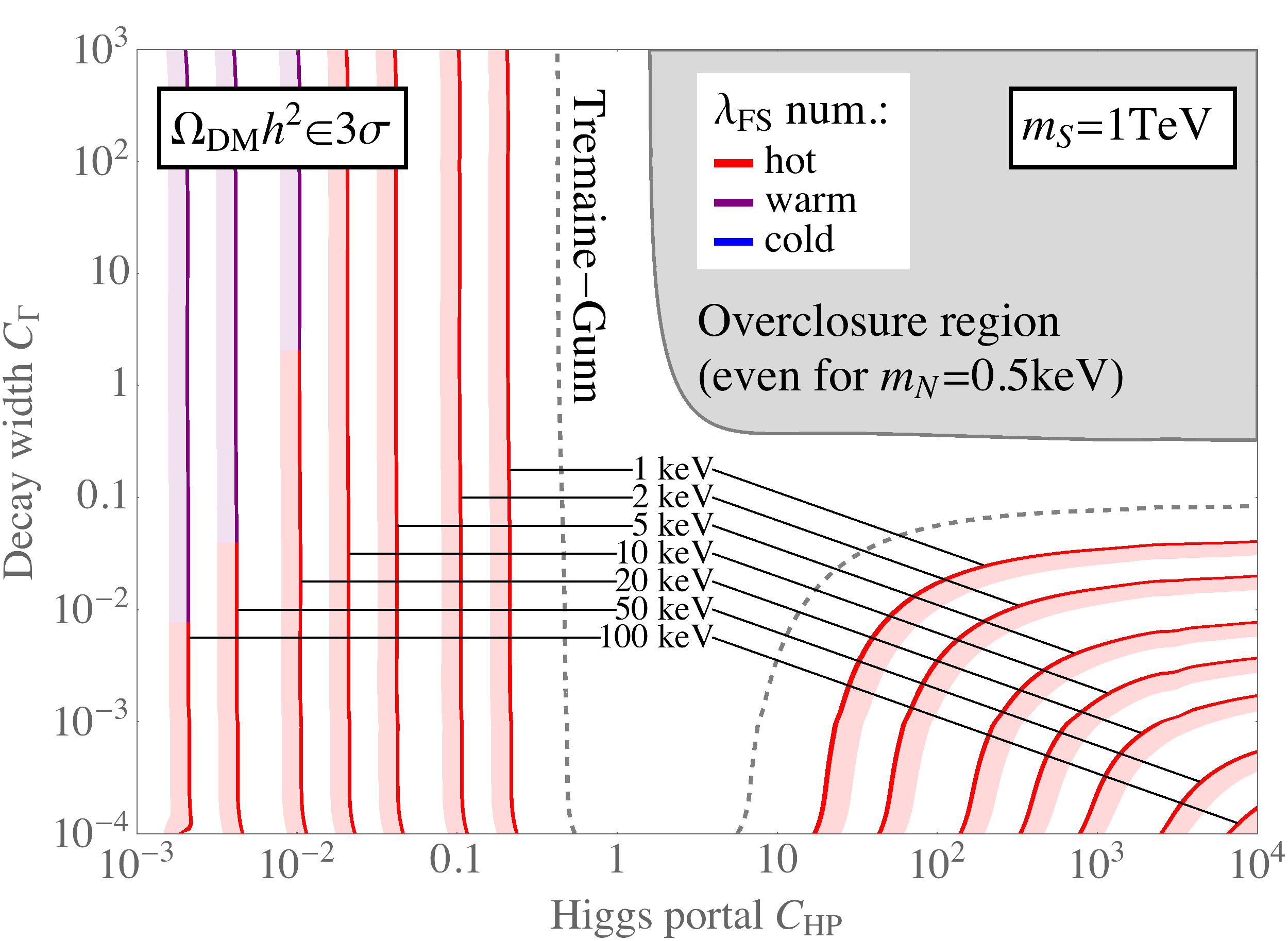}
  \caption{Numerical results for the free-streaming horizon.}
  \label{subfig:FreeStreamingNumerical}
 \end{subfigure}\\
 \begin{subfigure}{0.7 \textwidth}
  \centering
  \includegraphics[width=\textwidth]{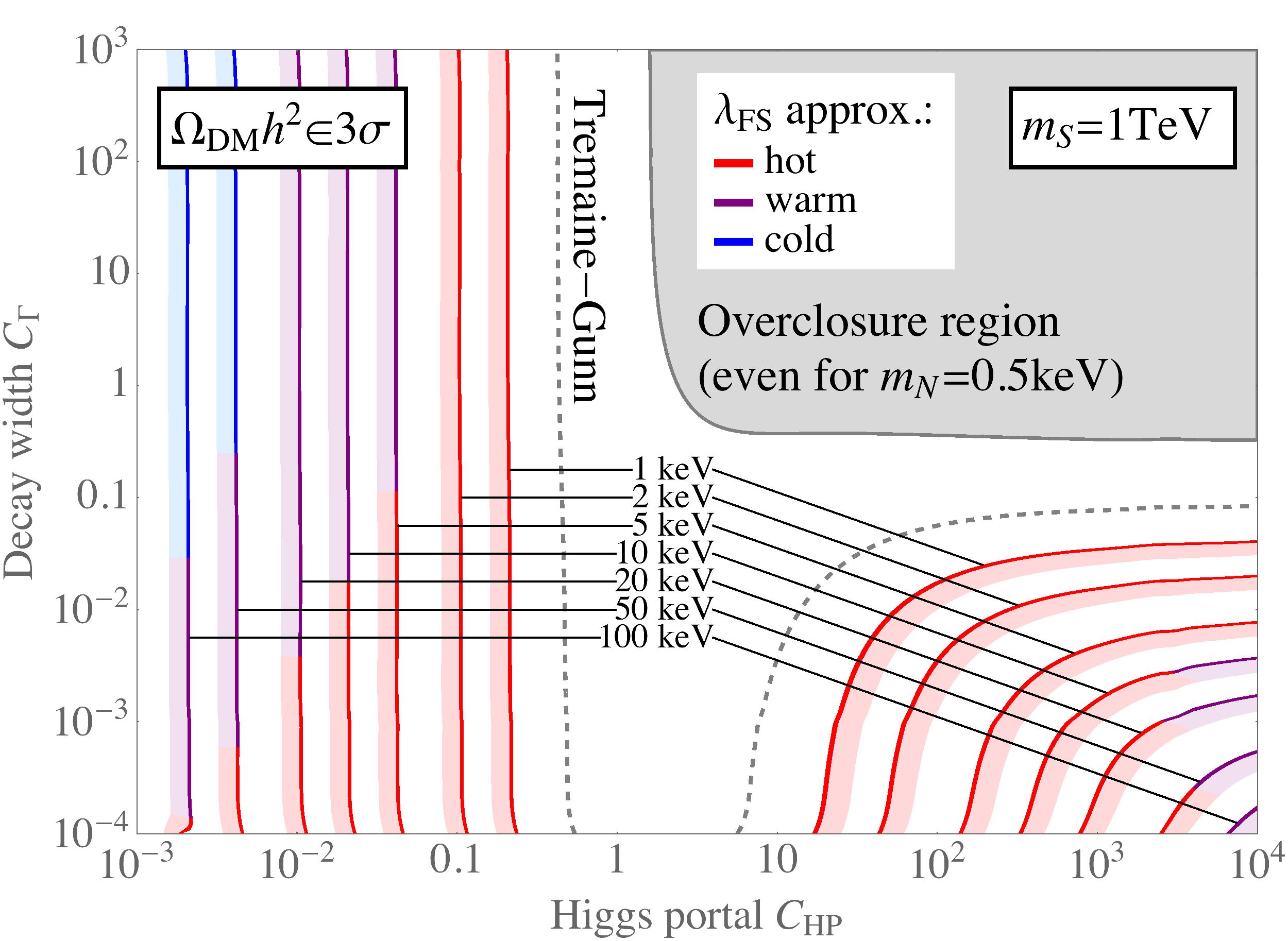}
  \caption{Same as \Figref{subfig:FreeStreamingNumerical} but using the analytical estimate.}
  \label{subfig:FreeStreamingEstimate}
 \end{subfigure}
 \caption{\label{fig:FreeStreaming}Comparison of numerical results and analytical estimates of the free-streaming horizon, i.e., with and without taking into account the evolution of $a(t)$ in the computation of the integral for $\sub{\lambda}{FS}$. Note that the strong dependence of these results on $m_S$ is hidden in the definition of $\CHP$ and $\CGamma$. However, there is a residual weak dependence on $m_S$ since the time-temperature relation used to calculate the free-streaming horizon is sensitive to the absolute temperature scale (as opposed to $\CHP$ and $\CGamma$). Hence we explicitly state the benchmark value of $m_S=\unit{1}{TeV}$ even though the result will not change dramatically for scalar masses varying by a factor of a few. For this reason and for the sake of being able to compare to the other plots more easily, we also renounce labelling the axes with Lagrangian level couplings.}
\end{figure}

As discussed in \Secref{subsec:DefinitionFreeStreaming} we present the free-streaming scale 1) as calculated by our numerical approach, fully taking into account the evolution of $a(t)$ (cf.~\Appref{app:DetailsrFS}), and 2) following the estimates put forward in \cite[Eq.~(20)]{Merle:2013wta}. In \Figref{fig:FreeStreaming}, we display again the bands in the \CHP-\CGamma\ plane reproducing the correct relic abundance (just as in Figs.~\ref{fig:ClosureRegionsCGammaCHP} and~\ref{fig:Abundances}), but this time colour-coding whether the sterile neutrino is \emph{hot} (red), \emph{warm} (purple), or \emph{cold} (blue) according to the definitions in \Secref{subsec:DefinitionFreeStreaming}. The scalar mass is chosen to be $m_S=\unit{1}{TeV}$, however, the results depend only mildly on the mass of the scalar~\cite{Merle:2013wta,Petraki:2008ef,Adulpravitchai:2014xna}. This is true since in our computation the only dependence of the scalar mass enters through the effective number of d.o.f.\ which are a function of physical temperature. The strongest physical dependence on the mass of the scalar is still implicit in the definition of the parameters $\CHP$ and $\CGamma$. If we lowered the scalar mass to some hundred GeV, the result would be altered only by a few percent.

In \Figref{fig:FreeStreaming} it is clearly visible that our numerical results disfavour more of the parameter space than the analytical estimates do. In particular, everything but the FIMP case is under tension in that analysis. However, we want to emphasise once more that these results should be interpreted with care. First of all, the numerical approach also suffers from (mainly systematical) uncertainties: the simplified truncation of the time-temperature relation into two distinct regimes (either purely radiation dominated or purely matter dominated) will differ from the exact time-temperature relation, see \Appref{app:DetailsrFS} for details. Second, as we will see in Sec.~\ref{subsec:AssessingEvolutionDOF}, in the region where the sterile neutrinos are particularly hot (i.e., for small decay constants), the approximation of a constant number of d.o.f.\ during production becomes less reliable, which also affects the calculation of $\lambda_{\rm FS}$. Third, as discussed in Sec.~\ref{subsec:DefinitionFreeStreaming}, the free-streaming horizon -- only taking into account \emph{average} properties of the spectrum -- cannot capture all features of structure formation and can hence only serve as an indication. More detailed analyses can be done using the so-called \emph{transfer function}, i.e.~the square root of the linear power spectrum of matter perturbations, which can in turn be constrained by data from the Lyman-$\alpha$ measurements. Such studies have been performed in \cite{Merle:2014xpa} for sterile neutrino dark matter produced from the DW-mechanism, from the SF-mechanism, and from a simplified version of scalar decay, using the Boltzmann solver \code{CLASS}~\cite{Blas:2011rf} to obtain the transfer functions from an extended Press-Schechter approach. A similar study taking into account the subtleties of sterile neutrino DM from scalar decay as discussed in this paper is subject of on-going work~\cite{Merle:Proj:StructureFormation}. First indications of this analysis seem to confirm the conclusions drawn from Fig.~\ref{subfig:FreeStreamingNumerical} rather than those of Fig.~\ref{subfig:FreeStreamingEstimate}, which puts the scenario of freezing-out scalars under severe tension and might also indicate a comparatively large lower mass limit of the sterile neutrino of roughly $\unit{20}{keV}$ for the case of freeze-in, which might be an interesting finding in particular in the context of the tentative $3.5$-keV line~\cite{Bulbul:2014sua,Boyarsky:2014jta}. However, in order to give a definitive answer, we will extend the current study~\cite{Merle:Proj:RefinedScalarProd&DW,Merle:Proj:StructureFormation}, as already discussed in the introduction.

\subsection{\label{subsec:DarkRadiation}The dark radiation bound}

Somewhat related to the bound from structure formation is the bound on the effective number of neutrinos, $\sub{N}{eff}$. In the SM, this number is equal to is 3.046, the small deviation from 3 arising due to the effects of the reheating at $e^+e^-$ decoupling~\cite{Dolgov:2002wy}. However, if the sterile neutrinos in our setting are too hot, they effectively act as radiation and could in that case also contribute to the deviation $\Delta \sub{N}{eff}$ of $\sub{N}{eff}$ from its standard value.

We can calculate the contribution of the sterile neutrinos to $\Delta \sub{N}{eff}$ by comparing the 
kinetic part of their energy density to the energy density $\ssscript{\rho}{term}{ferm}$ of a perfectly relativistic (i.e.~massless) fermionic species
at the same temperature in equilibrium:\footnote{Note that this is slightly different to the standard case found in the literature, where the dark radiation component is typically highly relativistic, see e.g.~\cite{Vogel:2013raa,Hasenkamp:2011em}, whereas in our case we do not know a priori whether this is the case and have to take this subtlety into account by subtracting the rest energy which is negligibly small in the highly relativistic limit. Alternatively, one can estimate the contribution of non-thermal DM to dark radiation by using the Lorentz factor~\cite{Hooper:2011aj}.}
\begin{equation}
 \Delta \sub{N}{eff}\left(T\right) \equiv \frac{\rho - n m_N}{2 \ssscript{\rho}{therm}{ferm}}
 =\frac{60}{7\pi^4} \left(\frac{T}{T_\nu}\right)^4  \frac{m_N}{T}\int_{0}^{\infty}{\diffd x\, x^2 \left(\sqrt{1+\left(\frac{x}{m_N/T}\right)^2}-1\right)f_N(x,T)} .
 \label{eq:DefDeltaNeff}
\end{equation}
The factor of $2$ in the denominator of the first term is due to the fact that our distribution function already contains both particle and antiparticle while $\sub{N}{eff}$ is constructed in a way to reproduce the number of \emph{families}, i.e.~3 (up to the aforementioned small corrections) and not a value of 6. Note also, that the factor $\left(T/T_\nu\right)^4$ accounts for the fact that, once the reaction $e^+e^- \leftrightarrow \gamma \gamma$ freezes out, the photons get reheated while the neutrinos have already decoupled from the plasma, such that no energy is transferred to neutrinos by the annihilation of the electron-positron pairs. This is also true for sterile neutrinos, however, the temperature of their distribution (if one can define this quantity at all, given that the distribution may be highly non-thermal) is implicitly contained in the quantity $f_N(x,T)$. Yet, the relative reheating of the photons compared to sterile neutrinos is the same as that compared to active neutrinos, at least if the small corrections due to weak interactions are neglected. Thus we can simply use the factor $\left(T/T_\nu\right)^4$ in Eq.~\eqref{eq:DefDeltaNeff}, where $T_\nu$ is indeed the temperature of the \emph{active} neutrinos. In this case, the value of $\left(T/T_\nu\right)^{4}$ rises from unity to $\left(11/4\right)^{4/3} \approx 3.85$. Hence we include the factor of $3.85$ for late times after electron-positron-annihilation while we drop it for temperatures above about $\unit{1}{MeV}$.

Using \equref{eq:DefDeltaNeff}, we can -- for every point $(\CHP, \CGamma)$ -- simply fix the sterile neutrino mass such as to reproduce the correct relic abundance and then calculate $\Delta \sub{N}{eff}$ at any given temperature $T$. Again, all the information lies in the distribution function $f_N(x,T)$, where $x \equiv p/T$: the higher the distribution peaks at large momenta, the bigger the contribution to the extra radiation will be. If, on the other hand, the sterile neutrino abundance was tiny, this would also be reflected in $f_N(x,T)$ and the integral in Eq.~\eqref{eq:DefDeltaNeff} would yield a vanishingly small result.

\begin{figure}
\centering
 \begin{subfigure}{0.7 \textwidth}
  \centering
  \includegraphics[width=\textwidth]{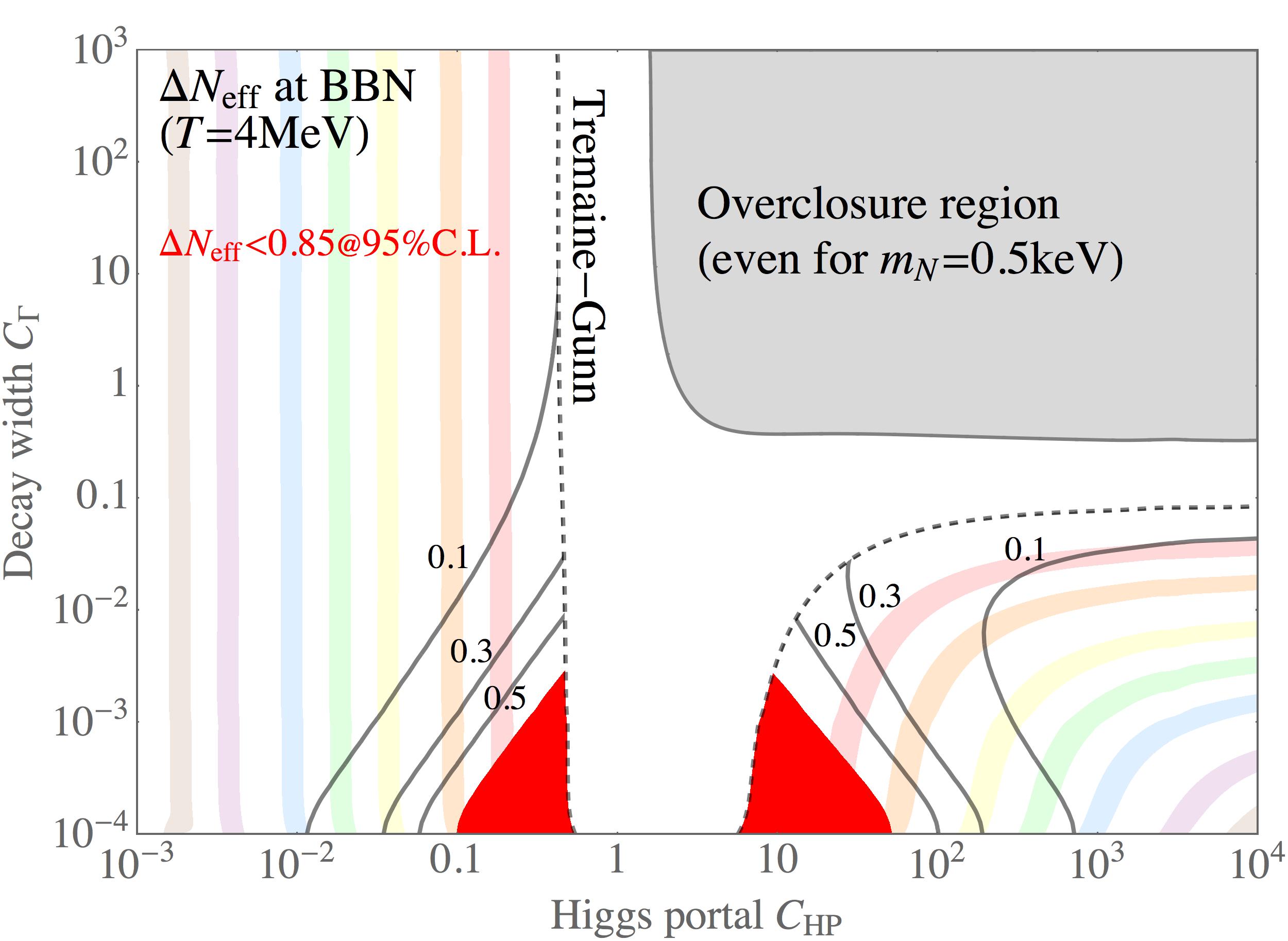}
  \caption{Deviation of the effective number of neutrinos from the SM value at BBN and resulting excluded region (red patches).}
  \label{subfig:DeltaNeffBBN}
 \end{subfigure}\\
 \begin{subfigure}{0.7 \textwidth}
  \centering
  \includegraphics[width=\textwidth]{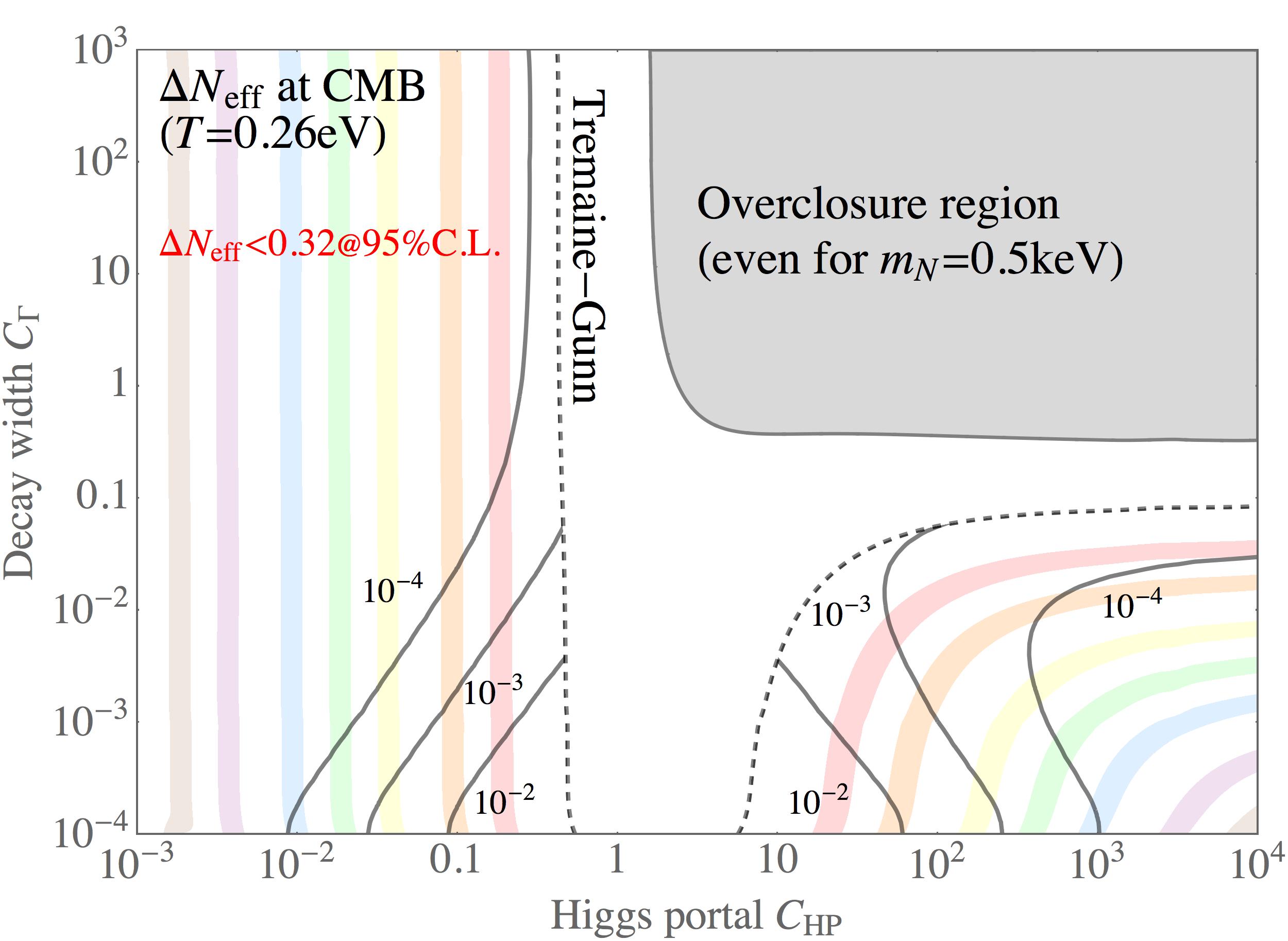}
  \caption{Same as \Figref{subfig:DeltaNeffBBN} at the temperature of CMB decoupling.}
  \label{subfig:DeltaNeffCMB}
 \end{subfigure}
 \caption{Deviation of the effective number of neutrinos from its SM value of $3.046$.}
 \label{fig:DeltaNeff}
\end{figure}

In general, we have information on $\Delta \sub{N}{eff}$ from two different epochs in the history of the Universe: during big bang nucleosynthesis (BBN), at $\sub{T}{BBN} = \unit{4}{MeV}$,\footnote{The beginning of BBN happens at a temperature of a few MeV~\cite{Kneller:2004jz}, and we take 4~MeV as example which is known to ``reset'' conditions to how they were prior to BBN~\cite{Kawasaki:2000en,Hannestad:2004px}. Given that the sterile neutrinos keep cooling down until the end of BBN and even further, taking such an early temperature corresponds to some extend to a pretty conservative limit. However, we also have to take into account that the main temperature dependence is factored out of $\Delta \sub{N}{eff}$ per definition, so that $\sub{T}{BBN} = \unit{4}{MeV}$ is in fact not much more conservative than taking a value of $\unit{1}{MeV}$, or similar.} the formation rate of nuclei depends on the overall expansion rate of the Universe which, in turn, depends on its overall radiation content~\cite{Sarkar:1995dd,Fields:2014uja,Agashe:2014kda}. Thus, if we do not want to spoil BBN, we have to respect an upper bound on the amount of extra radiation at BBN. While the \emph{Particle Data Group} still cites a relatively old bound, $\Delta \sub{N}{eff}^{\rm BBN} < 1.5@95\%$~C.L.~\cite{Cyburt:2004yc,Lisi:1999ng}, newer versions exist: $\Delta \sub{N}{eff}^{\rm BBN} < 1@95\%$~C.L.~\cite{Mangano:2011ar}, $\Delta \sub{N}{eff}^{\rm BBN}< 0.93@95\%$~C.L.~\cite{Izotov:2014fga}, and $\Delta \sub{N}{eff}^{\rm BBN}< 0.85@95\%$~C.L.~\cite{Cooke:2013cba}. We have in our plots adopted the most stringent limit, to illustrate that not even the strongest constraint does influence our results in a significant way. A seemingly more stringent constraint can be obtained from the measurement of the cosmic microwave background (CMB), which decouples at  $\sub{T}{CMB}\approx \unit{0.26}{eV}$\cite{Kolb:1990vq}, since the CMB spectrum also depends on the expansion rate of the Universe and thus on the radiation content~\cite{Archidiacono:2013fha}. The bound for that time is $\Delta \sub{N}{eff}^{\rm CMB} < 0.32@95\%$~C.L.~\cite{Planck:2015xua}.

In our analysis, we take into account both bounds and calculate $\Delta \sub{N}{eff}\left(\sub{T}{BBN}\right)$ as well as $\Delta \sub{N}{eff}\left(\sub{T}{CMB}\right)$. Although the BBN bound on  $\Delta \sub{N}{eff}$ appears to be weaker than the one from the CMB measurement, one has to take into account that BBN happens much earlier than the CMB decoupling. Thus, given that the sterile neutrinos produced from scalar decays cool down as time goes by, it may very well be that their contribution to the radiation content at BBN is much more significant than later on (and, indeed, this will turn out to be the case here). Such settings with a type of dark radiation that contributes differently at BBN time than later on are known in other contexts, too (see, e.g., Refs.~\cite{Fischler:2010xz,Foot:2011ve,Menestrina:2011mz,DiBari:2013dna}). Some example contour lines for $\Delta \sub{N}{eff}$ are displayed in the \CHP-\CGamma\ plane in Fig.~\ref{fig:DeltaNeff}, both for BBN (upper panel) and CMB (lower panel). As a guide for the eye, we have again displayed the lines of correct abundance, cf.~Fig.~\ref{fig:Abundances}, since we fix the mass of the sterile neutrino entering the computation of $\Delta \sub{N}{eff}$ via \equref{eq:DefDeltaNeff} by the constraint of reproducing the observed relic abundance. As can be seen, at BBN there could be a non-negligible contribution of the sterile neutrinos to $\Delta \sub{N}{eff}\left(\sub{T}{BBN}\right)$, and there is even a small excluded region which would violate the bound (marked by the red areas at the bottom centre of the plot). This region of a too large contribution arises only for very small decay widths \CGamma, i.e., the scalars must be very long-lived and inject highly energetic sterile neutrinos into the Universe at relatively late times. However, we would in any case exclude this region from any serious consideration because it would, trivially, fall far into the region where the DM is hot in any case, cf.\ Fig.~\ref{fig:FreeStreaming}. At CMB, on the other hand, there is not even a serious constraint left since the sterile neutrinos have cooled down by then and only have comparatively small momenta. 

Thus, even though there is in principle a contribution of the sterile neutrinos to $\Delta \sub{N}{eff}$, no strong constraint arises from it and there is no threat to our production mechanism.

\subsection{\label{subsec:Bounds}Other bounds and constraints}

In this section, we will discuss the two remaining conditions which may affect the DM production mechanism proposed. It turns out that both of them are no actual problems: the first one would only affect regions so far away from the interesting part of the parameter space that they do not play a role in practice. The second problem is more of a ``theoretical'' nature, i.e., while it may be important to take into account, it can be easily circumvented in concrete settings and may only appear to be problematic if the Lagrangian presented in Eq.~\eqref{eq:ModelLagrangian} was viewed as ``theory of everything'' valid up to the Planck scale. However, for completeness we would like to at least briefly mention these points.

In general, collider bounds could also restrict the parameter space of our model, since after all the Higgs portal coupling $\lambda$ could be used to produce two singlet scalars from two SM-like Higgses. Using the limits on the mixing between a scalar singlet and the Higgs boson as proposed in~\cite{Lopez-Val:2014jva}, the bounds on $\lambda$ are far above even the largest value of \CHP\ we show in our plots. This holds true even if the VEV of the scalar singlet is larger than its mass by two orders of magnitude. This behaviour can be intuitively understood by taking into consideration that even a value of $\CHP=10^4$ corresponds to rather small couplings $\lambda$, due to the large factor $M_0/m_S$ involved in its definition. Thus, in practice, we do not need to be bothered by any current bounds from colliders. However, at least in principle, there is an upper bound on \CHP\ which may become important if our study was extended to considerably larger values of the Higgs portal coupling. This conclusion still holds when confronted with updated analyses \cite{Robens:2015gla}.

There is an orthogonal problem which is related to the symmetry breaking resulting from the scalar potential in Eq.~\eqref{eq:ScalarPotentialLagrangian}. As we had already explained, the shape of this potential is determined by an underlying discrete $\mathbb{Z}_4$ symmetry. While we only use this symmetry for simplicity and we could even skip it without too drastic consequences, at least in principle it would be broken by a VEV $\left\langle S\right\rangle$ of the scalar field. Since the potential has two perfectly degenerate minima, different parts of the Universe could then have their ground states in different minima, thereby leading to so-called  \emph{domain walls}~\cite{Zeldovich:1974uw}. The existence of such objects would have considerably changed the history of the Universe and they are hence problematic. However, there are many possible solutions to this problem, since the slightest difference in energies of the two vacua could be generated by all kinds of physics, in which case the walls would decay exponentially quickly~\cite{Preskill:1991kd,Riva:2010jm,Dvali:1995cc,Dvali:1996zr,Larsson:1996sp}. This can also happen in the case at hand~\cite{Merle:2013wta}.

\subsection{\label{subsec:AssessingEvolutionDOF}Assessing the validity of approximating the relativistic d.o.f.\ as constant}

As stated at the very beginning, we followed the assumption that $g_*,g_{*S}$ are constant during the DM production process. This assumption impacts on the form of the spectrum, cf.~\equref{eq:LiouvilleOperator_r_x}, and thereby also on the implications for structure formation. Even in the analytical estimate of the free-streaming horizon, it makes a difference if the dilution factor $\xi^{1/3}$ is calculated from a starting point of $g_*=106.75$ or at some lower value. In order to assess this approximation, we have for every point in the \CHP-\CGamma\ plane computed the temperature when the production of sterile neutrinos is completed (i.e., when the abundances surpasses 95\% of its maximum value) and we plot the comparison to the SM number of d.o.f.\ of the primordial plasma at that temperature. To this end, we again assume a scalar mass of $\unit{1}{TeV}$.

\begin{figure}[t]
 \centering
 \includegraphics[width=0.7 \textwidth]{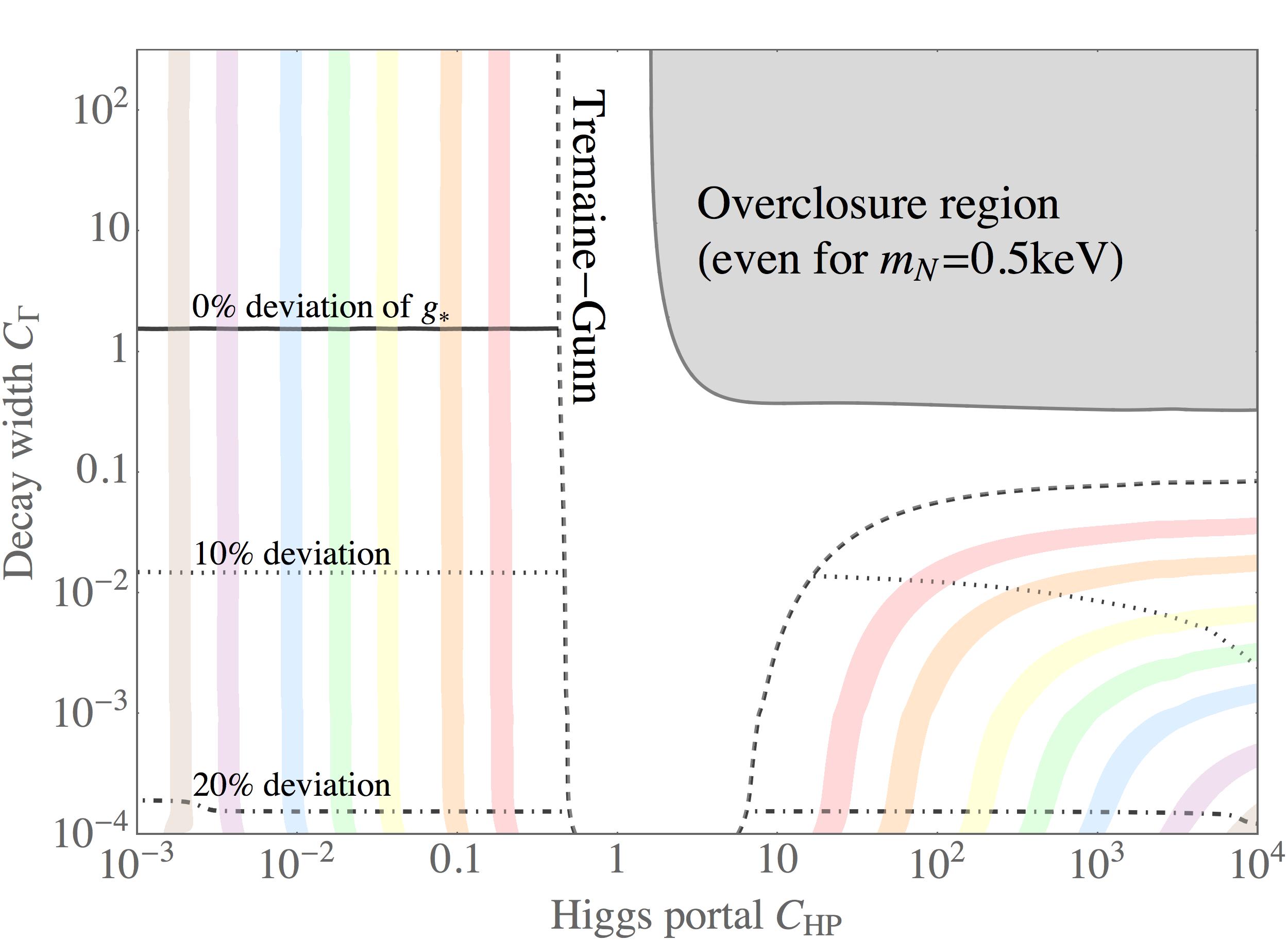}
 \caption{\label{fig:AssessingEvolutionDOF}Deviation of the number of d.o.f.\ at the time of DM production from the maximum value of $g_*=106.75$. This deviation can be interpreted as a check of the goodness of the approximation of $g_*=\mathrm{const.}$ during production, which was used to simplify \equref{eq:LiouvilleOperator_r_x}.}
\end{figure}

In our case, the extension of the SM will never contribute more than $1+2\times7/8=22/8 = 2.75$ (i.e., less than $3\%$ of the SM value of 106.75) units to the count of d.o.f.\ since this would be the maximum possible contribution of scalars and sterile neutrinos both being present with a thermal abundance and relativistic velocities. Since we expect the SM d.o.f.\ to be much larger during the time of production, we can interpret the further d.o.f.\ as small additional perturbation in the background of the SM. 

Hence the approximation of constant d.o.f.\ can be assessed by checking the evolution of the SM background. If the number of SM d.o.f.\ at end of production is still close to the maximum value of $g_*=106.75$, the approximation can be seen as adequate. \Figref{fig:AssessingEvolutionDOF} shows the deviation of the number of d.o.f.\ at time of production from the maximum value of $106.75$. Comparing to Figs.~\ref{fig:FreeStreaming}, in most of the interesting parameter space our approximation is not too bad. In fact, for the cooler regions in the FIMP case, the approximation is even excellent. Only for the relatively hot regions the estimate of the deviation compared to the exact treatment is larger than 10\%, but this region is in any case not the favoured one. Since, after all, the Boltzmann treatment of DM production in any case cannot be expected to yield sub-percent accuracy~\cite{Hamaguchi:2011jy}, this means that the approximation is in fact not too bad.

\section{\label{sec:Conclusions}Conclusions \& outlook}

We have presented a fully comprehensive study of the production of keV sterile neutrino Dark Matter in the early Universe by singlet scalar decays. The current paper lays the foundation for several follow-up considerations. Aiming at a clear overview first, we have applied some approximations that enabled us to present analytical results in addition to a detailed numerical computation, which we used to further back up certain limiting cases. Based on these initial considerations, we have derived the system of Boltzmann equations to be solved at the level of momentum distribution functions, and we have furthermore introduced a very efficient parametrisation to do so. After presenting our analytical results and discussing some aspects related to cosmological structure formation, we turned to present our numerical results. Our numerical solutions for the distribution functions have not only provided a comprehensive picture of where the observed Dark Matter abundance can be obtained in the large parameter space investigated, but they have also allowed us to account for all bounds applicable. This is the first time that such detailed results have been obtained for the production mechanism at hand, and we have shown how to fully exploit the information contained in the distributions. We have in particular found situations in which highly non-trivial distribution functions featuring more than one momentum scale can result from the simple decay mechanism presented, which could be very interesting for cosmological structure formation. In such cases, the simple minded estimate of structure formation properties using only the free-streaming horizon will fail, as we have explained by an illustrative example. While preliminary results of our on-going work (beyond what is presented in this paper) indicate that the numerical estimate of the free-streaming horizon yields a rather comprehensive picture despite the difficulties involved, we nevertheless have to postpone a definitive conclusion to the pending results.

In general, we have not only found very good agreement between our analytical and numerical results, but we have also shown that the assumptions applied (in particular the effective number of degrees of freedom being constant during Dark Matter production and the singlet scalar having a mass larger than that of the Higgs boson) are good in a significant fraction of the parameter space, although of course certain regimes exist in which the error introduced gets unacceptably large. Thus, the natural next step will be to extend our (numerical) considerations to the regimes in which the approximations applied are not valid~\cite{Merle:Proj:RefinedScalarProd&DW}, which will in particular allow us to treat considerably smaller singlet scalar masses. We will furthermore extend our studies to investigate the detailed implications for structure formation~\cite{Merle:Proj:StructureFormation}, which we have clearly shown to require a more advanced machinery than the free-streaming horizon. Ultimately, we aim to provide a fully comprehensive study of keV sterile neutrino Dark Matter production by scalar decays, so that this production mechanism can be put to the acid test to determine how good an alternative to resonant production it truly is.

\section*{Acknowledgements}

We would like to thank Ninetta Saviano for useful discussions and we are particularly grateful to Aurel Schneider for giving us many insights into the details of cosmological structure formation and for kindly providing us with preliminary results of our collaborative work. AM furthermore acknowledges partial support by the European Union FP7 ITN-INVISIBLES (Marie Curie Actions, PITN-GA-2011-289442). MT acknowledges financial support from the Studienstiftung des deutschen Volkes.

\appendix

\section{\label{app:DetailsKineticEquations}Appendix: Details on the kinetic equations}
\renewcommand{\theequation}{A-\arabic{equation}}
\setcounter{equation}{0}  

The collision term for a species $\Phi$ in contact with other species via some interaction of the generic type $\Phi+a+b+... \leftrightarrow \alpha + \beta + ...$ is given by 
\begin{dmath}
  C\left[f_\Phi\right] = \frac{1}{2 E_{p_\Phi}} \int \left[ \diffd P_a \diffd P_b ... \diffd P_\alpha \diffd P_\beta ... \times \twopin{4} \deltadistn{4}{p_\Phi + p_a +p_b +... -p_\alpha - p_\beta - ...} \times \Msquared \nonumber \\ \times \left[f_\alpha f_\beta ... f_\Psi \left(1\pm f_a\right) \left(1\pm f_b\right) ... - f_a f_b ... \left(1 \pm f_\alpha\right) \left(1\pm f_\beta\right) ...\left(1 \pm f_\Psi\right)\right] \right] \,.
 \label{eq:CollisionTermBoltzmann}
\end{dmath}
Some remarks about \equref{eq:CollisionTermBoltzmann} are in order:
\begin{enumerate}
    \item The quantity $E_{p_x}$ denotes the energy of particle $x$ and is hence given by $E_x=\sqrt{p_x^2+m_x^2}$.
    \item The internal degrees of freedom of a species are denoted by $g_x$. 
    \item We have introduced a symbol for the invariant phase-space element: 
      \begin{equation}
      \diffd P_x = g_x \frac{\diffd^3 p_x}{2E_{p_x} \twopin{3}} \,.
      \label{eq:DefInvariantPhaseSpace}
      \end{equation}
     \item The plus signs apply in the case of bosons and the minus signs in the case of fermions.
     \item The squared matrix element is defined following the convention in \cite[chapter 5]{Kolb:1990vq}, i.e.~the squared matrix element contains all relevant symmetry factors and averages over \emph{both} the initial and the final state spins.
\end{enumerate}

\subsection{\label{subapp:DetailsKineticEquationsScalar}Kinetic equation for the scalar}

In this appendix, we derive the kinematic functions $\mathcal{F}$ and $\mathcal{G}$ in their exact form. We also illustrate some pedagogical steps to ease the derivation of the relevant collision terms.

Let us start by constructing the collision terms in the variables $x$ and $r$, cf.\ Eq.~\eqref{eq:variables}, under the further simplification that we approximate all factors in \equref{eq:CollisionTermBoltzmann} arising from the final states by unity, i.e.\ we neglect the bosonic/fermionic nature of the particles, which is a good approximation provided that their energy is much larger than the temperature of the plasma. There are several processes that can contribute to the production of the singlet scalars in the early Universe.

We start with the collision term describing the production of a pair of scalars $S S$ from a pair of SM-like Higgses $hh$:
\begin{dmath}
  C_{hh\rightarrow SS}^S\left(q\right) =  \frac{1}{2 E_q} \iiint \invps{q'} \invps{p} \invps{p'}\ 4 \lambda^2 \twopin{4} \\
  \times \deltadist{E_q + E_{q'} -E_p -E_{p'}} \deltadistn{3}{\vec{q}+\vec{q}'-\vec{p}-\vec{p}'} \ssscriptupper{f}{h}{eq}\left(q\right) \ssscriptupper{f}{h}{eq}\left(q'\right)  \, .
  \label{eq:BoltzmannScalarRaw}
\end{dmath}
In \equref{eq:BoltzmannScalarRaw}, the momenta $q$ and $q'$ ($p$ and $p'$) belong to the Higgs bosons in the initial state (to the scalars $S$ in the final state). In \equref{eq:BoltzmannScalarRaw}, we have explicitly inserted the squared matrix element $\Msquared = 4 \lambda^2$. 

Using an argument based on detailed balance~\cite{Gondolo1991145}, we can make the following replacement in 
\equref{eq:BoltzmannScalarRaw}:
\begin{equation}
 \ssscriptupper{f}{h}{eq}\left(q\right) \ssscriptupper{f}{h}{eq}\left(q'\right) 
 = \ssscriptupper{f}{S}{eq}\left(p\right) \ssscriptupper{f}{S}{eq}\left(p'\right) \,.
 \label{eq:BoltzmannScalarEnergyConservation}
\end{equation}
Note that this can be shown quite easily in an explicit way in the case of a Maxwell-Boltzmann approximation that we are using, exploiting only energy conservation.

Accordingly, we can also use the explicit form of a Maxwellian distribution to simplify \equref{eq:BoltzmannScalarRaw} further. Integrating out the phase space in $p$ and $p'$, we obtain
\begin{dmath}
  C_{hh\rightarrow SS}^S\left(q\right) = \frac{4\lambda^2}{8\pi} \frac{1}{2E_q}\int \invps{q'}\sqrt{\frac{\left(E_q + E_{q'}\right)^2 -qq ' \cos \theta - 4 m_H^2}{\left(E_q + E_{q'}\right)^2 -qq ' \cos \theta}}\  e^{-(E_q + E_{q'})/T} \,.
  \label{eq:BoltzmannScalarStep2}
\end{dmath}
In order to arrive at \equref{eq:BoltzmannScalarStep2} by integrating out the phase spaces in $p$ and $p'$ in \equref{eq:BoltzmannScalarRaw}, we have used the standard phase space integral:
\begin{equation}
 \iint \invps{p} \invps{p'} \twopin{4} \deltadistn{4}{q +q' - p -p'} = \int \frac{\diffd \sub{\Omega}{CM}}{4 \pi} \frac{1}{8\pi} \left(\frac{2 \sub{p}{CM}}{\sub{E}{CM}}\right) \, .
\end{equation}
In our case, the centre-of-mass velocity is given by
\begin{equation}
 \frac{\sub{p}{CM}}{\sub{E}{CM}} = \frac{1}{2} 
 \sqrt{\frac{\left(E_q + E_{q'}\right)^2 -qq ' \cos \theta - 4 m_h^2}{\left(E_q + E_{q'}\right)^2 -qq ' \cos \theta}} \,.
\end{equation}
With this it is straightforward to arrive at \equref{eq:BoltzmannScalarStep2} from where the definition of $\mathcal{F}$ can directly be read after changing to the variables $x$ and $r$:
\begin{dmath}
\mathcal{F}\left(x,r,\xi \right) \equiv 2\pi \int\limits_0^\infty \diffd \hat{x}\, \hat{x}^2 \int\limits_{-1}^{\sub{\alpha}{max}} \diffd \cos \theta \frac{e^{-\sqrt{\hat{x}^2+r^2}}}{\sqrt{\hat{x}^2+r^2}} \times \sqrt{\frac{\left(\sqrt{\hat{x}^2+r^2} + \sqrt{x^2+r^2}\right)^2 -x\hat{x} \cos \theta - 4 \xi^2 r^2}{\left(\sqrt{\hat{x}^2+r^2} + \sqrt{x^2+r^2}\right)^2 -x\hat{x} \cos \theta}} \, ,
\label{eq:KinematicIntegral_hhToSS}
\end{dmath}
where $\xi \equiv m_h/m_S$ (cf.~\Secref{sec:AnalyticalResults}). 
The maximum allowed value for the cosine of $\theta$ comes from the trivial constraint that the argument in the square root must be non-negative. This leads to
\begin{align}
 \sub{\alpha}{max} = \mathrm{min}\left[1, \mathrm{max} \left[-1, \frac{\left(\sqrt{x^2 +r^2}+\sqrt{\hat{x}^2 +r^2}\right)^2 - 4\xi^2 r^2}{x \hat{x}} \right]\right]\,.
 \label{eq:DefCosThetaMax}
\end{align}

For scalar masses much larger than the Higgs mass, i.e.\ $\xi \ll 1$, we can simplify \equref{eq:KinematicIntegral_hhToSS}:
\begin{equation}
 \mathcal{F}\left(x,r,\xi\ll1\right) = 4\pi K_1\left(r\right)\,,
 \label{eq:KinematicIntegral_hhToSS_xi0}
\end{equation}
where $K_1$ is the first modified Bessel function of second kind.

With this result at hand, the first part of the kinetic equation for the scalar, cf.~\equref{eq:DynamicsScalar}, reads:
\begin{equation}
 \partialdd{f_S^{hh\rightarrow SS}\left(r,x\right)}{r} \equiv \DD{T}{r} \DD{t}{T} C_{hh\rightarrow SS}^{S} = \frac{M_0}{m_S} \frac{1}{\sqrt{x^2+r^2}} \frac{\lambda^2}{64\pi^4} \exp\left(-\sqrt{x^2+r^2}\right) \mathcal{F}\left(x,r\right) \,.
 \label{eq:CollisionTerm_hhToSS}
\end{equation}

For the process of a pair of scalars annihilating into a pair of Higgs bosons, the collision term reads: 
\begin{align}
{C}_{SS \rightarrow hh}^{S}\left(q\right) &=  \left(-\frac{1}{2E_q}\right) \int \invps{q'} \invps{p} \invps{p'} \frac{4\lambda^2}{8 \pi} \nonumber \\
 &\times \twopin{4} \deltadistn{4}{q+q'-p-p'}f_S\left(q\right) f_S\left(q'\right) \nonumber \\
&= -\frac{f_S\left(q\right)}{2E_q}\frac{4\lambda^2}{16 \pi} \int \invps{q'} f_S\left(q'\right) \sqrt{\frac{\left(E_q +E_{q'}\right)^2 -qq'\cos \theta - 4m_h^2}{\left(E_q +E_{q'}\right)^2 -qq'\cos \theta}} \,.
\label{eq:BoltzmannScalarAnn}
\end{align}
Again, we can directly infer the definition of $\mathcal{G}$ as in \equref{eq:CollisionTerm_SSTohh}:
\begin{dmath}
 \mathcal{G}\left(x,r,\xi\right) \equiv 2\pi \int\limits_0^\infty \diffd \hat{x}\, \hat{x}^2 \int\limits_{-1}^{\sub{\alpha}{max}} \diffd \cos \theta \frac{1}{\sqrt{\hat{x}^2+r^2}} \times \sqrt{\frac{\left(\sqrt{\hat{x}^2+r^2} + \sqrt{x^2+r^2}\right)^2 -x\hat{x} \cos \theta - 4 \xi^2 r^2}{\left(\sqrt{\hat{x}^2+r^2} + \sqrt{x^2+r^2}\right)^2 -x\hat{x} \cos \theta}} \, .
  \label{eq:KinematicIntegral_SSTohh}
\end{dmath}
Note that the only difference between Eqs.~\eqref{eq:KinematicIntegral_SSTohh} and~\eqref{eq:KinematicIntegral_hhToSS} is the exponential factor in the integral, stemming from the equilibrium distribution in \equref{eq:BoltzmannScalarRaw}. The entity $\sub{\alpha}{max}$ is again defined as in \equref{eq:DefCosThetaMax}.

Thus, the second part of the kinetic equation of the scalar (again cf.~\equref{eq:DynamicsScalar}) reads:
\begin{equation}
 \partialdd{f_S^{SS\rightarrow hh}\left(r,x\right)}{r} \equiv \DD{T}{r} \DD{t}{T} \: C_{SS\rightarrow hh}^{S} = - \frac{M_0}{m_S} \frac{1}{\sqrt{x^2+r^2}} \frac{\lambda^2}{64\pi^4} f_S\left(x,r\right) \int{\diffd^3 \hat{x} f_S\left(\hat{x},r\right) \mathcal{G}\left(\hat{x},r\right)} \,.
 \label{eq:CollisionTerm_SSTohh}
\end{equation}
Note that the collision term in \equref{eq:CollisionTerm_SSTohh} contains the actual distribution function of the scalar on the right-hand side, yielding an integro-differential equation for the distribution function $f_S$ we are interested in.

The term describing the decay of a scalar into a pair of sterile neutrinos is constructed as:
\begin{align}
 C_{S\rightarrow NN}^{S}\left(q\right) &= - \frac{1}{2 E_q} \iint \frac{2\diffd ^3 p}{\left(2\pi\right)^3 2 E_{p}}  \frac{2\diffd ^3 p'}{\left(2\pi\right)^3 2 E_{p'}} \frac{1}{2} y^2 E_p E_{p'} \left[1 - \frac{\vec{p} \cdot \vec{p'}}{E_p E_{p'}}\right] \twopin{4} \deltadistn{4}{q -p -p'} f_s\left(q\right)\nonumber \\
&= - f_S\left(q\right) \Gamma \frac{m_S}{E_q} \, ,
\label{eq:CollisionTermS:SToNNDerivation}
\end{align}
where we have explicitly inserted the decay width $\Gamma = \frac{y^2 m_S}{16\pi}$. This collision term can be interpreted intuitively: it is clear that the rate at which scalars are depleted due to decays must be proportional to the time-dilated decay width $\Gamma \frac{m_S}{E_q}$, which involves an additional boost factor owing to the physical momentum $q$ of the scalar, and that it must be proportional to the amount of scalars present at that particular momentum, i.e.\ to $f_S\left(q\right)$.

With this, the third part of the kinetic equation of the scalar, is then given by:
\begin{equation}
 \partialdd{f_S^{S\rightarrow NN}\left(r,x\right)}{r} \equiv \DD{T}{r} \DD{t}{T}\: C_{S\rightarrow NN}^{S} = -  \frac{M_0}{m_S} \frac{1}{\sqrt{x^2+r^2}} r^2 \frac{\Gamma}{m_S} f_S\left(x,r\right)\,.
 \label{eq:CollisionTerm_SToNN_S}
\end{equation}

\subsection{\label{subapp:DetailsKineticEquationsSN}Kinetic equation for the sterile neutrino}

We also want to show the most import steps to derive at the kinetic equation of the sterile neutrino. The source term looks like:
\begin{align}
C_{S\rightarrow NN}^{N} = 2\times \frac{1}{2E_p} \iint  \frac{2\diffd ^3 p'}{\left(2\pi\right)^3 2 E_{p'}} \invps{p_S} \twopin{4} \deltadist{E_{p_S} - E_p -E_{p'}} \deltadistn{3}{\vec{p}_S -\vec{p} -\vec{p'}} 2 \Msquared f_S\left(p_S,t \right) \,,
 \label{eq:CSNNN}
\end{align}
where the matrix element for the decay reads:
\begin{align}
 \Msquared = \frac{1}{2}y^2 p \cdot p'= \frac{1}{2} y^2 E_p E_{p'} \left[1- \frac{\vec{p} \cdot \vec{p'}}{E_p E_{p'}}\right]\,.
\end{align}
According to our conventions for symmetry factors, we average over initial \emph{and} final states.

Integrating out the phase space in $p'$, we can get rid of the spatial $\delta$-distribution in Eq.~\eqref{eq:CSNNN}. Doing so, we can replace $\vec{p} \cdot \vec{p'}$ by $\vec{p} \cdot \vec{p}_S -p^2$. The scalar product $\vec{p} \cdot \vec{p}_S$ is restricted by the kinematics of the process. Using $-1\leq \cos [\sphericalangle (\vec{p},\vec{p}_S) ] \leq 1$, where $\sphericalangle (\vec{a},\vec{b})$ is the angle between the 3-vectors $\vec{a}$ and $\vec{b}$, this gives the constraint 
\begin{align}
 p_S \geq \left| p - \frac{m_S^2}{4p} \right| \equiv \sub{p}{min} \,,
\end{align}
which implies the lower boundary in \equref{eq:CollisionTerm_SToNN_N}.

Hence, using \equref{eq:DynamicsSN}, the kinetic equation of the sterile neutrino is finally given by
\begin{align}
 \partialdd{f_N^{S\rightarrow NN} \left(x,r\right)}{r}=2 \CGamma \frac{r^2}{x^2} \int_{\sub{\hat{x}}{min}}^{\infty}{\diffd \hat{x} \frac{\hat{x}}{\sqrt{\hat{x}^2 + r^2}} f_S\left(\hat{x},r\right)}
 \label{eq:CollisionTerm_SToNN_N} \,,
\end{align}
which we use as our master equation, cf.\ Eq.~\eqref{eq:KineticEquationSNDoubleIntegral}.

\section{\label{app:DetailsrFS}Appendix: Some background on the numerical calculation of the free-streaming horizon}
\renewcommand{\theequation}{B-\arabic{equation}}
\setcounter{equation}{0}  

This appendix briefly summarises some technical details implemented in our numerics in order to evaluate the integral occurring in the definition of the free-streaming horizon, cf.~\equref{eq:DefVMean}. We treat the thermal history of the Universe in a way where there are two distinct periods of interest, namely a purely radiation-dominated one which is followed by an immediate turnover into complete matter domination at $\left(\sub{T}{eq},\sub{t}{eq}\right)=\left(\unit{1.48}{eV}, \unit{6.04 \times 10^{11}}{s}\right)$. Note that these quantities stemming from our rather crude approximation agree fairly well with the values that can be found in the literature.

During radiation dominance, we can infer the time-temperature relation from~\equref{eq:DefM0} using the evolution of the d.o.f.\ as given in~\cite{PhysRevD.82.123508}. A relation between the scale factor and the temperature can be established by solving for $T$ in the equation of conservation of comoving entropy density,
\begin{align}
\frac{2\pi}{45} g_{*S}T^3 a^3 =\mathrm{const.}=s_0,
\label{eq:ScaleTimeRelation}
\end{align}
normalising today's scale factor to unity. In our numerics, we implement the numerical evolution of the d.o.f.\ as presented in~\cite{PhysRevD.82.123508}.

During matter dominance, integrating the Friedmann equation gives a relation between the cosmological time and the scale factor which reads
\begin{equation}
 t= \sub{t}{eq} + \sqrt{\frac{3}{8\pi}}\frac{2}{3} \sub{M}{Pl} \left(\rho_M^0\right)^{-1/2}\left(a^{3/2}-\sub{a}{eq}^{3/2}\right)\,.
 \label{eq:TimeScaleRelationMatterDominance}
\end{equation}
This can be converted into a time-temperature relation by virtue of \equref{eq:ScaleTimeRelation}.

We have checked that our numerical treatment of the time-temperature relation reproduces other known benchmark points (like the time of CMB-decoupling) to a very reasonable accuracy.

\bibliographystyle{./apsrev}
\bibliography{ScalProdBIB}

\begin{thebibliography}{10}
\expandafter\ifx\csname bibnamefont\endcsname\relax
  \def\bibnamefont#1{#1}\fi
\expandafter\ifx\csname bibfnamefont\endcsname\relax
  \def\bibfnamefont#1{#1}\fi
\expandafter\ifx\csname url\endcsname\relax
  \def\url#1{\texttt{#1}}\fi
\expandafter\ifx\csname urlprefix\endcsname\relax\def\urlprefix{URL }\fi
\providecommand{\bibinfo}[2]{#2}
\providecommand{\eprint}[2][]{\url{#2}}

\bibitem{Ade:2013zuv}
\bibinfo{author}{\bibfnamefont{P.~A.~R.} \bibnamefont{Ade}} \emph{et~al.}
  (\bibinfo{collaboration}{Planck Collaboration}), \bibinfo{journal}{Astron.
  Astrophys.} \textbf{\bibinfo{volume}{571}}, \bibinfo{pages}{A16}
  (\bibinfo{year}{2014}), \eprint{1303.5076}.

\bibitem{Planck:2015xua}
\bibinfo{author}{\bibfnamefont{P.}~\bibnamefont{Ade}} \emph{et~al.}
  (\bibinfo{collaboration}{Planck})  (\bibinfo{year}{2015}),
  \eprint{1502.01589}.

\bibitem{Aprile:2012nq}
\bibinfo{author}{\bibfnamefont{E.}~\bibnamefont{Aprile}} \emph{et~al.}
  (\bibinfo{collaboration}{XENON100 Collaboration}), \bibinfo{journal}{Phys.
  Rev. Lett.} \textbf{\bibinfo{volume}{109}}, \bibinfo{pages}{181301}
  (\bibinfo{year}{2012}), \eprint{1207.5988}.

\bibitem{Akerib:2013tjd}
\bibinfo{author}{\bibfnamefont{D.~S.} \bibnamefont{Akerib}} \emph{et~al.}
  (\bibinfo{collaboration}{LUX Collaboration}), \bibinfo{journal}{Phys. Rev.
  Lett.} \textbf{\bibinfo{volume}{112}}(\bibinfo{number}{9}),
  \bibinfo{pages}{091303} (\bibinfo{year}{2014}), \eprint{1310.8214}.

\bibitem{Agnese:2014aze}
\bibinfo{author}{\bibfnamefont{R.}~\bibnamefont{Agnese}} \emph{et~al.}
  (\bibinfo{collaboration}{SuperCDMS Collaboration}), \bibinfo{journal}{Phys.
  Rev. Lett.} \textbf{\bibinfo{volume}{112}}(\bibinfo{number}{24}),
  \bibinfo{pages}{241302} (\bibinfo{year}{2014}), \eprint{1402.7137}.

\bibitem{Angloher:2014myn}
\bibinfo{author}{\bibfnamefont{G.}~\bibnamefont{Angloher}} \emph{et~al.}
  (\bibinfo{collaboration}{CRESST-II Collaboration}), \bibinfo{journal}{Eur.
  Phys. J.} \textbf{\bibinfo{volume}{C74}}(\bibinfo{number}{12}),
  \bibinfo{pages}{3184} (\bibinfo{year}{2014}), \eprint{1407.3146}.

\bibitem{Baer:2014eja}
\bibinfo{author}{\bibfnamefont{H.}~\bibnamefont{Baer}},
  \bibinfo{author}{\bibfnamefont{K.-Y.} \bibnamefont{Choi}},
  \bibinfo{author}{\bibfnamefont{J.~E.} \bibnamefont{Kim}}, \bibnamefont{and}
  \bibinfo{author}{\bibfnamefont{L.}~\bibnamefont{Roszkowski}},
  \bibinfo{journal}{Phys. Rept.} \textbf{\bibinfo{volume}{555}},
  \bibinfo{pages}{1} (\bibinfo{year}{2015}), \eprint{1407.0017}.

\bibitem{Duffy:2009ig}
\bibinfo{author}{\bibfnamefont{L.~D.} \bibnamefont{Duffy}} \bibnamefont{and}
  \bibinfo{author}{\bibfnamefont{K.}~\bibnamefont{van Bibber}},
  \bibinfo{journal}{New J. Phys.} \textbf{\bibinfo{volume}{11}},
  \bibinfo{pages}{105008} (\bibinfo{year}{2009}), \eprint{0904.3346}.

\bibitem{Steffen:2006hw}
\bibinfo{author}{\bibfnamefont{F.~D.} \bibnamefont{Steffen}},
  \bibinfo{journal}{JCAP} \textbf{\bibinfo{volume}{0609}}, \bibinfo{pages}{001}
  (\bibinfo{year}{2006}), \eprint{hep-ph/0605306}.

\bibitem{Choi:2013lwa}
\bibinfo{author}{\bibfnamefont{K.-Y.} \bibnamefont{Choi}},
  \bibinfo{author}{\bibfnamefont{J.~E.} \bibnamefont{Kim}}, \bibnamefont{and}
  \bibinfo{author}{\bibfnamefont{L.}~\bibnamefont{Roszkowski}},
  \bibinfo{journal}{J. Korean Phys. Soc.} \textbf{\bibinfo{volume}{63}},
  \bibinfo{pages}{1685} (\bibinfo{year}{2013}), \eprint{1307.3330}.

\bibitem{Kolb:1998ki}
\bibinfo{author}{\bibfnamefont{E.~W.} \bibnamefont{Kolb}},
  \bibinfo{author}{\bibfnamefont{D.~J.} \bibnamefont{Chung}}, \bibnamefont{and}
  \bibinfo{author}{\bibfnamefont{A.}~\bibnamefont{Riotto}} pp.
  \bibinfo{pages}{91--105} (\bibinfo{year}{1998}), \eprint{hep-ph/9810361}.

\bibitem{Abazajian:2012ys}
\bibinfo{author}{\bibfnamefont{K.~N.} \bibnamefont{Abazajian}},
  \bibinfo{author}{\bibfnamefont{M.~A.} \bibnamefont{Acero}},
  \bibinfo{author}{\bibfnamefont{S.~K.} \bibnamefont{Agarwalla}},
  \bibinfo{author}{\bibfnamefont{A.~A.} \bibnamefont{Aguilar-Arevalo}},
  \bibinfo{author}{\bibfnamefont{C.~H.} \bibnamefont{Albright}}, \emph{et~al.}
  (\bibinfo{year}{2012}), \eprint{1204.5379}.

\bibitem{Merle:2013gea}
\bibinfo{author}{\bibfnamefont{A.}~\bibnamefont{Merle}}, \bibinfo{journal}{Int.
  J. Mod. Phys.} \textbf{\bibinfo{volume}{D22}}, \bibinfo{pages}{1330020}
  (\bibinfo{year}{2013}), \eprint{1302.2625}.

\bibitem{Dodelson:1993je}
\bibinfo{author}{\bibfnamefont{S.}~\bibnamefont{Dodelson}} \bibnamefont{and}
  \bibinfo{author}{\bibfnamefont{L.~M.} \bibnamefont{Widrow}},
  \bibinfo{journal}{Phys. Rev. Lett.} \textbf{\bibinfo{volume}{72}},
  \bibinfo{pages}{17} (\bibinfo{year}{1994}), \eprint{hep-ph/9303287}.

\bibitem{Colombi:1995ze}
\bibinfo{author}{\bibfnamefont{S.}~\bibnamefont{Colombi}},
  \bibinfo{author}{\bibfnamefont{S.}~\bibnamefont{Dodelson}}, \bibnamefont{and}
  \bibinfo{author}{\bibfnamefont{L.~M.} \bibnamefont{Widrow}},
  \bibinfo{journal}{Astrophys. J.} \textbf{\bibinfo{volume}{458}},
  \bibinfo{pages}{1} (\bibinfo{year}{1996}), \eprint{astro-ph/9505029}.

\bibitem{Boyarsky:2008xj}
\bibinfo{author}{\bibfnamefont{A.}~\bibnamefont{Boyarsky}},
  \bibinfo{author}{\bibfnamefont{J.}~\bibnamefont{Lesgourgues}},
  \bibinfo{author}{\bibfnamefont{O.}~\bibnamefont{Ruchayskiy}},
  \bibnamefont{and} \bibinfo{author}{\bibfnamefont{M.}~\bibnamefont{Viel}},
  \bibinfo{journal}{JCAP} \textbf{\bibinfo{volume}{0905}}, \bibinfo{pages}{012}
  (\bibinfo{year}{2009}), \eprint{0812.0010}.

\bibitem{Canetti:2012kh}
\bibinfo{author}{\bibfnamefont{L.}~\bibnamefont{Canetti}},
  \bibinfo{author}{\bibfnamefont{M.}~\bibnamefont{Drewes}},
  \bibinfo{author}{\bibfnamefont{T.}~\bibnamefont{Frossard}}, \bibnamefont{and}
  \bibinfo{author}{\bibfnamefont{M.}~\bibnamefont{Shaposhnikov}},
  \bibinfo{journal}{Phys. Rev.}
  \textbf{\bibinfo{volume}{D87}}(\bibinfo{number}{9}), \bibinfo{pages}{093006}
  (\bibinfo{year}{2013}), \eprint{1208.4607}.

\bibitem{Merle:2013ibc}
\bibinfo{author}{\bibfnamefont{A.}~\bibnamefont{Merle}} \bibnamefont{and}
  \bibinfo{author}{\bibfnamefont{V.}~\bibnamefont{Niro}},
  \bibinfo{journal}{Phys. Rev.}
  \textbf{\bibinfo{volume}{D88}}(\bibinfo{number}{11}), \bibinfo{pages}{113004}
  (\bibinfo{year}{2013}), \eprint{1302.2032}.

\bibitem{Bulbul:2014sua}
\bibinfo{author}{\bibfnamefont{E.}~\bibnamefont{Bulbul}},
  \bibinfo{author}{\bibfnamefont{M.}~\bibnamefont{Markevitch}},
  \bibinfo{author}{\bibfnamefont{A.}~\bibnamefont{Foster}},
  \bibinfo{author}{\bibfnamefont{R.~K.} \bibnamefont{Smith}},
  \bibinfo{author}{\bibfnamefont{M.}~\bibnamefont{Loewenstein}}, \emph{et~al.},
  \bibinfo{journal}{Astrophys. J.} \textbf{\bibinfo{volume}{789}},
  \bibinfo{pages}{13} (\bibinfo{year}{2014}), \eprint{1402.2301}.

\bibitem{Boyarsky:2014jta}
\bibinfo{author}{\bibfnamefont{A.}~\bibnamefont{Boyarsky}},
  \bibinfo{author}{\bibfnamefont{O.}~\bibnamefont{Ruchayskiy}},
  \bibinfo{author}{\bibfnamefont{D.}~\bibnamefont{Iakubovskyi}},
  \bibnamefont{and} \bibinfo{author}{\bibfnamefont{J.}~\bibnamefont{Franse}},
  \bibinfo{journal}{Phys. Rev. Lett.} \textbf{\bibinfo{volume}{113}},
  \bibinfo{pages}{251301} (\bibinfo{year}{2014}), \eprint{1402.4119}.

\bibitem{Riemer-Sorensen:2014yda}
\bibinfo{author}{\bibfnamefont{S.}~\bibnamefont{Riemer-Sorensen}}
  (\bibinfo{year}{2014}), \eprint{1405.7943}.

\bibitem{Anderson:2014tza}
\bibinfo{author}{\bibfnamefont{M.~E.} \bibnamefont{Anderson}},
  \bibinfo{author}{\bibfnamefont{E.}~\bibnamefont{Churazov}}, \bibnamefont{and}
  \bibinfo{author}{\bibfnamefont{J.~N.} \bibnamefont{Bregman}}
  (\bibinfo{year}{2014}), \eprint{1408.4115}.

\bibitem{Boyarsky:2014ska}
\bibinfo{author}{\bibfnamefont{A.}~\bibnamefont{Boyarsky}},
  \bibinfo{author}{\bibfnamefont{J.}~\bibnamefont{Franse}},
  \bibinfo{author}{\bibfnamefont{D.}~\bibnamefont{Iakubovskyi}},
  \bibnamefont{and}
  \bibinfo{author}{\bibfnamefont{O.}~\bibnamefont{Ruchayskiy}}
  (\bibinfo{year}{2014}), \eprint{1408.2503}.

\bibitem{Jeltema:2014qfa}
\bibinfo{author}{\bibfnamefont{T.~E.} \bibnamefont{Jeltema}} \bibnamefont{and}
  \bibinfo{author}{\bibfnamefont{S.}~\bibnamefont{Profumo}}
  (\bibinfo{year}{2014}), \eprint{1408.1699}.

\bibitem{Boyarsky:2014paa}
\bibinfo{author}{\bibfnamefont{A.}~\bibnamefont{Boyarsky}},
  \bibinfo{author}{\bibfnamefont{J.}~\bibnamefont{Franse}},
  \bibinfo{author}{\bibfnamefont{D.}~\bibnamefont{Iakubovskyi}},
  \bibnamefont{and}
  \bibinfo{author}{\bibfnamefont{O.}~\bibnamefont{Ruchayskiy}}
  (\bibinfo{year}{2014}), \eprint{1408.4388}.

\bibitem{Bulbul:2014ala}
\bibinfo{author}{\bibfnamefont{E.}~\bibnamefont{Bulbul}},
  \bibinfo{author}{\bibfnamefont{M.}~\bibnamefont{Markevitch}},
  \bibinfo{author}{\bibfnamefont{A.~R.} \bibnamefont{Foster}},
  \bibinfo{author}{\bibfnamefont{R.~K.} \bibnamefont{Smith}},
  \bibinfo{author}{\bibfnamefont{M.}~\bibnamefont{Loewenstein}}, \emph{et~al.}
  (\bibinfo{year}{2014}), \eprint{1409.4143}.

\bibitem{Malyshev:2014xqa}
\bibinfo{author}{\bibfnamefont{D.}~\bibnamefont{Malyshev}},
  \bibinfo{author}{\bibfnamefont{A.}~\bibnamefont{Neronov}}, \bibnamefont{and}
  \bibinfo{author}{\bibfnamefont{D.}~\bibnamefont{Eckert}},
  \bibinfo{journal}{Phys. Rev.}
  \textbf{\bibinfo{volume}{D90}}(\bibinfo{number}{10}), \bibinfo{pages}{103506}
  (\bibinfo{year}{2014}), \eprint{1408.3531}.

\bibitem{Shi:1998km}
\bibinfo{author}{\bibfnamefont{X.-D.} \bibnamefont{Shi}} \bibnamefont{and}
  \bibinfo{author}{\bibfnamefont{G.~M.} \bibnamefont{Fuller}},
  \bibinfo{journal}{Phys. Rev. Lett.} \textbf{\bibinfo{volume}{82}},
  \bibinfo{pages}{2832} (\bibinfo{year}{1999}), \eprint{astro-ph/9810076}.

\bibitem{Abazajian:2014gza}
\bibinfo{author}{\bibfnamefont{K.~N.} \bibnamefont{Abazajian}},
  \bibinfo{journal}{Phys. Rev. Lett.}
  \textbf{\bibinfo{volume}{112}}(\bibinfo{number}{16}), \bibinfo{pages}{161303}
  (\bibinfo{year}{2014}), \eprint{1403.0954}.

\bibitem{Popa:2015eta}
\bibinfo{author}{\bibfnamefont{L.~A.} \bibnamefont{Popa}},
  \bibinfo{author}{\bibfnamefont{A.}~\bibnamefont{Caramete}}, \bibnamefont{and}
  \bibinfo{author}{\bibfnamefont{D.}~\bibnamefont{Tonoiu}}
  (\bibinfo{year}{2015}), \eprint{1501.06355}.

\bibitem{Merle:2014xpa}
\bibinfo{author}{\bibfnamefont{A.}~\bibnamefont{Merle}} \bibnamefont{and}
  \bibinfo{author}{\bibfnamefont{A.}~\bibnamefont{Schneider}}
  (\bibinfo{year}{2014}), \eprint{1409.6311}.

\bibitem{Bezrukov:2009th}
\bibinfo{author}{\bibfnamefont{F.}~\bibnamefont{Bezrukov}},
  \bibinfo{author}{\bibfnamefont{H.}~\bibnamefont{Hettmansperger}},
  \bibnamefont{and} \bibinfo{author}{\bibfnamefont{M.}~\bibnamefont{Lindner}},
  \bibinfo{journal}{Phys. Rev.} \textbf{\bibinfo{volume}{D81}},
  \bibinfo{pages}{085032} (\bibinfo{year}{2010}), \eprint{0912.4415}.

\bibitem{Nemevsek:2012cd}
\bibinfo{author}{\bibfnamefont{M.}~\bibnamefont{Nemevsek}},
  \bibinfo{author}{\bibfnamefont{G.}~\bibnamefont{Senjanovic}},
  \bibnamefont{and} \bibinfo{author}{\bibfnamefont{Y.}~\bibnamefont{Zhang}},
  \bibinfo{journal}{JCAP} \textbf{\bibinfo{volume}{1207}}, \bibinfo{pages}{006}
  (\bibinfo{year}{2012}), \eprint{1205.0844}.

\bibitem{King:2012wg}
\bibinfo{author}{\bibfnamefont{S.~F.} \bibnamefont{King}} \bibnamefont{and}
  \bibinfo{author}{\bibfnamefont{A.}~\bibnamefont{Merle}},
  \bibinfo{journal}{JCAP} \textbf{\bibinfo{volume}{1208}}, \bibinfo{pages}{016}
  (\bibinfo{year}{2012}), \eprint{1205.0551}.

\bibitem{Shaposhnikov:2006xi}
\bibinfo{author}{\bibfnamefont{M.}~\bibnamefont{Shaposhnikov}}
  \bibnamefont{and} \bibinfo{author}{\bibfnamefont{I.}~\bibnamefont{Tkachev}},
  \bibinfo{journal}{Phys. Lett.} \textbf{\bibinfo{volume}{B639}},
  \bibinfo{pages}{414} (\bibinfo{year}{2006}), \eprint{hep-ph/0604236}.

\bibitem{Bezrukov:2009yw}
\bibinfo{author}{\bibfnamefont{F.}~\bibnamefont{Bezrukov}} \bibnamefont{and}
  \bibinfo{author}{\bibfnamefont{D.}~\bibnamefont{Gorbunov}},
  \bibinfo{journal}{JHEP} \textbf{\bibinfo{volume}{1005}}, \bibinfo{pages}{010}
  (\bibinfo{year}{2010}).

\bibitem{Kusenko:2006rh}
\bibinfo{author}{\bibfnamefont{A.}~\bibnamefont{Kusenko}},
  \bibinfo{journal}{Phys. Rev. Lett.} \textbf{\bibinfo{volume}{97}},
  \bibinfo{pages}{241301} (\bibinfo{year}{2006}), \eprint{hep-ph/0609081}.

\bibitem{Petraki:2007gq}
\bibinfo{author}{\bibfnamefont{K.}~\bibnamefont{Petraki}} \bibnamefont{and}
  \bibinfo{author}{\bibfnamefont{A.}~\bibnamefont{Kusenko}},
  \bibinfo{journal}{Phys. Rev.} \textbf{\bibinfo{volume}{D77}},
  \bibinfo{pages}{065014} (\bibinfo{year}{2008}), \eprint{0711.4646}.

\bibitem{Merle:2013wta}
\bibinfo{author}{\bibfnamefont{A.}~\bibnamefont{Merle}},
  \bibinfo{author}{\bibfnamefont{V.}~\bibnamefont{Niro}}, \bibnamefont{and}
  \bibinfo{author}{\bibfnamefont{D.}~\bibnamefont{Schmidt}},
  \bibinfo{journal}{JCAP} \textbf{\bibinfo{volume}{1403}}, \bibinfo{pages}{028}
  (\bibinfo{year}{2014}), \eprint{1306.3996}.

\bibitem{Adulpravitchai:2014xna}
\bibinfo{author}{\bibfnamefont{A.}~\bibnamefont{Adulpravitchai}}
  \bibnamefont{and} \bibinfo{author}{\bibfnamefont{M.~A.}
  \bibnamefont{Schmidt}}, \bibinfo{journal}{JHEP}
  \textbf{\bibinfo{volume}{1501}}, \bibinfo{pages}{006} (\bibinfo{year}{2015}),
  \eprint{1409.4330}.

\bibitem{Kang:2014cia}
\bibinfo{author}{\bibfnamefont{Z.}~\bibnamefont{Kang}}  (\bibinfo{year}{2014}),
  \eprint{1411.2773}.

\bibitem{Frigerio:2014ifa}
\bibinfo{author}{\bibfnamefont{M.}~\bibnamefont{Frigerio}} \bibnamefont{and}
  \bibinfo{author}{\bibfnamefont{C.~E.} \bibnamefont{Yaguna}},
  \bibinfo{journal}{Eur.~Phys.~J.}
  \textbf{\bibinfo{volume}{C75}}(\bibinfo{number}{1}), \bibinfo{pages}{31}
  (\bibinfo{year}{2015}), \eprint{1409.0659}.

\bibitem{Lello:2014yha}
\bibinfo{author}{\bibfnamefont{L.}~\bibnamefont{Lello}} \bibnamefont{and}
  \bibinfo{author}{\bibfnamefont{D.}~\bibnamefont{Boyanovsky}},
  \bibinfo{journal}{Phys.~Rev.}
  \textbf{\bibinfo{volume}{D91}}(\bibinfo{number}{6}), \bibinfo{pages}{063502}
  (\bibinfo{year}{2015}), \eprint{1411.2690}.

\bibitem{Abada:2014zra}
\bibinfo{author}{\bibfnamefont{A.}~\bibnamefont{Abada}},
  \bibinfo{author}{\bibfnamefont{G.}~\bibnamefont{Arcadi}}, \bibnamefont{and}
  \bibinfo{author}{\bibfnamefont{M.}~\bibnamefont{Lucente}},
  \bibinfo{journal}{Journal of Cosmology and Astroparticle Physics}
  \textbf{\bibinfo{volume}{2014}}(\bibinfo{number}{10}), \bibinfo{pages}{001}
  (\bibinfo{year}{2014}).

\bibitem{Boyanovsky:2008nc}
\bibinfo{author}{\bibfnamefont{D.}~\bibnamefont{Boyanovsky}},
  \bibinfo{journal}{Phys. Rev.} \textbf{\bibinfo{volume}{D78}},
  \bibinfo{pages}{103505} (\bibinfo{year}{2008}), \eprint{0807.0646}.

\bibitem{Shuve:2014doa}
\bibinfo{author}{\bibfnamefont{B.}~\bibnamefont{Shuve}} \bibnamefont{and}
  \bibinfo{author}{\bibfnamefont{I.}~\bibnamefont{Yavin}},
  \bibinfo{journal}{Phys. Rev.}
  \textbf{\bibinfo{volume}{D89}}(\bibinfo{number}{11}), \bibinfo{pages}{113004}
  (\bibinfo{year}{2014}), \eprint{1403.2727}.

\bibitem{Petraki:2008ef}
\bibinfo{author}{\bibfnamefont{K.}~\bibnamefont{Petraki}},
  \bibinfo{journal}{Phys. Rev.} \textbf{\bibinfo{volume}{D77}},
  \bibinfo{pages}{105004} (\bibinfo{year}{2008}), \eprint{0801.3470}.

\bibitem{Bezrukov:2014qda}
\bibinfo{author}{\bibfnamefont{F.}~\bibnamefont{Bezrukov}} \bibnamefont{and}
  \bibinfo{author}{\bibfnamefont{D.}~\bibnamefont{Gorbunov}}
  (\bibinfo{year}{2014}), \eprint{1412.1341}.

\bibitem{Merle:Proj:RefinedScalarProd&DW}
\bibinfo{author}{\bibfnamefont{A.}~\bibnamefont{Merle}},
  \bibinfo{author}{\bibfnamefont{A.}~\bibnamefont{Schneider}},
  \bibnamefont{and} \bibinfo{author}{\bibfnamefont{M.}~\bibnamefont{Totzauer}}
  \bibinfo{note}{{\emph{A Fully Numerical Investigation of keV Sterile Neutrino
  Dark Matter produced from Singlet Scalar Decays (Work in progress)}}}.

\bibitem{Merle:Proj:StructureFormation}
\bibinfo{author}{\bibfnamefont{A.}~\bibnamefont{Merle}},
  \bibinfo{author}{\bibfnamefont{A.}~\bibnamefont{Schneider}},
  \bibnamefont{and} \bibinfo{author}{\bibfnamefont{M.}~\bibnamefont{Totzauer}}
  \bibinfo{note}{{\emph{Structure Formation Properties of keV Sterile Neutrino
  Dark Matter produced from Singlet Scalar Decays (Work in progress)}}}.

\bibitem{McDonald:2001vt}
\bibinfo{author}{\bibfnamefont{J.}~\bibnamefont{McDonald}},
  \bibinfo{journal}{Phys. Rev. Lett.} \textbf{\bibinfo{volume}{88}},
  \bibinfo{pages}{091304} (\bibinfo{year}{2002}), \eprint{hep-ph/0106249}.

\bibitem{Hall:2009bx}
\bibinfo{author}{\bibfnamefont{L.~J.} \bibnamefont{Hall}},
  \bibinfo{author}{\bibfnamefont{K.}~\bibnamefont{Jedamzik}},
  \bibinfo{author}{\bibfnamefont{J.}~\bibnamefont{March-Russell}},
  \bibnamefont{and} \bibinfo{author}{\bibfnamefont{S.~M.} \bibnamefont{West}},
  \bibinfo{journal}{JHEP} \textbf{\bibinfo{volume}{1003}}, \bibinfo{pages}{080}
  (\bibinfo{year}{2010}), \eprint{0911.1120}.

\bibitem{Klypin:1999uc}
\bibinfo{author}{\bibfnamefont{A.~A.} \bibnamefont{Klypin}},
  \bibinfo{author}{\bibfnamefont{A.~V.} \bibnamefont{Kravtsov}},
  \bibinfo{author}{\bibfnamefont{O.}~\bibnamefont{Valenzuela}},
  \bibnamefont{and} \bibinfo{author}{\bibfnamefont{F.}~\bibnamefont{Prada}},
  \bibinfo{journal}{Astrophys. J.} \textbf{\bibinfo{volume}{522}},
  \bibinfo{pages}{82} (\bibinfo{year}{1999}), \eprint{astro-ph/9901240}.

\bibitem{BoylanKolchin:2011dk}
\bibinfo{author}{\bibfnamefont{M.}~\bibnamefont{Boylan-Kolchin}},
  \bibinfo{author}{\bibfnamefont{J.~S.} \bibnamefont{Bullock}},
  \bibnamefont{and}
  \bibinfo{author}{\bibfnamefont{M.}~\bibnamefont{Kaplinghat}},
  \bibinfo{journal}{Mon. Not. Roy. Astron. Soc.}
  \textbf{\bibinfo{volume}{422}}, \bibinfo{pages}{1203} (\bibinfo{year}{2012}),
  \eprint{1111.2048}.

\bibitem{Navarro:1996gj}
\bibinfo{author}{\bibfnamefont{J.~F.} \bibnamefont{Navarro}},
  \bibinfo{author}{\bibfnamefont{C.~S.} \bibnamefont{Frenk}}, \bibnamefont{and}
  \bibinfo{author}{\bibfnamefont{S.~D.} \bibnamefont{White}},
  \bibinfo{journal}{Astrophys. J.} \textbf{\bibinfo{volume}{490}},
  \bibinfo{pages}{493} (\bibinfo{year}{1997}), \eprint{astro-ph/9611107}.

\bibitem{Ferrero:2011au}
\bibinfo{author}{\bibfnamefont{I.}~\bibnamefont{Ferrero}},
  \bibinfo{author}{\bibfnamefont{M.~G.} \bibnamefont{Abadi}},
  \bibinfo{author}{\bibfnamefont{J.~F.} \bibnamefont{Navarro}},
  \bibinfo{author}{\bibfnamefont{L.~V.} \bibnamefont{Sales}}, \bibnamefont{and}
  \bibinfo{author}{\bibfnamefont{S.}~\bibnamefont{Gurovich}},
  \bibinfo{journal}{Mon. Not. Roy. Astron. Soc.}
  \textbf{\bibinfo{volume}{425}}, \bibinfo{pages}{2817} (\bibinfo{year}{2012}),
  \eprint{1111.6609}.

\bibitem{Agashe:2014kda}
\bibinfo{author}{\bibfnamefont{K.~A.} \bibnamefont{Olive}} \emph{et~al.}
  (\bibinfo{collaboration}{Particle Data Group}), \bibinfo{journal}{Chin.
  Phys.} \textbf{\bibinfo{volume}{C38}}, \bibinfo{pages}{090001}
  (\bibinfo{year}{2014}).

\bibitem{Abazajian:2004zh}
\bibinfo{author}{\bibfnamefont{K.}~\bibnamefont{Abazajian}},
  \bibinfo{author}{\bibfnamefont{E.~R.} \bibnamefont{Switzer}},
  \bibinfo{author}{\bibfnamefont{S.}~\bibnamefont{Dodelson}},
  \bibinfo{author}{\bibfnamefont{K.}~\bibnamefont{Heitmann}}, \bibnamefont{and}
  \bibinfo{author}{\bibfnamefont{S.}~\bibnamefont{Habib}},
  \bibinfo{journal}{Phys. Rev.} \textbf{\bibinfo{volume}{D71}},
  \bibinfo{pages}{043507} (\bibinfo{year}{2005}), \eprint{astro-ph/0411552}.

\bibitem{dePutter:2012sh}
\bibinfo{author}{\bibfnamefont{R.}~\bibnamefont{de~Putter}},
  \bibinfo{author}{\bibfnamefont{O.}~\bibnamefont{Mena}},
  \bibinfo{author}{\bibfnamefont{E.}~\bibnamefont{Giusarma}},
  \bibinfo{author}{\bibfnamefont{S.}~\bibnamefont{Ho}},
  \bibinfo{author}{\bibfnamefont{A.}~\bibnamefont{Cuesta}}, \emph{et~al.},
  \bibinfo{journal}{Astrophys. J.} \textbf{\bibinfo{volume}{761}},
  \bibinfo{pages}{12} (\bibinfo{year}{2012}), \eprint{1201.1909}.

\bibitem{Carmeli:2005if}
\bibinfo{author}{\bibfnamefont{M.}~\bibnamefont{Carmeli}},
  \bibinfo{author}{\bibfnamefont{J.~G.} \bibnamefont{Hartnett}},
  \bibnamefont{and} \bibinfo{author}{\bibfnamefont{F.~J.}
  \bibnamefont{Oliveira}}, \bibinfo{journal}{Found. Phys. Lett.}
  \textbf{\bibinfo{volume}{19}}, \bibinfo{pages}{277} (\bibinfo{year}{2006}),
  \eprint{gr-qc/0506079}.

\bibitem{PhysRevD.82.123508}
\bibinfo{author}{\bibfnamefont{O.}~\bibnamefont{Wantz}} \bibnamefont{and}
  \bibinfo{author}{\bibfnamefont{E.~P.~S.} \bibnamefont{Shellard}},
  \bibinfo{journal}{Phys. Rev.} \textbf{\bibinfo{volume}{D82}},
  \bibinfo{pages}{123508} (\bibinfo{year}{2010}), \eprint{0910.1066}.

\bibitem{Colin:2000dn}
\bibinfo{author}{\bibfnamefont{P.}~\bibnamefont{Colin}},
  \bibinfo{author}{\bibfnamefont{V.}~\bibnamefont{Avila-Reese}},
  \bibnamefont{and}
  \bibinfo{author}{\bibfnamefont{O.}~\bibnamefont{Valenzuela}},
  \bibinfo{journal}{Astrophys. J.} \textbf{\bibinfo{volume}{542}},
  \bibinfo{pages}{622} (\bibinfo{year}{2000}), \eprint{astro-ph/0004115}.

\bibitem{Lin:2000qq}
\bibinfo{author}{\bibfnamefont{W.~B.} \bibnamefont{Lin}},
  \bibinfo{author}{\bibfnamefont{D.~H.} \bibnamefont{Huang}},
  \bibinfo{author}{\bibfnamefont{X.}~\bibnamefont{Zhang}}, \bibnamefont{and}
  \bibinfo{author}{\bibfnamefont{R.~H.} \bibnamefont{Brandenberger}},
  \bibinfo{journal}{Phys. Rev. Lett.} \textbf{\bibinfo{volume}{86}},
  \bibinfo{pages}{954} (\bibinfo{year}{2001}), \eprint{astro-ph/0009003}.

\bibitem{Viel:2005qj}
\bibinfo{author}{\bibfnamefont{M.}~\bibnamefont{Viel}},
  \bibinfo{author}{\bibfnamefont{J.}~\bibnamefont{Lesgourgues}},
  \bibinfo{author}{\bibfnamefont{M.~G.} \bibnamefont{Haehnelt}},
  \bibinfo{author}{\bibfnamefont{S.}~\bibnamefont{Matarrese}},
  \bibnamefont{and} \bibinfo{author}{\bibfnamefont{A.}~\bibnamefont{Riotto}},
  \bibinfo{journal}{Phys. Rev.} \textbf{\bibinfo{volume}{D71}},
  \bibinfo{pages}{063534} (\bibinfo{year}{2005}), \eprint{astro-ph/0501562}.

\bibitem{Das:2010ts}
\bibinfo{author}{\bibfnamefont{S.}~\bibnamefont{Das}} \bibnamefont{and}
  \bibinfo{author}{\bibfnamefont{K.}~\bibnamefont{Sigurdson}},
  \bibinfo{journal}{Phys. Rev.} \textbf{\bibinfo{volume}{D85}},
  \bibinfo{pages}{063510} (\bibinfo{year}{2012}), \eprint{1012.4458}.

\bibitem{Schneider:2014rda}
\bibinfo{author}{\bibfnamefont{A.}~\bibnamefont{Schneider}}
  (\bibinfo{year}{2014}), \eprint{1412.2133}.

\bibitem{Tremaine:1979we}
\bibinfo{author}{\bibfnamefont{S.}~\bibnamefont{Tremaine}} \bibnamefont{and}
  \bibinfo{author}{\bibfnamefont{J.~E.} \bibnamefont{Gunn}},
  \bibinfo{journal}{Phys. Rev. Lett.} \textbf{\bibinfo{volume}{42}},
  \bibinfo{pages}{407} (\bibinfo{year}{1979}).

\bibitem{Boyarsky:2008ju}
\bibinfo{author}{\bibfnamefont{A.}~\bibnamefont{Boyarsky}},
  \bibinfo{author}{\bibfnamefont{O.}~\bibnamefont{Ruchayskiy}},
  \bibnamefont{and}
  \bibinfo{author}{\bibfnamefont{D.}~\bibnamefont{Iakubovskyi}},
  \bibinfo{journal}{JCAP} \textbf{\bibinfo{volume}{0903}}, \bibinfo{pages}{005}
  (\bibinfo{year}{2009}), \eprint{0808.3902}.

\bibitem{Blas:2011rf}
\bibinfo{author}{\bibfnamefont{D.}~\bibnamefont{Blas}},
  \bibinfo{author}{\bibfnamefont{J.}~\bibnamefont{Lesgourgues}},
  \bibnamefont{and} \bibinfo{author}{\bibfnamefont{T.}~\bibnamefont{Tram}},
  \bibinfo{journal}{JCAP} \textbf{\bibinfo{volume}{1107}}, \bibinfo{pages}{034}
  (\bibinfo{year}{2011}), \eprint{1104.2933}.

\bibitem{Dolgov:2002wy}
\bibinfo{author}{\bibfnamefont{A.~D.} \bibnamefont{Dolgov}},
  \bibinfo{journal}{Phys. Rept.} \textbf{\bibinfo{volume}{370}},
  \bibinfo{pages}{333} (\bibinfo{year}{2002}), \eprint{hep-ph/0202122}.

\bibitem{Vogel:2013raa}
\bibinfo{author}{\bibfnamefont{H.}~\bibnamefont{Vogel}} \bibnamefont{and}
  \bibinfo{author}{\bibfnamefont{J.}~\bibnamefont{Redondo}},
  \bibinfo{journal}{JCAP} \textbf{\bibinfo{volume}{1402}}, \bibinfo{pages}{029}
  (\bibinfo{year}{2014}), \eprint{1311.2600}.

\bibitem{Hasenkamp:2011em}
\bibinfo{author}{\bibfnamefont{J.}~\bibnamefont{Hasenkamp}},
  \bibinfo{journal}{Phys. Lett.} \textbf{\bibinfo{volume}{B707}},
  \bibinfo{pages}{121} (\bibinfo{year}{2012}), \eprint{1107.4319}.

\bibitem{Hooper:2011aj}
\bibinfo{author}{\bibfnamefont{D.}~\bibnamefont{Hooper}},
  \bibinfo{author}{\bibfnamefont{F.~S.} \bibnamefont{Queiroz}},
  \bibnamefont{and} \bibinfo{author}{\bibfnamefont{N.~Y.}
  \bibnamefont{Gnedin}}, \bibinfo{journal}{Phys. Rev.}
  \textbf{\bibinfo{volume}{D85}}, \bibinfo{pages}{063513}
  (\bibinfo{year}{2012}), \eprint{1111.6599}.

\bibitem{Kneller:2004jz}
\bibinfo{author}{\bibfnamefont{J.~P.} \bibnamefont{Kneller}} \bibnamefont{and}
  \bibinfo{author}{\bibfnamefont{G.}~\bibnamefont{Steigman}},
  \bibinfo{journal}{New J. Phys.} \textbf{\bibinfo{volume}{6}},
  \bibinfo{pages}{117} (\bibinfo{year}{2004}), \eprint{astro-ph/0406320}.

\bibitem{Kawasaki:2000en}
\bibinfo{author}{\bibfnamefont{M.}~\bibnamefont{Kawasaki}},
  \bibinfo{author}{\bibfnamefont{K.}~\bibnamefont{Kohri}}, \bibnamefont{and}
  \bibinfo{author}{\bibfnamefont{N.}~\bibnamefont{Sugiyama}},
  \bibinfo{journal}{Phys. Rev.} \textbf{\bibinfo{volume}{D62}},
  \bibinfo{pages}{023506} (\bibinfo{year}{2000}), \eprint{astro-ph/0002127}.

\bibitem{Hannestad:2004px}
\bibinfo{author}{\bibfnamefont{S.}~\bibnamefont{Hannestad}},
  \bibinfo{journal}{Phys. Rev.} \textbf{\bibinfo{volume}{D70}},
  \bibinfo{pages}{043506} (\bibinfo{year}{2004}), \eprint{astro-ph/0403291}.

\bibitem{Sarkar:1995dd}
\bibinfo{author}{\bibfnamefont{S.}~\bibnamefont{Sarkar}},
  \bibinfo{journal}{Rept. Prog. Phys.} \textbf{\bibinfo{volume}{59}},
  \bibinfo{pages}{1493} (\bibinfo{year}{1996}), \eprint{hep-ph/9602260}.

\bibitem{Fields:2014uja}
\bibinfo{author}{\bibfnamefont{B.~D.} \bibnamefont{Fields}},
  \bibinfo{author}{\bibfnamefont{P.}~\bibnamefont{Molaro}}, \bibnamefont{and}
  \bibinfo{author}{\bibfnamefont{S.}~\bibnamefont{Sarkar}},
  \bibinfo{journal}{Chin. Phys.} \textbf{\bibinfo{volume}{C38}}
  (\bibinfo{year}{2014}), \eprint{1412.1408}.

\bibitem{Cyburt:2004yc}
\bibinfo{author}{\bibfnamefont{R.~H.} \bibnamefont{Cyburt}},
  \bibinfo{author}{\bibfnamefont{B.~D.} \bibnamefont{Fields}},
  \bibinfo{author}{\bibfnamefont{K.~A.} \bibnamefont{Olive}}, \bibnamefont{and}
  \bibinfo{author}{\bibfnamefont{E.}~\bibnamefont{Skillman}},
  \bibinfo{journal}{Astropart. Phys.} \textbf{\bibinfo{volume}{23}},
  \bibinfo{pages}{313} (\bibinfo{year}{2005}), \eprint{astro-ph/0408033}.

\bibitem{Lisi:1999ng}
\bibinfo{author}{\bibfnamefont{E.}~\bibnamefont{Lisi}},
  \bibinfo{author}{\bibfnamefont{S.}~\bibnamefont{Sarkar}}, \bibnamefont{and}
  \bibinfo{author}{\bibfnamefont{F.~L.} \bibnamefont{Villante}},
  \bibinfo{journal}{Phys. Rev.} \textbf{\bibinfo{volume}{D59}},
  \bibinfo{pages}{123520} (\bibinfo{year}{1999}), \eprint{hep-ph/9901404}.

\bibitem{Mangano:2011ar}
\bibinfo{author}{\bibfnamefont{G.}~\bibnamefont{Mangano}} \bibnamefont{and}
  \bibinfo{author}{\bibfnamefont{P.~D.} \bibnamefont{Serpico}},
  \bibinfo{journal}{Phys. Lett.} \textbf{\bibinfo{volume}{B701}},
  \bibinfo{pages}{296} (\bibinfo{year}{2011}), \eprint{1103.1261}.

\bibitem{Izotov:2014fga}
\bibinfo{author}{\bibfnamefont{Y.~I.} \bibnamefont{Izotov}},
  \bibinfo{author}{\bibfnamefont{T.~X.} \bibnamefont{Thuan}}, \bibnamefont{and}
  \bibinfo{author}{\bibfnamefont{N.~G.} \bibnamefont{Guseva}},
  \bibinfo{journal}{Mon. Not. Roy. Astron. Soc.}
  \textbf{\bibinfo{volume}{445}}, \bibinfo{pages}{778} (\bibinfo{year}{2014}),
  \eprint{1408.6953}.

\bibitem{Cooke:2013cba}
\bibinfo{author}{\bibfnamefont{R.}~\bibnamefont{Cooke}},
  \bibinfo{author}{\bibfnamefont{M.}~\bibnamefont{Pettini}},
  \bibinfo{author}{\bibfnamefont{R.~A.} \bibnamefont{Jorgenson}},
  \bibinfo{author}{\bibfnamefont{M.~T.} \bibnamefont{Murphy}},
  \bibnamefont{and} \bibinfo{author}{\bibfnamefont{C.~C.}
  \bibnamefont{Steidel}}, \bibinfo{journal}{Astrophys. J.}
  \textbf{\bibinfo{volume}{781}}, \bibinfo{pages}{31} (\bibinfo{year}{2014}),
  \eprint{1308.3240}.

\bibitem{Kolb:1990vq}
\bibinfo{author}{\bibfnamefont{E.~W.} \bibnamefont{Kolb}} \bibnamefont{and}
  \bibinfo{author}{\bibfnamefont{M.~S.} \bibnamefont{Turner}},
  \bibinfo{journal}{Front. Phys.} \textbf{\bibinfo{volume}{69}},
  \bibinfo{pages}{1} (\bibinfo{year}{1990}).

\bibitem{Archidiacono:2013fha}
\bibinfo{author}{\bibfnamefont{M.}~\bibnamefont{Archidiacono}},
  \bibinfo{author}{\bibfnamefont{E.}~\bibnamefont{Giusarma}},
  \bibinfo{author}{\bibfnamefont{S.}~\bibnamefont{Hannestad}},
  \bibnamefont{and} \bibinfo{author}{\bibfnamefont{O.}~\bibnamefont{Mena}},
  \bibinfo{journal}{Adv. High Energy Phys.} \textbf{\bibinfo{volume}{2013}},
  \bibinfo{pages}{191047} (\bibinfo{year}{2013}), \eprint{1307.0637}.

\bibitem{Fischler:2010xz}
\bibinfo{author}{\bibfnamefont{W.}~\bibnamefont{Fischler}} \bibnamefont{and}
  \bibinfo{author}{\bibfnamefont{J.}~\bibnamefont{Meyers}},
  \bibinfo{journal}{Phys. Rev.} \textbf{\bibinfo{volume}{D83}},
  \bibinfo{pages}{063520} (\bibinfo{year}{2011}), \eprint{1011.3501}.

\bibitem{Foot:2011ve}
\bibinfo{author}{\bibfnamefont{R.}~\bibnamefont{Foot}}, \bibinfo{journal}{Phys.
  Lett.} \textbf{\bibinfo{volume}{B711}}, \bibinfo{pages}{238}
  (\bibinfo{year}{2012}), \eprint{1111.6366}.

\bibitem{Menestrina:2011mz}
\bibinfo{author}{\bibfnamefont{J.~L.} \bibnamefont{Menestrina}}
  \bibnamefont{and} \bibinfo{author}{\bibfnamefont{R.~J.}
  \bibnamefont{Scherrer}}, \bibinfo{journal}{Phys. Rev.}
  \textbf{\bibinfo{volume}{D85}}, \bibinfo{pages}{047301}
  (\bibinfo{year}{2012}), \eprint{1111.0605}.

\bibitem{DiBari:2013dna}
\bibinfo{author}{\bibfnamefont{P.}~\bibnamefont{Di~Bari}},
  \bibinfo{author}{\bibfnamefont{S.~F.} \bibnamefont{King}}, \bibnamefont{and}
  \bibinfo{author}{\bibfnamefont{A.}~\bibnamefont{Merle}},
  \bibinfo{journal}{Phys. Lett.} \textbf{\bibinfo{volume}{B724}},
  \bibinfo{pages}{77} (\bibinfo{year}{2013}), \eprint{1303.6267}.

\bibitem{Lopez-Val:2014jva}
\bibinfo{author}{\bibfnamefont{D.}~\bibnamefont{Lopez-Val}} \bibnamefont{and}
  \bibinfo{author}{\bibfnamefont{T.}~\bibnamefont{Robens}},
  \bibinfo{journal}{Phys. Rev.} \textbf{\bibinfo{volume}{D90}},
  \bibinfo{pages}{114018} (\bibinfo{year}{2014}), \eprint{1406.1043}.

\bibitem{Robens:2015gla}
\bibinfo{author}{\bibfnamefont{T.}~\bibnamefont{Robens}} \bibnamefont{and}
  \bibinfo{author}{\bibfnamefont{T.}~\bibnamefont{Stefaniak}},
  \bibinfo{journal}{Eur. Phys. J.}
  \textbf{\bibinfo{volume}{C75}}(\bibinfo{number}{3}), \bibinfo{pages}{104}
  (\bibinfo{year}{2015}), \eprint{1501.02234}.

\bibitem{Zeldovich:1974uw}
\bibinfo{author}{\bibfnamefont{Y.~B.} \bibnamefont{Zeldovich}},
  \bibinfo{author}{\bibfnamefont{I.~Y.} \bibnamefont{Kobzarev}},
  \bibnamefont{and} \bibinfo{author}{\bibfnamefont{L.~B.} \bibnamefont{Okun}},
  \bibinfo{journal}{Zh. Eksp. Teor. Fiz.} \textbf{\bibinfo{volume}{67}},
  \bibinfo{pages}{3} (\bibinfo{year}{1974}).

\bibitem{Preskill:1991kd}
\bibinfo{author}{\bibfnamefont{J.}~\bibnamefont{Preskill}},
  \bibinfo{author}{\bibfnamefont{S.~P.} \bibnamefont{Trivedi}},
  \bibinfo{author}{\bibfnamefont{F.}~\bibnamefont{Wilczek}}, \bibnamefont{and}
  \bibinfo{author}{\bibfnamefont{M.~B.} \bibnamefont{Wise}},
  \bibinfo{journal}{Nucl. Phys.} \textbf{\bibinfo{volume}{B363}},
  \bibinfo{pages}{207} (\bibinfo{year}{1991}).

\bibitem{Riva:2010jm}
\bibinfo{author}{\bibfnamefont{F.}~\bibnamefont{Riva}}, \bibinfo{journal}{Phys.
  Lett.} \textbf{\bibinfo{volume}{B690}}, \bibinfo{pages}{443}
  (\bibinfo{year}{2010}), \eprint{1004.1177}.

\bibitem{Dvali:1995cc}
\bibinfo{author}{\bibfnamefont{G.~R.} \bibnamefont{Dvali}} \bibnamefont{and}
  \bibinfo{author}{\bibfnamefont{G.}~\bibnamefont{Senjanovic}},
  \bibinfo{journal}{Phys. Rev. Lett.} \textbf{\bibinfo{volume}{74}},
  \bibinfo{pages}{5178} (\bibinfo{year}{1995}), \eprint{hep-ph/9501387}.

\bibitem{Dvali:1996zr}
\bibinfo{author}{\bibfnamefont{G.~R.} \bibnamefont{Dvali}},
  \bibinfo{author}{\bibfnamefont{A.}~\bibnamefont{Melfo}}, \bibnamefont{and}
  \bibinfo{author}{\bibfnamefont{G.}~\bibnamefont{Senjanovic}},
  \bibinfo{journal}{Phys. Rev.} \textbf{\bibinfo{volume}{D54}},
  \bibinfo{pages}{7857} (\bibinfo{year}{1996}), \eprint{hep-ph/9601376}.

\bibitem{Larsson:1996sp}
\bibinfo{author}{\bibfnamefont{S.~E.} \bibnamefont{Larsson}},
  \bibinfo{author}{\bibfnamefont{S.}~\bibnamefont{Sarkar}}, \bibnamefont{and}
  \bibinfo{author}{\bibfnamefont{P.~L.} \bibnamefont{White}},
  \bibinfo{journal}{Phys. Rev.} \textbf{\bibinfo{volume}{D55}},
  \bibinfo{pages}{5129} (\bibinfo{year}{1997}), \eprint{hep-ph/9608319}.

\bibitem{Hamaguchi:2011jy}
\bibinfo{author}{\bibfnamefont{K.}~\bibnamefont{Hamaguchi}},
  \bibinfo{author}{\bibfnamefont{T.}~\bibnamefont{Moroi}}, \bibnamefont{and}
  \bibinfo{author}{\bibfnamefont{K.}~\bibnamefont{Mukaida}},
  \bibinfo{journal}{JHEP} \textbf{\bibinfo{volume}{1201}}, \bibinfo{pages}{083}
  (\bibinfo{year}{2012}), \eprint{1111.4594}.

\bibitem{Gondolo1991145}
\bibinfo{author}{\bibfnamefont{P.}~\bibnamefont{Gondolo}} \bibnamefont{and}
  \bibinfo{author}{\bibfnamefont{G.}~\bibnamefont{Gelmini}},
  \bibinfo{journal}{Nuclear Physics B}
  \textbf{\bibinfo{volume}{360}}(\bibinfo{number}{1}), \bibinfo{pages}{145 }
  (\bibinfo{year}{1991}).

\end{thebibliography}

\end{document}